\documentclass[aps,prd,preprint,superscriptaddress,amssymb,amsmath,nofootinbib]{revtex4}

\usepackage{color}
\usepackage{epsfig}
\usepackage{graphicx}
\usepackage{float}
\usepackage{epstopdf}
\usepackage{hyperref} 
\usepackage{comment}
\usepackage[usenames,dvipsnames]{xcolor}
\usepackage{multirow}
\usepackage[normalem]{ulem}
\usepackage{slashed}
\usepackage{feynmf}
\usepackage{multirow}

\begin{document}

\preprint{}

\title{Two-Higgs-doublet model effective field theory}

\author{Radovan Dermisek}
\email[]{dermisek@iu.edu}
\affiliation{Physics Department, Indiana University, Bloomington, Indiana 47405, USA}

\author{Keith Hermanek}
\email[]{khermane@iu.edu}
\affiliation{Physics Department, Indiana University, Bloomington, Indiana 47405, USA}


\date{August 25, 2024}

\begin{abstract}

We construct the general two-Higgs doublet model effective field theory where  the effects of additional new physics are parametrized by operators up to mass dimension-six. We further transform this effective theory to the Higgs basis and provide matching of the Wilson coefficients between the two descriptions.  We illustrate the advantages of the Higgs basis which include the separation of operators that modify standard model couplings and masses from operators that contribute to scattering processes only, transparent correlations between scattering processes resulting from the same operator, and derivation of correlations between different operators in specific UV completions. For completeness, we also construct specific versions corresponding to four types of two-Higgs-doublet models: type-I, -II, -X, and -Y, distinguished by $Z_2$ symmetries which restrict the couplings of the Higgs doublets to standard model fermions. Furthermore, we derive general vacuum and stability conditions of the scalar potential in the presence of higher-dimensional terms.
\end{abstract}


\maketitle

\section{Introduction}

The effective field theories are very useful for general explorations of effects of new physics without relying on specific complete models. Among these, the standard model effective field theory (SMEFT) describes possible effects of new physics, provided that the low energy degrees of freedom are the known particles in the standard model (SM) and the new physics enters at a new scale, $\Lambda$, significantly above the electroweak (EW) scale. Below $\Lambda$, heavy degrees of freedom are integrated out and their effects are parametrized by nonrenormalizable, higher-dimensional operators with Wilson coefficients $C_i \propto \Lambda^{4-d}$, where $d > 4$ is the mass dimension of the operator~\cite{Appelquist:1974tg}. Constructing all operators up to mass dimension-six~\cite{Grzadkowski:2010es} and up to mass dimension-eight~\cite{Murphy:2020rsh} has led to a broad theoretical and experimental effort to constrain possible effects of new physics.

The two-Higgs-doublet model (2HDM) is one of the simplest extensions of the SM \cite{Gildener:1976ih}. While the new particles (two neutral and a pair of charged Higgs bosons) may be very heavy and thus their effects would be well described by the SMEFT resulting from such a model, they could also be light, within the reach of the Large Hadron Collider (LHC) or future colliders, in which case the new particles should be kept in the low energy theory. However, the 2HDM could by further extended by new particles, for example new quarks and leptons, an even richer scalar sector, or new gauge bosons. Just like for the SMEFT, it is possible to parametrize the effects of these extra particles by operators of higher dimensions, provided that the extra particles are above the scale of additional Higgs bosons. 

In this paper, we construct the general 2HDM effective field theory (2HDM EFT) where the low-energy theory is that of the 2HDM and the effects of additional new physics, possibly entering at a scale far above the masses of new Higgs bosons,  are parametrized by operators of mass dimension-five and -six. We assume the $SU(2)_L \times U(1)_Y$ theory is linear and all parameters in the scalar potential are  complex. We also construct specific versions of 2HDM EFTs, corresponding to four types of 2HDMs: type-I, -II, -X, and -Y, distinguished by $Z_2$ symmetries that restrict the couplings of Higgs doublets to SM fermions.  We will show that each model allows a unique set of operators as a consequence of specific $Z_2$ charge assignments for fermions. Furthermore, we transform the general 2HDM EFT to the Higgs basis and provide matching of the Wilson coefficients between the two descriptions. In the Higgs basis the SM degrees of freedom are contained in one doublet, $H_1$,  and all additional Higgses are in another doublet, $H_2$~\cite{Lavoura:1994fv,Lavoura:1994yu,Botella:1994cs,Davidson:2005cw}. We illustrate the advantages of working with the 2HDM EFT in the Higgs basis that include the separation of operators that modify SM couplings and masses from operators that contribute to scattering processes only, transparent correlations between scattering processes resulting from the same operator, and derivation of correlations between different operators in specific UV completions. For completeness, we also derive general vacuum and stability conditions of the scalar potential in the presence of higher-dimensional terms.

 Previously, the 2HDM EFT  has been studied in the case of a $CP$-conserving Higgs potential~\cite{Crivellin:2016ihg}. In this paper, we find agreement with the results therein, with the exception that in all four types of $CP$-conserving 2HDMs, we find twice as many operators which modify quark and lepton  masses, as well as operators involving covariant derivatives acting on either doublet contracted with right-handed quark currents.
 
 Other groups have extended this study toward a general 2HDM EFT~\cite{Karmakar:2017yek,Anisha:2019nzx}, where Ref.~\cite{Anisha:2019nzx}  has also counted the independent operators through the Hilbert series method. We note however several disagreements between these references, Ref. \cite{Crivellin:2016ihg}, and this paper, notably on constructing independent derivative operators involving the Higgs doublets. The 2HDM EFT in the Higgs basis has not been previously discussed.
 
 Other formalisms include a nonlinear type of EFT, called Higgs effective field theory, which assumes the Higgs is a SM gauge singlet and all additional SM scalars such as $G$ and $G^{\pm}$, longitudinal Goldstone modes of the $Z$ and $W^{\pm}$, are placed in an $SU(2)$ triplet~\cite{Feruglio:1992wf,Contino:2010mh}. Comparisons between this and the SM effective field theory have been collected in~Ref.~\cite{Brivio:2017vri}. More recently, a 2HDM equivalent of this has been studied \cite{Buchalla:2023hqk}. 
 Furthermore, SMEFT resulting from theories with extended Higgs sectors has been presented in Refs.~\cite{Gorbahn:2015gxa,Belusca-Maito:2016dqe,DasBakshi:2024krs}. Integrating out  the additional Higgses is accomplished by working in the Higgs basis where the mass scale of heavy Higgses  is the  scale of new physics, $\Lambda$. In this paper, we make no assumptions about the masses of new Higgses that can be anywhere between the EW scale and the UV cutoff $\Lambda$.

This paper is organized as follows. In Sec.~\ref{sec:eft_cons}, we define all parameters and conventions, an construct the entire set of effective operators in the general 2HDM up to mass dimension-six. In Sec.~\ref{sec:eft_2hdm}, we discuss the four specific kinds of 2HDM EFTs resulting from specific $Z_2$ charge assignments  and discuss what operators are common to all 2HDMs and what are unique for a specified model. We transform the general 2HDM EFT into the Higgs basis, and discuss advantages of working in the Higgs basis in Sec.~\ref{sec:eft_hb}. We conclude in Sec.~\ref{sec:conc}. We also provide two appendixes: in Appendix~\ref{sec:scalar_cor} we derive general vacuum and stability conditions of the scalar potential in the presence of higher-dimensional terms; and in Appendix~\ref{sec:higgs_basis} we list the complete expressions for all dimension-four, -five, and -six terms translated to the Higgs basis.

\section{Constructing the Effective Field Theory}
\label{sec:eft_cons}
\subsection{The standard basis definitions and parameters}
To start, we adopt the conventions of \cite{Grzadkowski:2010es} and construct the basis of linearly independent operators in a general 2HDM similar to that of the SMEFT Warsaw basis. For a given UV completion, where heavy degrees of freedom are decoupled at some high scale $\Lambda$, the effects of new physics are parametrized by higher-dimensional terms appearing alongside the fully renormalizable two-Higgs doublet model. Up to mass dimension-six, the Lagrangian is defined as 
\begin{equation}
    \mathcal{L} = \mathcal{L}_{2HDM}^{(4)} + \sum_{i} C_{i}^{(5)} \mathcal{O}_i^{(5)} + \sum_{i} C_{i}^{(6)} \mathcal{O}_i^{(6)} + \mathcal{O}\left(\frac{1}{\Lambda^3} \right),
\end{equation}
where the superscript denotes the mass dimension of the interactions. $\mathcal{L}_{2HDM}^{(4)}$ is the renormalizable 2HDM Lagrangian, which contains gauge and Yukawa interactions of SM fermions, as well as the scalar potential. $C_i^{(5)}$ and $C_i^{(6)}$ are the Wilson coefficients of the mass dimension-five and -six operators, respectively, and for compactness, the mass scale is contained in the Wilson coefficients.

The renormalizable 2HDM Lagrangian (summarized in \cite{Branco:2011iw} except for the mixed kinetic term) is given by
\begin{equation}
\begin{split}
    \mathcal{L}_{2HDM}^{(4)} & = - \frac{1}{4} B_{\mu \nu} B^{\mu \nu} - \frac{1}{4} W_{\mu \nu}^a W^{a \mu \nu} - \frac{1}{4} G_{\mu \nu}^a G^{a \mu \nu} \\
    & + (D_{\mu} \Phi_1)^{\dagger} D^{\mu} \Phi_1 + (D_{\mu} \Phi_2)^{\dagger} D^{\mu} \Phi_2 + \left(\eta (D_{\mu} \Phi_1)^{\dagger} D^{\mu} \Phi_2 + h.c.\right) - V(\Phi_1, \Phi_2) \\
    & + i \overline{l}_L \slashed{D} l_L + i \overline{e}_R \slashed{D} e_R + i \overline{q}_L \slashed{D} q_L + i \overline{d}_R \slashed{D} d_R + i \overline{u}_R \slashed{D} u_R \\
    & - \left(y_e^{(1)} \overline{l}_L e_R \Phi_1 + y_e^{(2)} \overline{l}_L e_R \Phi_2 + y_d^{(1)} \overline{q}_L d_R \Phi_1 + y_d^{(2)} \overline{q}_L d_R \Phi_2 \right. \\
    & \left. + y_u^{(1)} \overline{q}_L u_R \cdot \Phi_1^{\dagger} + y_u^{(2)} \overline{q}_L u_R \cdot \Phi_2^{\dagger} + h.c. \right),
\end{split} 
\label{eq:sm2hdmL}
\end{equation}
where the lepton and quark doublets are defined as $l_L = (\nu_L, e_L)^T$ and $q_L = (u_L, d_L)^T$. The gauge covariant derivative acting on an object charged under $SU(3) \times SU(2)_L \times U(1)_Y$ is defined as 
\begin{equation}
   (D_{\mu} q)_{\alpha i} = \left[\delta_{\alpha \beta} \delta_{ij} \partial_{\mu} + \frac{ig}{2} \delta_{\alpha \beta} (\tau^a)_{ij} W_{\mu}^a + \frac{i g_s}{2} (\lambda^a)_{\alpha \beta} \delta_{ij} G_{\mu}^a + i g' Y_q B_{\mu} \delta_{\alpha \beta} \delta_{ij}\right] q_{\beta j},
\end{equation}
where $\tau^a/2$ and $\lambda^a/2$ are the generators of $SU(2)$ and $SU(3)$, respectively, with $\tau^a$ and $\lambda^a$ being the Pauli matrices and Gell-Mann matrices. $SU(2)$ and $SU(3)$ fundamental indices are written with $(ij)$ and $(\alpha \beta)$ where needed, respectively. The hypercharges of the particle content are provided in Table~\ref{tab:Q_numbers}. Field strength tensors of the gauge fields are defined as
\begin{equation}
    \begin{split}
        & B_{\mu \nu} = \partial_{\mu} B_{\nu} - \partial_{\nu} B_{\mu}, \\
        & W_{\mu \nu}^a = \partial_{\mu} W_{\nu}^a - \partial_{\nu} W_{\mu}^a - g \epsilon^{abc} W_{\mu}^b W_{\nu}^c, \\
        & G_{\mu \nu}^a = \partial_{\mu} G_{\nu}^a - \partial_{\nu} G_{\mu}^a - g_s f^{abc} G_{\mu}^b G_{\nu}^c,
    \end{split}
\end{equation}
and for later purposes, the dual field strength tensor is $\widetilde{X}_{\mu \nu} = \epsilon_{\mu \nu \rho \sigma} X^{\rho \sigma} / 2$ for each gauge field $ X = B,W,G$ and $\epsilon_{0123} = + 1$.

The two Higgs fields $\Phi_{1,2}$ are complex $SU(2)$ doublets, each with hypercharge $Y_{1,2}  = + 1/2$, defined as \cite{Davidson:2005cw}
\begin{equation}
    \Phi_1 = \begin{pmatrix}
    \Phi_1^+ \\
    \Phi_1^0
    \end{pmatrix}, \ \ \ \ \ \Phi_2 = \begin{pmatrix}
    \Phi_2^+ \\
    \Phi_2^0
    \end{pmatrix}.
\label{eq:standard_d}
\end{equation}
For $SU(2)$ invariant combinations of doublets involving $\textbf{2} \times \textbf{2}$ or $\bar{\textbf{2}} \times \bar{\textbf{2}}$, contraction of $SU(2)$ indices with antisymmetric $\epsilon_{ij}$ are noted with an explicit ``$\cdot$'', e.g. $\overline{q}_{L} u_R \cdot \Phi_{1,2}^{\dagger} = (\overline{q}_{L})_1 u_R \epsilon_{12} (\Phi_{1,2}^{\dagger})_{2} + (\overline{q}_{L})_2 u_R \epsilon_{21} (\Phi_{1,2}^{\dagger})_{1} = \overline{u}_L u_R \Phi_{1,2}^{0*} - \overline{d}_L u_R \Phi_{1,2}^-$, where $\epsilon_{12} = - \epsilon_{21} = + 1$.

The most general scalar potential in a 2HDM is 
\begin{equation}
\begin{split}
    V(\Phi_1, \Phi_2) & = m_1^2 \left( \Phi_1^{\dagger} \Phi_1 \right) + m_2^2 \left( \Phi_2^{\dagger} \Phi_2 \right) + \left(  m_{12}^2 \Phi_1^{\dagger} \Phi_2 + h.c. \right) \\
    & + \frac{1}{2} \lambda_1 \left(\Phi_1^{\dagger} \Phi_1\right)^2 + \frac{1}{2} \lambda_2 \left(\Phi_2^{\dagger} \Phi_2 \right)^2 + \lambda_3 \left(\Phi_1^{\dagger} \Phi_1 \right) \left(\Phi_2^{\dagger} \Phi_2 \right) + \lambda_4 \left(\Phi_1^{\dagger} \Phi_2 \right)  \left(\Phi_2^{\dagger}\Phi_1 \right) \\
    & + \left( \frac{1}{2} \lambda_5 \left(\Phi_1^{\dagger} \Phi_2 \right)^2 + \lambda_6 (\Phi_1^{\dagger} \Phi_1)\Phi_1^{\dagger} \Phi_2 + \lambda_7 (\Phi_2^{\dagger} \Phi_2)\Phi_1^{\dagger} \Phi_2 + h.c.\right),
\label{eq:2hdm_pot}
\end{split}
\end{equation}
where the parameters $m_1^2, \ m_2^2, \ \lambda_1, \ \lambda_2, \ \lambda_3,$ and $\lambda_4$ are real by Hermiticity, while $m_{12}^2, \ \lambda_5, \ \lambda_6,$ and $\lambda_7$ are in general complex and can introduce $CP$-violating interactions in the scalar sector. Moreover, specifying a $Z_2$ symmetry on $\Phi_1$ and $\Phi_2$ restricts the potential to containing only $\lambda_5$ (and $m_{12}^2$ if soft-breaking terms are allowed).\footnote{The softly broken term is vital as it allows for a large mass spectrum for additional Higgses while respecting unitarity and perturbativity. When $m_{12}^2 \gg 8 \pi v^2$ for sufficiently heavy masses, $\tan \beta$ becomes largely unconstrained by direct searches \cite{Dermisek:2023tgq}.} Enforcing a $Z_2$ symmetry will be discussed in detail in the next section. This scalar potential will be modified in the presence of higher-dimensional terms, affecting the vacuum conditions of the extrema, concavity, and stability of the potential for asymptotically large values in field space. These conditions are collected in Appendix~\ref{sec:scalar_cor} and should be compared to the corresponding ones resulting from Eq.~(\ref{eq:2hdm_pot}) only \cite{Gunion:2002zf}.

\begin{table}[t!]
\centering
 \begin{tabular}{||c | c c c c c c c||} 
     \hline
     & $l_L$ \ \ & $e_R$ \ \ & $q_L$ \ \ & $u_R$ \ \ & $d_R$ \ \ & $\Phi_1$ \ \ & $\Phi_2$ \\
    \hline
    $SU(3)$ & $\textbf{1}$ \ \ & $\textbf{1}$ \ \ & $\textbf{3}$ \ \ & $\textbf{3}$ \ \ & $\textbf{3}$ \ \ & $\textbf{1}$ \ \ & $\textbf{1}$ \\
    $SU(2)_L$ & $\textbf{2}$ \ \ & $\textbf{1}$ \ \ & $\textbf{2}$ \ \ & $\textbf{1}$ \ \ & $\textbf{1}$ \ \ & $\textbf{2}$ \ \ & $\textbf{2}$ \\
    $U(1)_Y $ & $ -\frac{1}{2}$ \ \ & $-1$ \ \ & $ \frac{1}{6}$ \ \ & $\frac{2}{3}$ \ \ & $ -\frac{1}{3}$ \ \ & $\frac{1}{2}$ \ \ & $\frac{1}{2}$ \\ [1ex] 
 \hline    
\end{tabular}
    \caption{$SU(3) \times SU(2)_L \times U(1)_Y$ quantum numbers of standard model leptons, quarks, and Higgs doublets. The electric charge generated after electroweak symmetry breaking is $Q = T^3 + Y$. }
    \label{tab:Q_numbers}
\end{table}
The mixed kinetic term containing $\eta$ can be rotated away via the nonunitary transformation 
\begin{equation}
    (\Phi_1, \Phi_2) \rightarrow \left(\frac{\sqrt{\eta^*} \Phi_1 + \sqrt{\eta} \Phi_2}{2 \sqrt{|\eta|(1 + |\eta|)}} \pm \frac{\sqrt{\eta^*} \Phi_1 - \sqrt{\eta} \Phi_2}{2 \sqrt{|\eta|(1 - |\eta|)}} \right);
\end{equation}
however, in the most general sense (e.g. in the $CP$-violating 2HDM where $m_{12}^2, \lambda_{5,6,7} \in \mathbb{C}$), $\eta$ will be complex and generated at the loop level even in this diagonal basis. The renormalization of the 2HDM with a softly broken $Z_2$ symmetry was discussed in great detail in \cite{Krause:2016oke}. A general 2HDM can include this kinetic term or not by choosing different renormalization schemes, where these are related by scale-dependent field redefinitions \cite{Bijnens:2018rqw}. This detail about considering the mixed kinetic term was not mentioned previously in \cite{Karmakar:2017yek,Anisha:2019nzx} and we do so here to make the reader aware as we choose to include it. Including the mixed kinetic term in the Lagrangian is relevant for the equations of motion of $\Phi_1$ and $\Phi_2$, and after electroweak symmetry breaking (EWSB), the proper canonical definitions of scalar and vector boson fields. 

After EWSB, the neutral components of the scalar doublets develop vacuum expectation values (VEVs) $\langle \Phi^0_1 \rangle = v_1$ and $\langle \Phi^0_2 \rangle = v_2$, whereby $v = \sqrt{v_1^2 + v_2^2} = 174$ GeV and their ratio is parametrized by $v_2 / v_1 = \tan \beta$. The angle $\hat{\alpha}$ rotates the $CP$-even scalars $h$ and $H$ to the physical basis, while $\hat{\beta}$ diagonalizes the $CP$-odd scalars $G$ and $A$, and $\hat{\beta}^{\pm}$ diagonalizes the charged fields $G^{\pm}$ and $H^{\pm}$, yielding the following doublet components\footnote{The presence of dimension-six terms affects the physical Higgs spectrum; canonical field redefinitions and corrections to their masses must be considered. The vacuum angle $\beta$ thus no longer diagonalizes the $CP$-odd scalars and charged fields in Eq.~(\ref{eq:components}) as in the dimension-four theory. The difference between the vacuum and each diagonalization angle is $\beta - \hat{\beta}^{(\pm)} \sim \mathcal{O}(v^4 / \Lambda^2 M^2)$, where $M^2$ is either $m_A^2$ or $m_{H^{\pm}}^2$ depending on the angle.}:
\begin{equation}
\begin{split}
    & \Phi_1^0 = v_1 + \frac{1}{\sqrt{2}} \left(-h \sin \hat{\alpha} + H \cos \hat{\alpha} \right) + \frac{i}{\sqrt{2}} \left(G \cos \hat{\beta} - A \sin \hat{\beta} \right), \\
    & \Phi_2^0 = v_2 + \frac{1}{\sqrt{2}} \left(h \cos \hat{\alpha} + H \sin \hat{\alpha} \right) + \frac{i}{\sqrt{2}} \left(G \sin \hat{\beta} + A \cos \hat{\beta} \right), \\
    & \Phi_1^{\pm} = G^{\pm}  \cos \hat{\beta}^{\pm} - H^{\pm} \sin \hat{\beta}^{\pm}, \\
    & \Phi_2^{\pm} = G^{\pm} \sin \hat{\beta}^{\pm} + H^{\pm} \cos \hat{\beta}^{\pm}, 
\label{eq:components}
\end{split}
\end{equation}
where $h$ is the SM Higgs field, $G$ and $G^{\pm}$ are the longitudinal Goldstone modes for the $Z$ and $W^{\pm}$ bosons, respectively, and $H,A$, and $H^{\pm}$ are the additional Higgs degrees of freedom. 

In the mass sector, note that the presence of two doublets does not guarantee a simultaneous diagonalization of the lepton and quark mass matrices $y_{(e,d,u)}^{(1)}$ and $y_{(e,d,u)}^{(2)}$. This introduces so-called flavor changing neutral currents of SM fermions, where tree-level vertices involving scalars are nondiagonal in flavor space. These can be large due to $\tan \beta $ enhancements, but are highly constrained by many experimental measurements. 

\subsection{The complete set of operators up to dimension-Six}
Given the quantum numbers of SM fields in addition to two new doublets defined in Table~\ref{tab:Q_numbers}, we now proceed to construct the entire set of independent operators. The procedure is a straightforward extension of \cite{Grzadkowski:2010es}, notably, one can imagine replacing the SM Higgs doublet with either $\Phi_1$ or $\Phi_2$ since they all have identical quantum numbers. However, this generalization introduces non-Hermitian terms which mix $\Phi_1$ and $\Phi_2$ currents, generating new singlet combinations.


In the SMEFT, there is only one dimension-five operator, the so-called Weinberg operator \cite{Weinberg:1979sa}. However, in the general 2HDM, there are three dimension-five operators present \cite{Liao:2010ny} which may affect the mass generation of the three neutrinos. They are defined as
\begin{equation}
\begin{split}
    & \mathcal{O}_{\nu \nu \Phi}^{(11)} = (\Phi_1 \cdot l_L)^T \textbf{C} (\Phi_1 \cdot l_L), \\
    & \mathcal{O}_{\nu \nu \Phi}^{(22)} = (\Phi_2 \cdot l_L)^T \textbf{C} (\Phi_2 \cdot l_L), \\
    & \mathcal{O}_{\nu \nu \Phi}^{(12)} = (\Phi_1 \cdot l_L)^T \textbf{C} (\Phi_2 \cdot l_L),
\end{split}
\end{equation}
where $\textbf{C}$ is the charge conjugation matrix in the Dirac representation, $\textbf{C} = i \gamma^2 \gamma^0$. The combination of $(\Phi_2 \cdot l_L)^T \textbf{C} (\Phi_1 \cdot l_L)$ is redundant with the last operator.

Now, we proceed with constructing operators at mass dimension-six. Operators involving fermions and either scalar doublets or vector bosons are provided in Table~\ref{tab:gen2hdmops1}, whereas purely bosonic operators are displayed in Tables~\ref{tab:gen2hdmops2} and ~\ref{tab:gen2hdmops3}, and four-fermion currents in Table~\ref{tab:gen2hdmops4}. There are a total of 80 + 34 + 76 + 38 = 228 operators (including Hermitian conjugation when needed); 144 more than the operators present in the SMEFT Warsaw basis. This agrees with the independent counting introduced via the 2HDM Hilbert series in Ref. \cite{Anisha:2019nzx}. The 38 four-fermion operators [including baryon-number ($B$)-violating terms] and 4 pure gauge operators are identical to the SMEFT Warsaw basis and are unaffected by the additional Higgs doublet. However, we will see in the next section that the four-fermion operators become restricted once couplings to fermions are specified by a $Z_2$ symmetry in different types of 2HDMs. 

While keeping the notation as close as possible to SMEFT operators in the Warsaw basis, we choose a simple labeling of operators where the superscript $(\cdot \ \cdot)$ represents the contraction of Higgs doublets, where the first entry corresponds to the doublet under Hermitian conjugation (except for operators containing $\epsilon$ in charged quark currents, e.g. $\mathcal{O}_{\Phi u d}^{(11)}$ in Table~\ref{tab:gen2hdmops2}). Labels containing square brackets denote operators contracting $SU(2)$ doublets with $\delta_{ij}$, $[1],$ or with $(\tau^a)_{ij}$, $[3]$. Subscripts labeled first with the lepton doublet, $l_L,$ or quark doublet, $q_L = (u_L, d_L)^T$ followed by a scalar doublet are contracted together in operators such as $\mathcal{O}_{(l,u,d) \Phi_1}^{(11)}$ in Table~\ref{tab:gen2hdmops1}. Furthermore, for simplicity, note that we suppress flavor indices in Tables~\ref{tab:gen2hdmops1},~\ref{tab:gen2hdmops2}, and~\ref{tab:gen2hdmops4}. Note that the counting of the 228 operators (as well as any subset of operators) does not include flavor indices.

Operators containing covariant derivatives acting on the Higgs doublets $\Phi_{1,2}$ are constructed in an antisymmetric way, i.e.
\begin{equation}
\begin{split}
    & \Phi_{1,2}^{\dagger}i \overleftrightarrow{D}_{\mu} \Phi_{1,2} \equiv i \left( \Phi_{1,2}^{\dagger} (D_{\mu} \Phi_{1,2}) - (D_{\mu} \Phi_{1,2})^{\dagger} \Phi_{1,2} \right), \\
    & \Phi_{1,2}^{\dagger}i \overleftrightarrow{D}^a_{\mu} \Phi_{1,2} \equiv i \left( \Phi_{1,2}^{\dagger} \tau^a (D_{\mu} \Phi_{1,2}) - (D_{\mu} \Phi_{1,2})^{\dagger} \tau^a \Phi_{1,2} \right).
\end{split}
\label{eq:lroperator}
\end{equation}
If the doublets in the operator are identical, the above equations are Hermitian. However, because $\Phi_1$ and $\Phi_2$ have the same quantum numbers, non-Hermitian combinations such as $\Phi_{1}^{\dagger}i \overleftrightarrow{D}_{\mu} \Phi_{2}$ are also permitted. Other combinations involving antisymmetric $\epsilon$ and covariant derivatives appear as $\Phi_{1,2} \cdot i \overleftrightarrow{D}_{\mu} \Phi_{1,2} = 2 \Phi_{1,2} \cdot i D_{\mu} \Phi_{1,2}$. The symmetric combination, $\left( \Phi_{1,2}^{\dagger} (D_{\mu} \Phi_{1,2}) + (D_{\mu} \Phi_{1,2})^{\dagger} \Phi_{1,2} \right) = \partial_{\mu} (\Phi_{1,2}^{\dagger} \Phi_{1,2})$, is used to construct the other independent combination of operators. All other conventions and definitions on objects other than the new Higgs doublets are defined in \cite{Grzadkowski:2010es}. 

\begin{table}[t!]
\centering
 \begin{tabular}{||l | l  l  l||} 
    \hline
    & $\mathcal{O}_{l \Phi_1}^{(11)} = \overline{l}_L e_R \Phi_1 (\Phi_1^{\dagger} \Phi_1)$ & $\mathcal{O}_{d \Phi_1}^{(11)} = \overline{q}_L d_R \Phi_1 (\Phi_1^{\dagger} \Phi_1)$ & $\mathcal{O}_{u \Phi_1}^{(11)} = \overline{q}_L u_R \cdot \Phi_1^{\dagger} (\Phi_1^{\dagger} \Phi_1)$ \\
    & $\mathcal{O}_{l \Phi_1}^{(22)} = \overline{l}_L e_R \Phi_1 (\Phi_2^{\dagger} \Phi_2)$ & $\mathcal{O}_{d \Phi_1}^{(22)} = \overline{q}_L d_R \Phi_1 (\Phi_2^{\dagger} \Phi_2)$ & $\mathcal{O}_{u \Phi_1}^{(22)} = \overline{q}_L u_R \cdot \Phi_1^{\dagger} (\Phi_2^{\dagger} \Phi_2)$ \\
    & $\mathcal{O}_{l \Phi_1}^{(21)} = \overline{l}_L e_R \Phi_1 (\Phi_2^{\dagger} \Phi_1)$ & $\mathcal{O}_{d \Phi_1}^{(21)} = \overline{q}_L d_R \Phi_1 (\Phi_2^{\dagger} \Phi_1)$ & $\mathcal{O}_{u \Phi_1}^{(21)} = \overline{q}_L u_R \cdot \Phi_1^{\dagger} (\Phi_2^{\dagger} \Phi_1)$ \\
    $\psi^2 \phi^3 $& $\mathcal{O}_{l \Phi_1}^{(12)} = \overline{l}_L e_R \Phi_1 (\Phi_1^{\dagger} \Phi_2)$ & $\mathcal{O}_{d \Phi_1}^{(12)} = \overline{q}_L d_R \Phi_1 (\Phi_1^{\dagger} \Phi_2)$ & $\mathcal{O}_{u \Phi_1}^{(12)} = \overline{q}_L u_R \cdot \Phi_1^{\dagger} (\Phi_1^{\dagger} \Phi_2)$ \\
    & $\mathcal{O}_{l \Phi_2}^{(22)} = \overline{l}_L e_R \Phi_2 (\Phi_2^{\dagger} \Phi_2)$ & $\mathcal{O}_{d \Phi_2}^{(22)} = \overline{q}_L d_R \Phi_2 (\Phi_2^{\dagger} \Phi_2)$ & $\mathcal{O}_{u \Phi_2}^{(22)} = \overline{q}_L u_R \cdot \Phi_2^{\dagger} (\Phi_2^{\dagger} \Phi_2)$ \\
    & $\mathcal{O}_{l \Phi_2}^{(11)} = \overline{l}_L e_R \Phi_2 (\Phi_1^{\dagger} \Phi_1)$ & $\mathcal{O}_{d \Phi_2}^{(11)} = \overline{q}_L d_R \Phi_2 (\Phi_1^{\dagger} \Phi_1)$ & $\mathcal{O}_{u \Phi_2}^{(11)} = \overline{q}_L u_R \cdot \Phi_2^{\dagger} (\Phi_1^{\dagger} \Phi_1)$ \\
    & $\mathcal{O}_{l \Phi_2}^{(21)} = \overline{l}_L e_R \Phi_2 (\Phi_2^{\dagger} \Phi_1)$ & $\mathcal{O}_{d \Phi_2}^{(21)} = \overline{q}_L d_R \Phi_2 (\Phi_2^{\dagger} \Phi_1)$ & $\mathcal{O}_{u \Phi_2}^{(21)} = \overline{q}_L u_R \cdot \Phi_2^{\dagger} (\Phi_2^{\dagger} \Phi_1)$ \\
    & $\mathcal{O}_{l \Phi_2}^{(12)} = \overline{l}_L e_R \Phi_2 (\Phi_1^{\dagger} \Phi_2)$ & $\mathcal{O}_{d \Phi_2}^{(12)} = \overline{q}_L d_R \Phi_2 (\Phi_1^{\dagger} \Phi_2)$ & $\mathcal{O}_{u \Phi_2}^{(12)} = \overline{q}_L u_R \cdot \Phi_2^{\dagger} (\Phi_1^{\dagger} \Phi_2)$ \\
    \hline
    & $\mathcal{O}_{l B \Phi_1} = \overline{l}_L \sigma^{\mu \nu} e_R \Phi_1 B_{\mu \nu}$ & $\mathcal{O}_{d B \Phi_1} = \overline{q}_L \sigma^{\mu \nu} d_R \Phi_1 B_{\mu \nu}$ & $\mathcal{O}_{u B \Phi_1} = \overline{q}_L \sigma^{\mu \nu} u_R \cdot \Phi_1^{\dagger} B_{\mu \nu}$ \\
    & $\mathcal{O}_{l W \Phi_1} = \overline{l}_L \sigma^{\mu \nu} e_R \tau^a \Phi_1 W^a_{\mu \nu}$ & $\mathcal{O}_{d W \Phi_1} = \overline{q}_L \sigma^{\mu \nu} d_R \tau^a \Phi_1 W^a_{\mu \nu}$ & $\mathcal{O}_{u W \Phi_1} = \overline{q}_L \sigma^{\mu \nu} u_R \tau^a \cdot \Phi_1^{\dagger} W^a_{\mu \nu}$ \\ 
    $\psi^2 X \phi$ & $\mathcal{O}_{l B \Phi_2} = \overline{l}_L \sigma^{\mu \nu} e_R \Phi_2 B_{\mu \nu}$ & $\mathcal{O}_{d G \Phi_1} = \overline{q}_L \sigma^{\mu \nu} \lambda^a d_R \Phi_1 G^a_{\mu \nu}$ & $\mathcal{O}_{u G \Phi_1} = \overline{q}_L \sigma^{\mu \nu} \lambda^a u_R \cdot \Phi_1^{\dagger} G^a_{\mu \nu}$ \\
    & $\mathcal{O}_{l W \Phi_2} = \overline{l}_L \sigma^{\mu \nu} e_R \tau^a \Phi_2 W^a_{\mu \nu}$ & $\mathcal{O}_{d B \Phi_2} = \overline{q}_L \sigma^{\mu \nu} d_R \Phi_2 B_{\mu \nu}$ & $\mathcal{O}_{u B \Phi_2} = \overline{q}_L \sigma^{\mu \nu} u_R \cdot \Phi_2^{\dagger} B_{\mu \nu}$ \\
    & & $\mathcal{O}_{d W \Phi_2} = \overline{q}_L \sigma^{\mu \nu} d_R \tau^a \Phi_2 W^a_{\mu \nu}$ & $\mathcal{O}_{u W \Phi_2} = \overline{q}_L \sigma^{\mu \nu} u_R \tau^a \cdot \Phi_2^{\dagger} W^a_{\mu \nu}$ \\
    & & $\mathcal{O}_{d G \Phi_2} = \overline{q}_L \sigma^{\mu \nu} \lambda^a d_R \Phi_2 G^a_{\mu \nu}$ & $\mathcal{O}_{u G \Phi_2} = \overline{q}_L \sigma^{\mu \nu} \lambda^a u_R \cdot \Phi_2^{\dagger} G^a_{\mu \nu}$ \\ [1ex]
      \hline
\end{tabular}
    \caption{Operators with left- and right-handed fermions with either scalar doublets or vector bosons in the standard basis for a general 2HDM. Each operator has a distinct Hermitian conjugate, giving a total of 80 operators. Dipole operators are defined with $\sigma^{\mu \nu} = i[\gamma^{\mu},\gamma^{\nu}]/2$.}
\label{tab:gen2hdmops1}
\end{table}
\begin{table}[H]
\centering
 \begin{tabular}{|| l | l | l||} 
    \hline
    & $\mathcal{O}_{\Phi e}^{(11)} = (\Phi_1^{\dagger} i \overleftrightarrow{D}_{\mu} \Phi_1)(\overline{e}_R \gamma^{\mu} e_R)$ & $\mathcal{O}_{\Phi d}^{(11)} = (\Phi_1^{\dagger} i \overleftrightarrow{D}_{\mu} \Phi_1)(\overline{d}_R \gamma^{\mu} d_R)$ \\
    & $\mathcal{O}_{\Phi e}^{(22)} =(\Phi_2^{\dagger} i \overleftrightarrow{D}_{\mu} \Phi_2)(\overline{e}_R \gamma^{\mu} e_R)$ & $\mathcal{O}_{\Phi d}^{(22)} =(\Phi_2^{\dagger} i \overleftrightarrow{D}_{\mu} \Phi_2)(\overline{d}_R \gamma^{\mu} d_R)$ \\
    & $\mathcal{O}_{\Phi e}^{(12)} =(\Phi_1^{\dagger} i \overleftrightarrow{D}_{\mu} \Phi_2)(\overline{e}_R \gamma^{\mu} e_R) + h.c.$ & $\mathcal{O}_{\Phi d}^{(12)} =(\Phi_1^{\dagger} i \overleftrightarrow{D}_{\mu} \Phi_2)(\overline{d}_R \gamma^{\mu} d_R) + h.c.$ \\
    & $\mathcal{O}_{\Phi l}^{(11) [1]} =(\Phi_1^{\dagger} i \overleftrightarrow{D}_{\mu} \Phi_1)(\overline{l}_L \gamma^{\mu} l_L)$ & $\mathcal{O}_{\Phi u}^{(11)} = (\Phi_1^{\dagger} i \overleftrightarrow{D}_{\mu} \Phi_1)(\overline{u}_R \gamma^{\mu} u_R)$ \\
    & $\mathcal{O}_{\Phi l}^{(22) [1]} =(\Phi_2^{\dagger} i \overleftrightarrow{D}_{\mu} \Phi_2)(\overline{l}_L \gamma^{\mu} l_L)$ & $\mathcal{O}_{\Phi u}^{(22)} = (\Phi_2^{\dagger} i \overleftrightarrow{D}_{\mu} \Phi_2)(\overline{u}_R \gamma^{\mu} u_R)$ \\
    & $\mathcal{O}_{\Phi l}^{(12) [1]} =(\Phi_1^{\dagger} i \overleftrightarrow{D}_{\mu} \Phi_2)(\overline{l}_L \gamma^{\mu} l_L) + h.c.$ & $\mathcal{O}_{\Phi u}^{(12)} = (\Phi_1^{\dagger} i \overleftrightarrow{D}_{\mu} \Phi_2)(\overline{u}_R \gamma^{\mu} u_R) + h.c.$ \\
    & $\mathcal{O}_{\Phi l}^{(11)[3]} =(\Phi_1^{\dagger} i \overleftrightarrow{D}_{\mu}^a \Phi_1)(\overline{l}_L \tau^a \gamma^{\mu} l_L)$ & $\mathcal{O}_{\Phi q}^{(11) [1]} =(\Phi_1^{\dagger} i \overleftrightarrow{D}_{\mu} \Phi_1)(\overline{q}_L \gamma^{\mu} q_L)$ \\
    $\psi^2 \phi^2 D$ & $\mathcal{O}_{\Phi l}^{(22)[3]} =(\Phi_2^{\dagger} i \overleftrightarrow{D}_{\mu}^a \Phi_2)(\overline{l}_L \tau^a \gamma^{\mu} l_L)$ &  $\mathcal{O}_{\Phi q}^{(22) [1]} =(\Phi_2^{\dagger} i \overleftrightarrow{D}_{\mu} \Phi_2)(\overline{q}_L \gamma^{\mu} q_L)$ \\
    & $\mathcal{O}_{\Phi l}^{(12) [3]} =(\Phi_1^{\dagger} i \overleftrightarrow{D}_{\mu}^a \Phi_2)(\overline{l}_L \tau^a \gamma^{\mu} l_L) + h.c.$ & $\mathcal{O}_{\Phi q}^{(12) [1]} =(\Phi_1^{\dagger} i \overleftrightarrow{D}_{\mu} \Phi_2)(\overline{q}_L \gamma^{\mu} q_L) + h.c.$ \\
    & & $\mathcal{O}_{\Phi q}^{(11) [3]} =(\Phi_1^{\dagger} i \overleftrightarrow{D}_{\mu}^a \Phi_1)(\overline{q}_L \tau^a \gamma^{\mu} q_L)$ \\
    & & $\mathcal{O}_{\Phi q}^{(22) [3]} =(\Phi_2^{\dagger} i \overleftrightarrow{D}_{\mu}^a \Phi_2)(\overline{q}_L \tau^a \gamma^{\mu} q_L)$ \\
    & & $\mathcal{O}_{\Phi q}^{(12) [3]} = (\Phi_1^{\dagger} i \overleftrightarrow{D}_{\mu}^a \Phi_2)(\overline{q}_L \tau^a \gamma^{\mu} q_L) + h.c.$ \\
    & & $\mathcal{O}_{\Phi u d}^{(11)} = (\Phi_1 \cdot i D_{\mu} \Phi_1)(\overline{u}_R \gamma^{\mu} d_R) + h.c.$ \\
    & & $\mathcal{O}_{\Phi u d}^{(22)} = (\Phi_2 \cdot i D_{\mu} \Phi_2)(\overline{u}_R \gamma^{\mu} d_R) + h.c.$ \\
    & & $\mathcal{O}_{\Phi u d}^{(21)} = (\Phi_2 i \cdot \overleftrightarrow{D}_{\mu} \Phi_1)(\overline{u}_R \gamma^{\mu} d_R) + h.c.$ \\ [1ex] 
 \hline    
\end{tabular}
    \caption{Operators with left- or right-handed fermion currents involving covariant derivatives in the standard basis for a general 2HDM. There are a total of 34 operators.}
\label{tab:gen2hdmops2}
\end{table}
\begin{table}[H]
\centering
 \begin{tabular}{||c | l l l||} 
    \hline
    $X^3$ & $\mathcal{O}_{W} = \epsilon^{abc} W_{\mu}^{a \nu} W_{\nu}^{b \sigma} W_{\sigma}^{c \mu}$ & $\mathcal{O}_{\widetilde{W}} = \epsilon^{abc} \widetilde{W}_{\mu}^{a \nu} W_{\nu}^{b \sigma} W_{\sigma}^{c \mu}$ & \\
    & $\mathcal{O}_{G} = f^{abc} G_{\mu}^{a \nu} G_{\nu}^{b \sigma} G_{\sigma}^{c \mu}$ & $\mathcal{O}_{\widetilde{G}} = f^{abc} \widetilde{G}_{\mu}^{a \nu} G_{\nu}^{b \sigma} G_{\sigma}^{c \mu}$ & \\
    \hline
    & $\mathcal{O}_{\Phi G}^{(11)} = (\Phi_1^{\dagger} \Phi_1) G_{\mu \nu}^a G^{a \mu \nu}$ & $\mathcal{O}_{\Phi W}^{(11)} = (\Phi_1^{\dagger} \Phi_1) W_{\mu \nu}^a W^{a \mu \nu}$ & \\
    & $\mathcal{O}_{\Phi G}^{(22)} = (\Phi_2^{\dagger} \Phi_2) G_{\mu \nu}^a G^{a \mu \nu}$ & $\mathcal{O}_{\Phi W}^{(22)} = (\Phi_2^{\dagger} \Phi_2) W_{\mu \nu}^a W^{a \mu \nu}$ & \\
    & $\mathcal{O}_{\Phi G}^{(21)} = (\Phi_2^{\dagger} \Phi_1) G_{\mu \nu}^a G^{a \mu \nu} + h.c.$ & $\mathcal{O}_{\Phi W}^{(21)} = (\Phi_2^{\dagger} \Phi_1) W_{\mu \nu}^a W^{a \mu \nu} + h.c.$ & \\
    & $\mathcal{O}_{\Phi \widetilde{G}}^{(11)} = (\Phi_1^{\dagger} \Phi_1) \widetilde{G}_{\mu \nu}^a G^{a \mu \nu}$ & $\mathcal{O}_{\Phi \widetilde{W}}^{(11)} = (\Phi_1^{\dagger} \Phi_1) \widetilde{W}_{\mu \nu}^a W^{a \mu \nu}$ & \\
    & $\mathcal{O}_{\Phi \widetilde{G}}^{(22)} = (\Phi_2^{\dagger} \Phi_2) \widetilde{G}_{\mu \nu}^a G^{a \mu \nu}$ & $\mathcal{O}_{\Phi \widetilde{W}}^{(22)} = (\Phi_2^{\dagger} \Phi_2) \widetilde{W}_{\mu \nu}^a W^{a \mu \nu}$ & \\
    $X^2 \phi^2$ & $\mathcal{O}_{\Phi \widetilde{G}}^{(21)} = (\Phi_2^{\dagger}\Phi_1) \widetilde{G}_{\mu \nu}^a G^{a \mu \nu} + h.c.$ & $\mathcal{O}_{\Phi \widetilde{W}}^{(21)} = (\Phi_2^{\dagger} \Phi_1) \widetilde{W}_{\mu \nu}^a W^{a \mu \nu} + h.c.$ & \\
    & $\mathcal{O}_{\Phi B}^{(11)} = (\Phi_1^{\dagger} \Phi_1) B_{\mu \nu} B^{\mu \nu}$ & $\mathcal{O}_{\Phi W B}^{(11)} = (\Phi_1^{\dagger} \tau^a \Phi_1) W_{\mu \nu}^a B^{\mu \nu}$ & \\
    & $\mathcal{O}_{\Phi B}^{(22)} = (\Phi_2^{\dagger} \Phi_2) B_{\mu \nu} B^{\mu \nu}$ & $\mathcal{O}_{\Phi W B}^{(22)} = (\Phi_2^{\dagger} \tau^a \Phi_2) W_{\mu \nu}^a B^{\mu \nu}$ & \\
    & $\mathcal{O}_{\Phi B}^{(21)} = (\Phi_2^{\dagger} \Phi_1) B_{\mu \nu} B^{\mu \nu} + h.c.$ & $\mathcal{O}_{\Phi W B}^{(21)} = (\Phi_2^{\dagger} \tau^a \Phi_1) W_{\mu \nu}^a B^{\mu \nu} + h.c.$ & \\
    & $\mathcal{O}_{\Phi \widetilde{B}}^{(11)} = (\Phi_1^{\dagger} \Phi_1) \widetilde{B}_{\mu \nu} B^{\mu \nu}$ & $\mathcal{O}_{\Phi \widetilde{W} B}^{(11)} = (\Phi_1^{\dagger} \tau^a \Phi_1) \widetilde{W}_{\mu \nu}^a B^{\mu \nu}$ & \\
    & $\mathcal{O}_{\Phi \widetilde{B}}^{(22)} = (\Phi_2^{\dagger} \Phi_2) \widetilde{B}_{\mu \nu} B^{\mu \nu}$ & $\mathcal{O}_{\Phi \widetilde{W} B}^{(22)} = (\Phi_2^{\dagger} \tau^a \Phi_2) \widetilde{W}_{\mu \nu}^a B^{\mu \nu}$ & \\
    & $\mathcal{O}_{\Phi \widetilde{B}}^{(21)} = (\Phi_2^{\dagger} \Phi_1) \widetilde{B}_{\mu \nu} B^{\mu \nu} + h.c.$ & $\mathcal{O}_{\Phi \widetilde{W} B}^{(21)} = (\Phi_2^{\dagger} \tau^a \Phi_1) \widetilde{W}_{\mu \nu}^a B^{\mu \nu} + h.c.$ & \\
    \hline
    & $\mathcal{O}_{\Phi \partial^2}^{(11)(11)} = \partial_{\mu} (\Phi_1^{\dagger} \Phi_1) \partial^{\mu} (\Phi_1^{\dagger} \Phi_1)$ & $\mathcal{O}_{\Phi D}^{(11)(11)} = (\Phi_1^{\dagger} \overleftrightarrow{D}_{\mu} \Phi_1) (\Phi_1^{\dagger} \overleftrightarrow{D}^{\mu} \Phi_1)$ & \\
    & $\mathcal{O}_{\Phi \partial^2}^{(22)(22)} = \partial_{\mu}(\Phi_2^{\dagger} \Phi_2) \partial^{\mu} (\Phi_2^{\dagger} \Phi_2)$ & $\mathcal{O}_{\Phi D}^{(22)(22)} = (\Phi_2^{\dagger} \overleftrightarrow{D}_{\mu} \Phi_2) (\Phi_2^{\dagger} \overleftrightarrow{D}^{\mu} \Phi_2)$ & \\
    & $\mathcal{O}_{\Phi \partial^2}^{(11)(22)} = \partial_{\mu} (\Phi_1^{\dagger} \Phi_1) \partial^{\mu} (\Phi_2^{\dagger} \Phi_2)$ & $\mathcal{O}_{\Phi D}^{(11)(22)} = (\Phi_1^{\dagger} \overleftrightarrow{D}_{\mu} \Phi_1) (\Phi_2^{\dagger} \overleftrightarrow{D}^{\mu} \Phi_2)$ & \\
    $\phi^4 D^2$ & $\mathcal{O}_{\Phi \partial^2}^{(21)(21)} = \partial_{\mu} (\Phi_2^{\dagger} \Phi_1) \partial^{\mu} (\Phi_2^{\dagger} \Phi_1) + h.c.$ & $\mathcal{O}_{\Phi D}^{(21)(21)} = (\Phi_2^{\dagger} \overleftrightarrow{D}_{\mu} \Phi_1) (\Phi_2^{\dagger} \overleftrightarrow{D}^{\mu} \Phi_1) + h.c.$ & \\
    & $\mathcal{O}_{\Phi \partial^2}^{(21)(12)} = \partial_{\mu} (\Phi_2^{\dagger} \Phi_1) \partial^{\mu} (\Phi_1^{\dagger} \Phi_2) $ & $\mathcal{O}_{\Phi D}^{(21)(12)} = (\Phi_2^{\dagger} \overleftrightarrow{D}_{\mu} \Phi_1) (\Phi_1^{\dagger} \overleftrightarrow{D}^{\mu} \Phi_2)$ & \\
    & $\mathcal{O}_{\Phi \partial^2}^{(21)(11)} = \partial_{\mu} (\Phi_2^{\dagger} \Phi_1) \partial^{\mu} (\Phi_1^{\dagger} \Phi_1) + h.c.$ & $\mathcal{O}_{\Phi D}^{(21)(11)} = (\Phi_2^{\dagger} \overleftrightarrow{D}_{\mu} \Phi_1) (\Phi_1^{\dagger} \overleftrightarrow{D}^{\mu} \Phi_1) + h.c.$ & \\
    & $\mathcal{O}_{\Phi \partial^2}^{(21)(22)} = \partial_{\mu} (\Phi_2^{\dagger} \Phi_1) \partial^{\mu} (\Phi_2^{\dagger} \Phi_2) + h.c.$ & $\mathcal{O}_{\Phi D}^{(21)(22)} = (\Phi_2^{\dagger} \overleftrightarrow{D}_{\mu} \Phi_1) (\Phi_2^{\dagger} \overleftrightarrow{D}^{\mu} \Phi_2) + h.c.$ & \\
    \hline
    & $\mathcal{O}_{\Phi}^{(11)(11)(11)} = (\Phi_1^{\dagger} \Phi_1)^3$ & $\mathcal{O}_{\Phi}^{(11)(21)(21)} = (\Phi_1^{\dagger} \Phi_1) (\Phi_2^{\dagger} \Phi_1)^2 + h.c.$ & \\
    & $\mathcal{O}_{\Phi}^{(11)(11)(22)} = (\Phi_1^{\dagger} \Phi_1)^2 (\Phi_2^{\dagger} \Phi_2)$ & $\mathcal{O}_{\Phi}^{(11)(21)(12)} = (\Phi_1^{\dagger} \Phi_1) (\Phi_2^{\dagger} \Phi_1) (\Phi_1^{\dagger} \Phi_2)$ & \\
    & $\mathcal{O}_{\Phi}^{(11)(22)(22)} = (\Phi_1^{\dagger} \Phi_1)(\Phi_2^{\dagger} \Phi_2)^2$ & $\mathcal{O}_{\Phi}^{(22)(21)(21)} = (\Phi_2^{\dagger} \Phi_2) (\Phi_2^{\dagger} \Phi_1)^2 + h.c.$ & \\
    $\phi^6$ & $\mathcal{O}_{\Phi}^{(11)(11)(21)} = (\Phi_1^{\dagger} \Phi_1)^2 (\Phi_2^{\dagger} \Phi_1) + h.c.$ & $\mathcal{O}_{\Phi}^{(22)(21)(12)} = (\Phi_2^{\dagger} \Phi_2) (\Phi_2^{\dagger} \Phi_1) (\Phi_1^{\dagger} \Phi_2)$ & \\
    & $\mathcal{O}_{\Phi}^{(22)(22)(21)} = (\Phi_2^{\dagger} \Phi_2)^2 (\Phi_2^{\dagger} \Phi_1) + h.c.$ & $\mathcal{O}_{\Phi}^{(21)(21)(21)} = (\Phi_2^{\dagger} \Phi_1)^3 + h.c.$ & \\
    & $\mathcal{O}_{\Phi}^{(22)(22)(22)} = (\Phi_2^{\dagger} \Phi_2)^3$ & $\mathcal{O}_{\Phi}^{(21)(21)(12)} = (\Phi_2^{\dagger} \Phi_1)^2 (\Phi_1^{\dagger} 
    \Phi_2) + h.c.$ & \\
    & & $\mathcal{O}_{\Phi}^{(11)(22)(21)} = (\Phi_1^{\dagger} \Phi_1) (\Phi_2^{\dagger} \Phi_2)(\Phi_2^{\dagger} \Phi_1) + h.c.$ & \\ [1ex] 
    \hline
    \end{tabular}
    \caption{All purely bosonic operators in the standard basis for a general 2HDM. There are a total of 76 operators.}
\label{tab:gen2hdmops3}
\end{table}
\begin{table}[H]
\centering
 \begin{tabular}{||l | l l||} 
      \hline
      & $\mathcal{O}_{ll} = (\overline{l}_L \gamma^{\mu} l_L)(\overline{l}_L \gamma_{\mu} l_L)$ & $\mathcal{O}_{ee} = (\overline{e}_R \gamma^{\mu} e_R)(\overline{e}_R \gamma_{\mu} e_R)$ \\
      & $\mathcal{O}_{qq}^{(1)} = (\overline{q}_L \gamma^{\mu} q_L)(\overline{q}_L \gamma_{\mu} q_L)$ & $\mathcal{O}_{dd} = (\overline{d}_R \gamma^{\mu} d_R)(\overline{d}_R \gamma_{\mu} d_R)$ \\
      & $\mathcal{O}_{qq}^{(3)} = (\overline{q}_L \tau^a \gamma^{\mu} q_L)(\overline{q}_L \tau^a \gamma_{\mu} q_L)$ & $\mathcal{O}_{uu} = (\overline{u}_R \gamma^{\mu} u_R)(\overline{u}_R \gamma_{\mu} u_R)$ \\
      & $\mathcal{O}_{lq}^{(1)} = (\overline{l}_L \gamma^{\mu} l_L)(\overline{q}_L \gamma_{\mu} q_L)$ & $\mathcal{O}_{ed} = (\overline{e}_R \gamma^{\mu} e_R)(\overline{d}_R \gamma_{\mu} d_R)$ \\
      & $\mathcal{O}_{le} = (\overline{l}_L \gamma^{\mu} l_L)(\overline{e}_R \gamma_{\mu} e_R)$ & $\mathcal{O}_{eu} = (\overline{e}_R \gamma^{\mu} e_R)(\overline{u}_R \gamma_{\mu} u_R)$ \\
      & $\mathcal{O}_{ld} = (\overline{l}_L \gamma^{\mu} l_L)(\overline{d}_R \gamma_{\mu} d_R)$ & $\mathcal{O}_{ud}^{(1)} = (\overline{u}_R \gamma^{\mu} u_R)(\overline{d}_R \gamma_{\mu} d_R)$ \\
      & $\mathcal{O}_{lu} = (\overline{l}_L \gamma^{\mu} l_L)(\overline{u}_R \gamma_{\mu} u_R)$ & $\mathcal{O}_{ud}^{(8)} = (\overline{u}_R \lambda^a \gamma^{\mu} u_R)(\overline{d}_R \lambda^a \gamma_{\mu} d_R)$ \\
      $\psi^4$ & $\mathcal{O}_{qe} = (\overline{q}_L \gamma^{\mu} q_L)(\overline{e}_R \gamma_{\mu} e_R)$ & $\mathcal{O}_{qd}^{(1)} = (\overline{q}_L \gamma^{\mu} q_L)(\overline{d}_R \gamma_{\mu} d_R)$ \\
      & $\mathcal{O}_{lq}^{(3)} = (\overline{l}_L \tau^a \gamma^{\mu} l_L)(\overline{q}_L \tau^a \gamma_{\mu} q_L)$ & $\mathcal{O}_{qd}^{(8)} = (\overline{q}_L \lambda^a \gamma^{\mu} q_L)(\overline{d}_R \lambda^a \gamma_{\mu} d_R)$ \\
      & & $\mathcal{O}_{qu}^{(1)} = (\overline{q}_L \gamma^{\mu} q_L)(\overline{u}_R \gamma_{\mu} u_R)$ \\
      & & $\mathcal{O}_{qu}^{(8)} = (\overline{q}_L \lambda^a \gamma^{\mu} q_L)(\overline{u}_R \lambda^a \gamma_{\mu} u_R)$ \\ 
      \hline
      & $\mathcal{O}_{ledq} = (\overline{l}_L e_R)(\overline{d}_R q_L) + h.c.$ & $\mathcal{O}_{duq} = \epsilon^{\alpha \beta \gamma} \epsilon_{jk}((d_R^{\alpha})^T \textbf{C} u_R^{\beta})((q_{L j}^{\gamma})^T \textbf{C} l_{L k}) + h.c.$ \\
      & $\mathcal{O}_{quqd}^{(1)} =(\overline{q}_L u_R) \cdot (\overline{q}_L d_R) + h.c.$ & $\mathcal{O}_{qqu} = \epsilon^{\alpha \beta \gamma} \epsilon_{jk}((q_{L j}^{\alpha})^T \textbf{C} q_{L k}^{\beta})((u_R^{\gamma})^T \textbf{C} e_R) + h.c.$ \\
      & $\mathcal{O}_{quqd}^{(8)} = (\overline{q}_L \lambda^a u_R) \cdot (\overline{q}_L \lambda^a d_R) + h.c.$ & $\mathcal{O}_{qqq} = \epsilon^{\alpha \beta \gamma} \epsilon_{jn} \epsilon_{km}((q_{L j}^{\alpha})^T \textbf{C} q_{L k}^{\beta})((q_{L m}^{\gamma})^T \textbf{C} l_{L n}) + h.c.$ \\
      & $\mathcal{O}_{lequ}^{(1)} = (\overline{l}_L e_R) \cdot (\overline{q}_L u_R) + h.c.$ & $\mathcal{O}_{duu} = \epsilon^{\alpha \beta \gamma} ((d_R^{\alpha})^T \textbf{C} u_R^{\beta})((u_R^{\gamma})^T \textbf{C} e_R) + h.c.$ \\
      & $\mathcal{O}_{lequ}^{(3)} = (\overline{l}_L \sigma^{\mu \nu} e_R) \cdot (\overline{q}_L \sigma_{\mu \nu} u_R) + h.c.$ & \\ [1ex] 
 \hline    
\end{tabular}
    \caption{All four-fermion operators for a general 2HDM, identical to the SMEFT Warsaw basis. There are 38 operators, where entries in the second column of the second row are $B$-violating terms.}
\label{tab:gen2hdmops4}
\end{table}
In general, the Wilson coefficients of operators in the class of $\psi^2 \phi^3$ and $\psi^2 X \phi$ are complex, and any operator labeled $\mathcal{O}_{\Phi}^{(21)}$ contained in $\psi^2 \phi^2 D$ is non-Hermitian. Derivative operators involving the same fermion object and same $\Phi_{1,2}$ currents are Hermitian. For purely bosonic operators $X^3$ and $ X^2 \phi^2,$ all operators are Hermitian with the exception of those labeled with $\mathcal{O}_{\Phi}^{(21)}$. $\phi^6$ operators labeled with an odd number of $\Phi_1$ or $\Phi_2$ in the label are complex. Lastly, $\psi^4$ operators with identical pairings of fermion fields in the first row of Table~\ref{tab:gen2hdmops4} are Hermitian. 

Before we move forward, some special care is needed to construct the $\phi^4 D^2$ operators. Notably, there seems to be disagreements between Refs. \cite{Karmakar:2017yek} and \cite{Anisha:2019nzx} for the general 2HDM EFT,  
while in the limit the 2HDM is $CP$-conserving, there is an additional disagreement with \cite{Crivellin:2016ihg} on those operators. Disagreements between Refs. \cite{Crivellin:2016ihg}, \cite{Karmakar:2017yek}, and \cite{Anisha:2019nzx} were discussed in the last reference. However, we find disagreement with all three references, and here we focus on the disagreements with Ref. \cite{Anisha:2019nzx}.

Through the counting of independent operators in the Hilbert series of the 2HDM, \cite{Anisha:2019nzx} has correctly counted the 14 + 6 (Hermitian conjugation) = 20 operators predicted. However, operators containing the $\Box = \partial^2$ operator called $\mathcal{O}_{\Box}^{11(22)} = (\Phi_1^{\dagger} \Phi_1) \Box (\Phi_2^{\dagger} \Phi_2)$ and $\mathcal{O}_{\Box}^{22(11)} = (\Phi_2^{\dagger} \Phi_2) \Box (\Phi_1^{\dagger} \Phi_1)$ therein are not independent from each other due to integration by parts, yielding a total derivative and the latter combination. Furthermore, their operator labeled $\mathcal{O}_{\phi D}^{1 2 (1)(2)} = (\Phi_1^{\dagger} \Phi_2) \left[(D^{\mu} \Phi_1)^{\dagger} (D_{\mu} \Phi_2) \right]$ can be rewritten via equations of motions in terms of box operator $(\Phi_1^{\dagger} \Phi_2) \Box (\Phi_1^{\dagger} \Phi_2)$, terms involving $\psi^2 \phi^3, \phi^6,$ and $\phi^4$, in which the first is the Hermitian conjugate of their $\mathcal{O}_{\Box}^{21(21)} = (\Phi_2^{\dagger} \Phi_1) \Box (\Phi_2^{\dagger} \Phi_1)$ operator. A proper way to construct linearly independent combinations is by separating $SU(2)$ invariant combinations through the derivative operator [Eq.~(\ref{eq:lroperator})] and its symmetric counterpart. We can begin with constructing all 16 of the following triplet combinations:
\begin{equation}
\begin{split}
    & (\Phi_1^{\dagger} \tau^a \Phi_1) ((D_{\mu} \Phi_1)^{\dagger} \tau^a (D^{\mu} \Phi_1)), \ \ \ \ \ \ \ \ \ \ \ \ \ \ \ \ \ \ (\Phi_2^{\dagger} \tau^a \Phi_2) ((D_{\mu} \Phi_2)^{\dagger} \tau^a (D^{\mu} \Phi_2)), \\
    & (\Phi_1^{\dagger} \tau^a \Phi_1) ((D_{\mu} \Phi_2)^{\dagger} \tau^a (D^{\mu} \Phi_2)), \ \ \ \ \ \ \ \ \ \ \ \ \ \ \ \ \ \ (\Phi_2^{\dagger} \tau^a \Phi_2) ((D_{\mu} \Phi_1)^{\dagger} \tau^a (D^{\mu} \Phi_1)), \\
    & (\Phi_2^{\dagger} \tau^a \Phi_1) ((D_{\mu} \Phi_2)^{\dagger} \tau^a (D^{\mu} \Phi_1)) + h.c., \ \ \ \ \ \ \ \ \ \ (\Phi_2^{\dagger} \tau^a \Phi_1) ((D_{\mu} \Phi_1)^{\dagger} \tau^a (D^{\mu} \Phi_2)) + h.c., \\
    & (\Phi_1^{\dagger} \tau^a \Phi_1) ((D_{\mu} \Phi_2)^{\dagger} \tau^a (D^{\mu} \Phi_1)) + h.c., \ \ \ \ \ \ \ \ \ \ (\Phi_2^{\dagger} \tau^a \Phi_2) ((D_{\mu} \Phi_2)^{\dagger} \tau^a (D^{\mu} \Phi_1)) + h.c., \\
    & (\Phi_2^{\dagger} \tau^a \Phi_1) ((D_{\mu} \Phi_1)^{\dagger} \tau^a (D^{\mu} \Phi_1)) + h.c., \ \ \ \ \ \ \ \ \ \ (\Phi_2^{\dagger} \tau^a \Phi_1) ((D_{\mu} \Phi_2)^{\dagger} \tau^a (D^{\mu} \Phi_2)) + h.c..
\end{split}
\end{equation}
By exploiting the $SU(2)$ relation $(\tau^a)_{ij} (\tau^a)_{kl} = 2 \delta_{il} \delta_{jk} - \delta_{ij} \delta_{kl}$, we obtain singlet terms in which by separating derivatives in symmetric and antisymmetric combinations, we then find the results listed in Table~\ref{tab:gen2hdmops3}. These are independent from one another and free of ambiguity from where the partial derivatives act on. Note that operators $\mathcal{O}_{\Phi \partial^2}^{(mn)(pq)}$ in the class of $\phi^4 D^2$, $\partial_{\mu} (\Phi^{\dagger}_m \Phi_n) \partial^{\mu} (\Phi^{\dagger}_p \Phi_q)$, can be equivalently written with box operator on \textit{either} singlet pair via a total derivative: $(\Phi^{\dagger}_m \Phi_n) \Box (\Phi^{\dagger}_p \Phi_q)$ or $(\Phi^{\dagger}_p \Phi_q) \Box (\Phi^{\dagger}_m \Phi_n)$. However, we favor our notation not only because it is the symmetric combination of covariant derivative operators which is by definition orthogonal to the antisymmetric combination, $\mathcal{O}_{\Phi D}^{(mn)(pq)}$, but also writing those kinds of operators in that form helps collect dimension-six terms contributing to field redefinitions of the physical scalar spectrum after electroweak symmetry breaking.

Mass operators, $\psi^2 \phi^3$, found in the first row of Table~\ref{tab:gen2hdmops1} now modify the masses of the SM fields, in addition to the dimension-four Yukawa terms in Eq.~(\ref{eq:sm2hdmL}). The mass matrices for each type of fermion with explicit flavor indices are given by
\begin{equation}
    \mathcal{L}_M = - \overline{e}_{L,a} (M_e)_{ab} e_{R,b} - \overline{d}_{L,a}(M_d)_{ab} d_{R,b} - \overline{u}_{L,a} (M_u)_{ab} u_{R,b} + h.c., \\
\end{equation}
where
\begin{equation}
\begin{split}
    (M_e)_{ab} & = v (y_{e, a}^{(1)} \cos \beta + y_{e, a}^{(2)} \sin \beta) \delta_{ab} \\
    & - v^3 \left( \cos^3 \beta (C_{l \Phi_1}^{(11)})_{ab} + \sin^2 \beta \cos \beta (C_{l \Phi_1}^{(22)})_{ab} + \sin \beta \cos^2 \beta (C_{l \Phi_1}^{(12)})_{ab} + \sin \beta \cos^2 \beta (C_{l \Phi_1}^{(21)})_{ab} \right. \\
    & \left. + \sin^3 \beta (C_{l \Phi_2}^{(22)})_{ab} + \sin \beta \cos^2 \beta (C_{l \Phi_2}^{(11)})_{ab} + \sin^2 \beta \cos \beta (C_{l \Phi_2}^{(12)})_{ab} + \sin^2 \beta \cos \beta (C_{l \Phi_2}^{(21)})_{ab} \right),
\end{split}
\end{equation}
\begin{equation}
\begin{split}
    (M_d)_{ab} & = v (y_{d, a}^{(1)} \cos \beta + y_{d, a}^{(2)} \sin \beta) \delta_{ab} \\
    & - v^3 \left(\cos^3 \beta (C_{d \Phi_1}^{(11)})_{ab} + \sin^2 \beta \cos \beta (C_{d \Phi_1}^{(22)})_{ab} + \sin \beta \cos^2 \beta (C_{d \Phi_1}^{(12)})_{ab} + \sin \beta \cos^2 \beta (C_{d \Phi_1}^{(21)})_{ab} \right. \\
    & \left. + \sin^3 \beta (C_{d \Phi_2}^{(22)})_{ab} + \sin \beta \cos^2 \beta (C_{d \Phi_2}^{(11)})_{ab} + \sin^2 \beta \cos \beta (C_{d \Phi_2}^{(12)})_{ab} + \sin^2 \beta \cos \beta (C_{d \Phi_2}^{(21)})_{ab} \right),
\end{split}
\end{equation}
and
\begin{equation}
\begin{split}
    (M_u)_{ab} & = v (y_{u, a}^{(1)} \cos \beta + y_{u, a}^{(2)} \sin \beta) \delta_{ab} \\
    & - v^3 \left(\cos^3 \beta (C_{u \Phi_1}^{(11)})_{ab} + \sin^2 \beta \cos \beta (C_{u \Phi_1}^{(22)})_{ab} + \sin \beta \cos^2 \beta (C_{u \Phi_1}^{(12)})_{ab} + \sin \beta \cos^2 \beta (C_{u \Phi_1}^{(21)})_{ab} \right. \\
    & \left. + \sin^3 \beta (C_{u \Phi_2}^{(22)})_{ab} + \sin \beta \cos^2 \beta (C_{u \Phi_2}^{(11)})_{ab} + \sin^2 \beta \cos \beta (C_{u \Phi_2}^{(12)})_{ab} + \sin^2 \beta \cos \beta (C_{u \Phi_2}^{(21)})_{ab} \right).
\end{split}
\end{equation}
Rotating the fields to the mass eigenstate basis can be done via bi-unitary transformations $\psi_L \rightarrow U_L \hat{\psi}_L$ and $\psi_R \rightarrow U_R \hat{\psi}_R$, whereby the unitary matrices $U_{L,R}$ are such that the diagonal entries of 
\begin{equation}
\begin{split}
    & \hat{M}_e = U_L^{e \dagger} M_e U_R^e = \textrm{diag}(m_e, m_{\mu}, m_{\tau}), \\
    & \hat{M}_d = U_L^{d \dagger} M_d U_R^d = \textrm{diag}(m_d, m_s, m_b), \\
    & \hat{M}_u = U_L^{u \dagger} M_u U_R^u = \textrm{diag}(m_u, m_c, m_t),
\end{split}
\end{equation}
are real and positive. To obtain the operators in Tables~\ref{tab:gen2hdmops1} and ~\ref{tab:gen2hdmops2} after field rotations, one can redefine the Yukawa couplings and Wilson coefficients to ones with carrots via 
\begin{equation}
\begin{split}
   & (\hat{y}_{(e,d,u)}^{(1,2)})_{ab} = (U_L^{(e,d,u) \dagger} y_{(e,d,u)}^{(1,2)} U_R^{(e,d,u)})_{ab}, \\
   & (\hat{C}_{\nu \nu \Phi}^{(11,22,12)})_{ab} = (U_L^{\nu T} C_{\nu \nu \Phi}^{(11,22,12)} U_L^{\nu})_{ab}, \\
   & (\hat{C}_{(l,d,u) \Phi_{(1,2)}}^{(11,22,12,21)})_{ab} = (U_L^{(e,d,u) \dagger} C^{(11,22,12,21)}_{(l,d,u) \Phi_{(1,2)}} U_R^{(e,d,u)})_{ab}, \\
   & (\hat{C}_{l (B,W) \Phi_{(1,2)}})_{ab} = (U_L^{e \dagger} C_{l (B,W) \Phi_{(1,2)}} U_R^{e})_{ab}, \\
   & (\hat{C}_{d (B,W,G) \Phi_{(1,2)}})_{ab} = (U_L^{d \dagger} C_{d (B,W,G) \Phi_{(1,2)}} U_R^{d})_{ab}, \\
   & (\hat{C}_{u (B,W,G) \Phi_{(1,2)}})_{ab} = (U_L^{u \dagger} C_{u (B,W,G) \Phi_{(1,2)}} U_R^{u})_{ab}, \\
   & (\hat{C}_{\Phi (e,d,u)}^{(11,22,12)})_{ab} = (U_R^{(e,d,u) \dagger} C_{\Phi (e,d,u)}^{(11,22,12)} U_R^{(e,d,u)})_{ab}, \\
   & (\hat{C}_{\Phi (l,q)}^{(11,22,12)[1]})_{ab} = (U_L^{(e,d) \dagger} C_{\Phi (l,q)}^{(11,22,12) [1]} U_L^{(e,d)})_{ab}, \\
   & (\hat{C}_{\Phi (l,q)}^{(11,22,12)[3]})_{ab} = (U_L^{(e,d) \dagger} C_{\Phi (l,q)}^{(11,22,12) [3]} U_L^{(e,d)})_{ab}, \\
   & (\hat{C}_{\Phi ud}^{(11,22,21)})_{ab} = (U_R^{u \dagger} C_{\Phi ud}^{(11,22,21)} U_R^{d})_{ab}. \\
\end{split}
\end{equation}
Four-fermion operators in \ref{tab:gen2hdmops4} are rotated just as in SMEFT and can be found in Ref. \cite{Dedes:2017zog}. 

Similarly, the Majorana neutrino mass matrix generated by the dimension-five operators is given by
\begin{equation}
    \mathcal{L}_M = - \frac{1}{2} \nu_{L,a}^{T} (M_{\nu})_{ab} \textbf{C} \nu_{L,b} + h.c.,
\end{equation}
where 
\begin{equation}
    (M_{\nu})_{ab} = - 2 v^2 \left( \cos^2 \beta (C_{\nu \nu \Phi}^{(11)})_{ab} + \sin^2 \beta (C_{\nu \nu \Phi}^{(22)})_{ab} + \sin \beta \cos \beta (C_{\nu \nu \Phi}^{(12)})_{ab} \right).
\end{equation}
The diagonal mass matrix is given by
\begin{equation}
    \hat{M}_{\nu} = U_L^{\nu T} M_{\nu} U_L^{\nu} = \textrm{diag}(m_{\nu_1}, m_{\nu_2}, m_{\nu_3}). \\
\end{equation}
We can also introduce $\overline{\nu}_L^c = \nu_L^T \textbf{C}$, in which case $\mathcal{L}_M$ can be written as $ - \overline{\nu}_L^c M_{\nu} \nu_{L} / 2 + h.c.$ (note that the Majorana condition is $\nu = \nu_L + \nu_L^c = \nu^c$).

\section{Specific Types of 2HDMs}
\label{sec:eft_2hdm}

Different types of 2HDMs are motivated by a variety of reasons. The type-I 2HDM is unique in the sense that all fermionic masses are generated by a single Higgs doublet $\Phi_2$, while the remaining $\Phi_1$ doublet affects the generation of $Z$ and $W^{\pm}$ masses \cite{Haber:1978jt}. 
The type-II 2HDM is among the most simple and well studied; phenomenologically popular \cite{Gunion:1989we} and well motivated in supersymmetric models \cite{Fayet:1974pd} where masses of the down quarks and leptons are generated from $\Phi_1$ while up-type quarks acquire masses from $\Phi_2$. The lepton-specific type of 2HDM (known also as type-X) has been attractive in explaining neutrino mass generation and dark matter \cite{Aoki:2008av,Goh:2009wg}. In this case, the doublet $\Phi_1$ is responsible for generating the lepton masses while the remaining $\Phi_2$ generates all quark masses. For example, both type-II and -X models have been known to generate enhanced lepton couplings by factors of $\tan \beta$, useful for explaining anomalous magnetic dipole moments of the electron and/or the muon \cite{Broggio:2014mna,Abe:2015oca,Dermisek:2020cod,Dermisek:2021ajd,Dermisek:2023tgq} (note that type-X models require a light scalar spectrum to accomplish this while simultaneously, couplings of additional Higgses to quarks are suppressed, allowing for weaker constraints than in type-II models). The last type of model is known as a flipped 2HDM (or type-Y), where $\Phi_2$ is responsible for generating the masses of the up and lepton sectors, while $\Phi_1$ couples only to down-type quarks \cite{Barnett:1983mm,Barnett:1984zy}. In this model, lepton and up-type couplings can be suppressed while down-type couplings can be simultaneously enhanced. For a comprehensive review of all types of 2HDMs and their phenomenology, see \cite{Branco:2011iw}.

In all types of 2HDMs, we assume the standard $Z_2$ symmetry on the doublets $\Phi_1 \rightarrow - \Phi_1$ and $\Phi_2 \rightarrow + \Phi_2$ \cite{Davidson:2005cw}. This requirement restricts terms involving $\lambda_6$ and $\lambda_7$ in the scalar potential [Eq.~(\ref{eq:2hdm_pot})] (and possible softly broken effects if enforced exactly). Note that there is a difference between Refs. \cite{Crivellin:2016ihg} and \cite{Karmakar:2017yek} with respect to resulting operators in $CP$-conserving 2HDMs, where the former mandates no mixed contracted pairs $(\Phi_{1,2}^{\dagger} \Phi_{2,1})$ and the latter uses the convention $\Phi_1 \rightarrow + \Phi_1$ and $\Phi_2 \rightarrow - \Phi_2$. Specifying how each of the doublets couple to SM fermions is provided in Table~\ref{tab:2hdmQ_numbers}, where we use the convention that $\Phi_2$ always couples to the up-type sector. In any of the four models, the $Z_2$ symmetry prevents flavor changing neutral currents among fermionic species at the tree level. The Yukawa interactions for each type of model are
\begin{table}[t!]
\centering
 \begin{tabular}{||c | c c c | c c c c c c c||} 
     \hline
     Model & $u$ & $d$ & $e$ & $l_L$ \ \ & $e_R$ \ \ & $q_L$ \ \ & $u_R$ \ \ & $d_R$ \ \ & $\Phi_1$ \ \ & $\Phi_2$ \\
    \hline
    Type-I & $\Phi_2$ & $\Phi_2$ & $\Phi_2$ & $+$ \ \ & $+$ \ \ & $+$ \ \ & $+$ \ \ & $+$ \ \ & $-$ \ \ & $+$ \\
    Type-II & $\Phi_2$ & $\Phi_1$ & $\Phi_1$ & $+$ \ \ & $-$ \ \ & $+$ \ \ & $+$ \ \ & $-$ \ \ & $-$ \ \ & $+$ \\
    Type-X (Lepton-specific) & $\Phi_2$ & $\Phi_2$ & $\Phi_1$ & $+$ \ \ & $-$ \ \ & $+$ \ \ & $+$ \ \ & $ + $ \ \ & $-$ \ \ & $+$ \\
    Type-Y (Flipped) & $\Phi_2$ & $\Phi_1$ & $\Phi_2$ & $+$ \ \ & $+$ \ \ & $+$ \ \ & $+$ \ \ & $-$ \ \ & $-$ \ \ & $+$ \\ [1ex] 
 \hline    
\end{tabular}
    \caption{$Z_2$ charge assignments of leptons, quarks, and Higgs doublets in each type of 2HDM. The middle column indicates which scalar doublet couples to each sector, where by convention, $\Phi_2$ always couples to the up-type sector, while $\Phi_1 \rightarrow - \Phi_1$ and $\Phi_2 \rightarrow + \Phi_2$.}
    \label{tab:2hdmQ_numbers}
\end{table}
\begin{equation}
\begin{split}
    & \textrm{Type-I}: \mathcal{L} \supset - y_e \overline{l}_L e_R \Phi_2 - y_d \overline{q}_L d_R \Phi_2 - y_u \overline{q}_L u_R \cdot \Phi_2^{\dagger} + h.c., \\ 
    & \textrm{Type-II}: \mathcal{L} \supset - y_e \overline{l}_L e_R \Phi_1 - y_d \overline{q}_L d_R \Phi_1 - y_u \overline{q}_L u_R \cdot \Phi_2^{\dagger} + h.c., \\
    & \textrm{Type-X}: \mathcal{L} \supset - y_e \overline{l}_L e_R \Phi_1 - y_d \overline{q}_L d_R \Phi_2 - y_u \overline{q}_L u_R \cdot \Phi_2^{\dagger} + h.c., \\
    & \textrm{Type-Y}: \mathcal{L} \supset - y_e \overline{l}_L e_R \Phi_2 - y_d \overline{q}_L d_R \Phi_1 - y_u \overline{q}_L u_R \cdot \Phi_2^{\dagger} + h.c., 
\end{split}
\end{equation}
where the superscripts $(1,2)$ on the Yukawa couplings are omitted since there is only one type of Yukawa matrix for each fermion. 
Under the $Z_2$ symmetry of $\Phi_1$ and $\Phi_2$, there are a total of 76 operators common in all 4 types of 2HDMs, collected in Table~\ref{tab:2hdm_common_ops1}, and includes operators involving left- or right-handed fermion currents. Note that operators of those kinds in the general 2HDM which involved odd pairs of $\Phi_1 \Phi_2^{(\dagger)}$ or $\Phi_1^{(\dagger)} \Phi_2$ are now forbidden. Model-specific operators are then collected in Table~\ref{tab:t1ops} for type-I couplings, Table~\ref{tab:t2ops} for type-II couplings, Table~\ref{tab:txops} for type-X couplings, and Table~\ref{tab:tyops} for type-Y couplings. A type-II 2HDM EFT is the most restrictive, containing $24 + h.c. = 48$ specific operators, whereas the type-I model is less restrictive having $31 + h.c. = 62$ specific operators. Note that all type-specific operators are non-Hermitian.

In the case that $m_{12}^2$ and $\lambda_5$ are real, the Higgs potential is $CP$-conserving, which was first studied in the context of a 2HDM EFT in \cite{Crivellin:2016ihg}. The authors comment on the couplings of doublets to SM fermions, restricting the couplings to only right-handed fields. After translating to their notation, we find agreement with the operators in Ref. \cite{Crivellin:2016ihg}, with the exception of several: $\mathcal{O}_{\Phi u d}^{(21)}$, which should be present in type-II and -Y models and moreover, they report only the mass operators $\mathcal{O}_{(l,d,u) \Phi_2}^{(22)}$ and $\mathcal{O}_{(l,d,u) \Phi_2}^{(11)}$ are present in a type-I model, $\mathcal{O}_{(l,d) \Phi_1}^{(11)}, \ \mathcal{O}_{(l,d) \Phi_1}^{(22)}, \ \mathcal{O}_{u \Phi_2}^{(22)}$, and $\mathcal{O}_{u \Phi_2}^{(11)}$ in a type-II, $\mathcal{O}_{(d,u) \Phi_2}^{(22)}, 
\ \mathcal{O}_{(d,u) \Phi_2}^{(11)}, \ \mathcal{O}_{l \Phi_1}^{(11)}$, and $\mathcal{O}_{l \Phi_1}^{(22)}$ in a type-X and finally in a type-Y model, $\mathcal{O}_{(l,u) \Phi_2}^{(22)}, \ \mathcal{O}_{(l,u) \Phi_2}^{(11)}, \ \mathcal{O}_{d \Phi_1}^{(11)}$, and $\mathcal{O}_{d \Phi_1}^{(22)}$. The additional $\psi^2 \phi^3$ operators we find were perhaps overlooked by mandating no mixed contracted pairs $(\Phi_{1,2}^{\dagger} \Phi_{2,1})$ in the operators. We find twice as many mass operators in every type of model. 

\begin{table}[H]
\centering
 \begin{tabular}{||c | l l||} 
      \hline
       & $\mathcal{O}_{\Phi \partial^2}^{(11)(11)} = \partial_{\mu} (\Phi_1^{\dagger} \Phi_1) \partial^{\mu} (\Phi_1^{\dagger} \Phi_1)$ & $\mathcal{O}_{\Phi \partial^2}^{(21)(12)} = \partial_{\mu}(\Phi_2^{\dagger} \Phi_1) \partial^{\mu} (\Phi_1^{\dagger} \Phi_2)$ \\
       & $\mathcal{O}_{\Phi \partial^2}^{(22)(22)} = \partial_{\mu} (\Phi_2^{\dagger} \Phi_2) \partial^{\mu} (\Phi_2^{\dagger} \Phi_2)$ & $\mathcal{O}_{\Phi \partial^2}^{(21)(21)} = \partial_{\mu} (\Phi_2^{\dagger} \Phi_1) \partial^{\mu} (\Phi_2^{\dagger} \Phi_1) + h.c.$ \\
       $\phi^4 D^2$ & $\mathcal{O}_{\Phi \partial^2}^{(11)(22)} = \partial_{\mu} (\Phi_1^{\dagger} \Phi_1) \partial^{\mu} (\Phi_2^{\dagger} \Phi_2)$ & \\
       & $\mathcal{O}_{\Phi D}^{(11)(11)} = (\Phi_1^{\dagger} \overleftrightarrow{D}_{\mu} \Phi_1) (\Phi_1^{\dagger} \overleftrightarrow{D}^{\mu} \Phi_1)$ & $\mathcal{O}_{\Phi D}^{(21)(12)} = (\Phi_2^{\dagger} \overleftrightarrow{D}_{\mu} \Phi_1) (\Phi_1^{\dagger} \overleftrightarrow{D}^{\mu} \Phi_2)$ \\
       & $\mathcal{O}_{\Phi D}^{(22)(22)} = (\Phi_2^{\dagger} \overleftrightarrow{D}_{\mu} \Phi_2) (\Phi_2^{\dagger} \overleftrightarrow{D}^{\mu} \Phi_2)$ & $\mathcal{O}_{\Phi D}^{(21)(21)} = (\Phi_2^{\dagger} \overleftrightarrow{D}_{\mu} \Phi_1) (\Phi_2^{\dagger} \overleftrightarrow{D}^{\mu} \Phi_1) + h.c.$ \\ 
       & $\mathcal{O}_{\Phi D}^{(11)(22)} = (\Phi_1^{\dagger} \overleftrightarrow{D}_{\mu} \Phi_1) (\Phi_2^{\dagger} \overleftrightarrow{D}^{\mu} \Phi_2)$ & \\
       \hline 
       & $\mathcal{O}_{\Phi}^{(11)(11)(11)} = (\Phi_1^{\dagger} \Phi_1)^3$ & $\mathcal{O}_{\Phi}^{(11)(21)(21)} = (\Phi_1^{\dagger} \Phi_1) (\Phi_2^{\dagger} \Phi_1)^2 + h.c. $ \\ 
       $\phi^6$ & $\mathcal{O}_{\Phi}^{(22)(22)(22)} = (\Phi_2^{\dagger} \Phi_2)^3$ & $\mathcal{O}_{\Phi}^{(22)(21)(21)} = (\Phi_2^{\dagger} \Phi_2) (\Phi_2^{\dagger} \Phi_1)^2 + h.c.$ \\
       & $\mathcal{O}_{\Phi}^{(11)(11)(22)} = (\Phi_1^{\dagger} \Phi_1)^2 (\Phi_2^{\dagger} \Phi_2)$ & $\mathcal{O}_{\Phi}^{(11)(21)(12)} = (\Phi_1^{\dagger} \Phi_1) (\Phi_2^{\dagger} \Phi_1) (\Phi_1^{\dagger} \Phi_2)$ \\
       & $\mathcal{O}_{\Phi}^{(11)(22)(22)} = (\Phi_1^{\dagger} \Phi_1)(\Phi_2^{\dagger} \Phi_2)^2$ & $\mathcal{O}_{\Phi}^{(22)(21)(12)} = (\Phi_2^{\dagger} \Phi_2) (\Phi_2^{\dagger} \Phi_1) (\Phi_1^{\dagger} \Phi_2)$ \\ 
       \hline 
       & $\mathcal{O}_{\Phi G}^{(11)} = (\Phi_1^{\dagger} \Phi_1) G_{\mu \nu}^a G^{a \mu \nu}$ & $\mathcal{O}_{\Phi \widetilde{G}}^{(11)} = (\Phi_1^{\dagger} \Phi_1) \widetilde{G}_{\mu \nu}^a G^{a \mu \nu}$ \\
       & $\mathcal{O}_{\Phi G}^{(22)} = (\Phi_2^{\dagger} \Phi_2) G_{\mu \nu}^a G^{a \mu \nu}$ & $\mathcal{O}_{\Phi \widetilde{G}}^{(22)} = (\Phi_2^{\dagger} \Phi_2) \widetilde{G}_{\mu \nu}^a G^{a \mu \nu}$ \\
       $X^2 \phi^2$ & $\mathcal{O}_{\Phi W}^{(11)} = (\Phi_1^{\dagger} \Phi_1) W_{\mu \nu}^a W^{a \mu \nu}$ & $\mathcal{O}_{\Phi \widetilde{W}}^{(11)} = (\Phi_1^{\dagger} \Phi_1) \widetilde{W}_{\mu \nu}^a W^{a \mu \nu}$ \\
       & $\mathcal{O}_{\Phi W}^{(22)} = (\Phi_2^{\dagger} \Phi_2) W_{\mu \nu}^a W^{a \mu \nu}$ & $\mathcal{O}_{\Phi \widetilde{W}}^{(22)} = (\Phi_2^{\dagger} \Phi_2) \widetilde{W}_{\mu \nu}^a W^{a \mu \nu}$ \\
       & $\mathcal{O}_{\Phi B}^{(11)} = (\Phi_1^{\dagger} \Phi_1) B_{\mu \nu} B^{\mu \nu}$ & $\mathcal{O}_{\Phi \widetilde{B}}^{(11)} = (\Phi_1^{\dagger} \Phi_1) \widetilde{B}_{\mu \nu} B^{\mu \nu}$ \\
       & $\mathcal{O}_{\Phi B}^{(22)} = (\Phi_2^{\dagger} \Phi_2) B_{\mu \nu} B^{\mu \nu}$ & $\mathcal{O}_{\Phi \widetilde{B}}^{(22)} = (\Phi_2^{\dagger} \Phi_2) \widetilde{B}_{\mu \nu} B^{\mu \nu}$ \\
       & $\mathcal{O}_{\Phi W B}^{(11)} = (\Phi_1^{\dagger} \tau^a \Phi_1) W_{\mu \nu}^a B^{\mu \nu}$ & $\mathcal{O}_{\Phi \widetilde{W} B}^{(11)} = (\Phi_1^{\dagger} \tau^a \Phi_1) \widetilde{W}_{\mu \nu}^a B^{\mu \nu}$ \\
       & $\mathcal{O}_{\Phi W B}^{(22)} = (\Phi_2^{\dagger} \tau^a \Phi_2) W_{\mu \nu}^a B^{\mu \nu}$ & $\mathcal{O}_{\Phi \widetilde{W} B}^{(22)} = (\Phi_2^{\dagger} \tau^a \Phi_2) \widetilde{W}_{\mu \nu}^a B^{\mu \nu}$ \\
       \hline
       & $\mathcal{O}_{\Phi e}^{(11)} = (\Phi_1^{\dagger} i \overleftrightarrow{D}_{\mu} \Phi_1)(\overline{e}_R \gamma^{\mu} e_R)$ & $\mathcal{O}_{\Phi l}^{(11)[1]} =(\Phi_1^{\dagger} i \overleftrightarrow{D}_{\mu} \Phi_1)(\overline{l}_L \gamma^{\mu} l_L)$ \\
       & $\mathcal{O}_{\Phi e}^{(22)} =(\Phi_2^{\dagger} i \overleftrightarrow{D}_{\mu} \Phi_2)(\overline{e}_R \gamma^{\mu} e_R)$ & $\mathcal{O}_{\Phi l}^{(22)[1]} =(\Phi_2^{\dagger} i \overleftrightarrow{D}_{\mu} \Phi_2)(\overline{l}_L \gamma^{\mu} l_L)$ \\ 
       $\psi^2 \phi^2 D$ & $\mathcal{O}_{\Phi d}^{(11)} = (\Phi_1^{\dagger} i \overleftrightarrow{D}_{\mu} \Phi_1)(\overline{d}_R \gamma^{\mu} d_R)$ & $\mathcal{O}_{\Phi q}^{(11)[1]} =(\Phi_1^{\dagger} i \overleftrightarrow{D}_{\mu} \Phi_1)(\overline{q}_L \gamma^{\mu} q_L)$ \\
       & $\mathcal{O}_{\Phi d}^{(22)} =(\Phi_2^{\dagger} i \overleftrightarrow{D}_{\mu} \Phi_2)(\overline{d}_R \gamma^{\mu} d_R)$ & $\mathcal{O}_{\Phi q}^{(22)[1]} =(\Phi_2^{\dagger} i \overleftrightarrow{D}_{\mu} \Phi_2)(\overline{q}_L \gamma^{\mu} q_L)$ \\
       & $\mathcal{O}_{\Phi u}^{(11)} = (\Phi_1^{\dagger} i \overleftrightarrow{D}_{\mu} \Phi_1)(\overline{u}_R \gamma^{\mu} u_R)$ & $\mathcal{O}_{\Phi l}^{(11)[3]} =(\Phi_1^{\dagger} i \overleftrightarrow{D}_{\mu}^a \Phi_1)(\overline{l}_L \tau^a \gamma^{\mu} l_L)$ \\
       & $\mathcal{O}_{\Phi u}^{(22)} = (\Phi_2^{\dagger} i \overleftrightarrow{D}_{\mu} \Phi_2)(\overline{u}_R \gamma^{\mu} u_R)$ & $\mathcal{O}_{\Phi l}^{(22)[3]} =(\Phi_2^{\dagger} i \overleftrightarrow{D}_{\mu}^a \Phi_2)(\overline{l}_L \tau^a \gamma^{\mu} l_L)$ \\
       & & $\mathcal{O}_{\Phi q}^{(11)[3]} =(\Phi_1^{\dagger} i \overleftrightarrow{D}_{\mu}^a \Phi_1)(\overline{q}_L \tau^a \gamma^{\mu} q_L)$ \\
       & & $\mathcal{O}_{\Phi q}^{(22)[3]} =(\Phi_2^{\dagger} i \overleftrightarrow{D}_{\mu}^a \Phi_2)(\overline{q}_L \tau^a \gamma^{\mu} q_L)$ \\ [1ex] 
      \hline
       \end{tabular}
    \caption{Common operators in all types of 2HDMs. Note that all $X^3$ (4) and the first row of $\psi^4$ operators (20) are present, giving a total of 76 common operators.}
    \label{tab:2hdm_common_ops1}
\end{table}
\begin{table}[H]
\centering
 \begin{tabular}{||c | c c c||} 
      \hline
       & & Type-I (31 + $h.c.$ = 62) & \\
      \hline
       & $\mathcal{O}_{l \Phi_1}^{(21)} = \overline{l}_L e_R \Phi_1 (\Phi_2^{\dagger} \Phi_1)$ & $\mathcal{O}_{d \Phi_1}^{(21)} = \overline{q}_L d_R \Phi_1 (\Phi_2^{\dagger} \Phi_1)$ & $\mathcal{O}_{u \Phi_1}^{(21)} = \overline{q}_L u_R \cdot \Phi_1^{\dagger} (\Phi_2^{\dagger} \Phi_1)$ \\
       $\psi^2 \phi^3$ & $\mathcal{O}_{l \Phi_1}^{(12)} = \overline{l}_L e_R \Phi_1 (\Phi_1^{\dagger} \Phi_2)$ & $\mathcal{O}_{d \Phi_1}^{(12)} = \overline{q}_L d_R \Phi_1 (\Phi_1^{\dagger} \Phi_2)$ & $\mathcal{O}_{u \Phi_1}^{(12)} = \overline{q}_L u_R \cdot \Phi_1^{\dagger} (\Phi_1^{\dagger} \Phi_2)$ \\
       & $\mathcal{O}_{l \Phi_2}^{(22)} = \overline{l}_L e_R \Phi_2 (\Phi_2^{\dagger} \Phi_2)$ & $\mathcal{O}_{d \Phi_2}^{(22)} = \overline{q}_L d_R \Phi_2 (\Phi_2^{\dagger} \Phi_2)$ & $\mathcal{O}_{u \Phi_2}^{(22)} = \overline{q}_L u_R \cdot \Phi_2^{\dagger} (\Phi_2^{\dagger} \Phi_2)$ \\
       & $\mathcal{O}_{l \Phi_2}^{(11)} = \overline{l}_L e_R \Phi_2 (\Phi_1^{\dagger} \Phi_1)$ & $\mathcal{O}_{d \Phi_2}^{(11)} = \overline{q}_L d_R \Phi_2 (\Phi_1^{\dagger} \Phi_1)$ & $\mathcal{O}_{u \Phi_2}^{(11)} = \overline{q}_L u_R \cdot \Phi_2^{\dagger} (\Phi_1^{\dagger} \Phi_1)$ \\ 
       \hline 
       & $\mathcal{O}_{l B \Phi_2} = \overline{l}_L \sigma^{\mu \nu} e_R \Phi_2 B_{\mu \nu}$ & $\mathcal{O}_{d B \Phi_2} = \overline{q}_L \sigma^{\mu \nu} d_R \Phi_2 B_{\mu \nu}$ & $\mathcal{O}_{u B \Phi_2} = \overline{q}_L \sigma^{\mu \nu} u_R \cdot \Phi_2^{\dagger} B_{\mu \nu}$ \\ 
       $\psi^2 X \phi$ & $\mathcal{O}_{l W \Phi_2} = \overline{l}_L \sigma^{\mu \nu} e_R \tau^a \Phi_2 W^a_{\mu \nu}$ & $\mathcal{O}_{d W \Phi_2} = \overline{q}_L \sigma^{\mu \nu} d_R \tau^a \Phi_2 W^a_{\mu \nu}$ & $\mathcal{O}_{u W \Phi_2} = \overline{q}_L \sigma^{\mu \nu} u_R \tau^a \cdot \Phi_2^{\dagger} W^a_{\mu \nu}$ \\
       & & $\mathcal{O}_{d G \Phi_2} = \overline{q}_L \sigma^{\mu \nu} \lambda^a d_R \Phi_2 G^a_{\mu \nu}$ & $\mathcal{O}_{u G \Phi_2} = \overline{q}_L \sigma^{\mu \nu} \lambda^a u_R \cdot \Phi_2^{\dagger} G^a_{\mu \nu}$ \\ 
       [1ex]
       \hline
       \end{tabular}
       \begin{tabular}{||c | c c ||} 
        \hline
        $\psi^2 \phi^2 D$ & $\mathcal{O}_{\Phi u d}^{(11)} = (\Phi_1 \cdot i D_{\mu} \Phi_1)(\overline{u}_R \gamma^{\mu} d_R)$ & \ \ \ \ \ $\mathcal{O}_{\Phi u d}^{(22)} = (\Phi_2 \cdot i D_{\mu} \Phi_2)(\overline{u}_R \gamma^{\mu} d_R)$ \\
        \hline 
        & $\mathcal{O}_{ledq} = (\overline{l}_L e_R)(\overline{d}_R q_L)$ & \ \ \ \ \ $\mathcal{O}_{duq} = \epsilon^{\alpha \beta \gamma} \epsilon_{jk}((d_R^{\alpha})^T \textbf{C} u_R^{\beta})((q_{L j}^{\gamma})^T \textbf{C} l_{L k})$ \\
       & $\mathcal{O}_{quqd}^{(1)} =(\overline{q}_L u_R) \cdot (\overline{q}_L d_R)$ & \ \ \ \ \ $\mathcal{O}_{qqu} = \epsilon^{\alpha \beta \gamma} \epsilon_{jk}((q_{L j}^{\alpha})^T \textbf{C} q_{L k}^{\beta})((u_R^{\gamma})^T \textbf{C} e_R)$ \\
       $\psi^4$ & $\mathcal{O}_{quqd}^{(8)} = (\overline{q}_L \lambda^a u_R) \cdot (\overline{q}_L \lambda^a d_R)$ & \ \ \ \ \ $\mathcal{O}_{qqq} = \epsilon^{\alpha \beta \gamma} \epsilon_{jn} ((q_{L j}^{\alpha})^T \textbf{C} q_{L}^{\beta})\cdot ((q_{L}^{\gamma})^T \textbf{C} l_{L n})$ \\
       & $\mathcal{O}_{lequ}^{(1)} = (\overline{l}_L e_R) \cdot (\overline{q}_L u_R)$ & \ \ \ \ \ $\mathcal{O}_{duu} = \epsilon^{\alpha \beta \gamma} ((d_R^{\alpha})^T \textbf{C} u_R^{\beta})((u_R^{\gamma})^T \textbf{C} e_R)$ \\
       & $\mathcal{O}_{lequ}^{(3)} = (\overline{l}_L \sigma^{\mu \nu} e_R) \cdot (\overline{q}_L \sigma_{\mu \nu} u_R)$ & \\ [1ex]
       \hline
       \end{tabular}
    \caption{Type-I 2HDM-specific operators, each with a distinct Hermitian conjugate.}
    \label{tab:t1ops}
\end{table}
\begin{table}[H]
\centering
 \begin{tabular}{||c | c c c||} 
      \hline
        & & Type-II (24 + $h.c.$ = 48) & \\
      \hline
       & $\mathcal{O}_{l \Phi_1}^{(11)} = \overline{l}_L e_R \Phi_1 (\Phi_1^{\dagger} \Phi_1)$ & $\mathcal{O}_{d \Phi_1}^{(11)} = \overline{q}_L d_R \Phi_1 (\Phi_1^{\dagger} \Phi_1)$ & $\mathcal{O}_{u \Phi_1}^{(21)} = \overline{q}_L u_R \cdot \Phi_1^{\dagger} (\Phi_2^{\dagger} \Phi_1)$ \\
       $\psi^2 \phi^3$ & $\mathcal{O}_{l \Phi_1}^{(22)} = \overline{l}_L e_R \Phi_1 (\Phi_2^{\dagger} \Phi_2)$ & $\mathcal{O}_{d \Phi_1}^{(22)} = \overline{q}_L d_R \Phi_1 (\Phi_2^{\dagger} \Phi_2)$ & $\mathcal{O}_{u \Phi_1}^{(12)} = \overline{q}_L u_R \cdot \Phi_1^{\dagger} (\Phi_1^{\dagger} \Phi_2)$ \\
       & $\mathcal{O}_{l \Phi_2}^{(21)} = \overline{l}_L e_R \Phi_2 (\Phi_2^{\dagger} \Phi_1)$ & $\mathcal{O}_{d \Phi_2}^{(21)} = \overline{q}_L d_R \Phi_2 (\Phi_2^{\dagger} \Phi_1)$ & $\mathcal{O}_{u \Phi_2}^{(22)} = \overline{q}_L u_R \cdot \Phi_2^{\dagger} (\Phi_2^{\dagger} \Phi_2)$ \\
       & $\mathcal{O}_{l \Phi_2}^{(12)} = \overline{l}_L e_R \Phi_2 (\Phi_1^{\dagger} \Phi_2)$ & $\mathcal{O}_{d \Phi_2}^{(12)} = \overline{q}_L d_R \Phi_2 (\Phi_1^{\dagger} \Phi_2)$ & $\mathcal{O}_{u \Phi_2}^{(11)} = \overline{q}_L u_R \cdot \Phi_2^{\dagger} (\Phi_1^{\dagger} \Phi_1)$ \\ 
       \hline 
       & $\mathcal{O}_{l B \Phi_1} = \overline{l}_L \sigma^{\mu \nu} e_R \Phi_1 B_{\mu \nu}$ & $\mathcal{O}_{d B \Phi_1} = \overline{q}_L \sigma^{\mu \nu} d_R \Phi_1 B_{\mu \nu}$ & $\mathcal{O}_{u B \Phi_2} = \overline{q}_L \sigma^{\mu \nu} u_R \cdot \Phi_2^{\dagger} B_{\mu \nu}$ \\ 
       $\psi^2 X \phi$ & $\mathcal{O}_{l W \Phi_1} = \overline{l}_L \sigma^{\mu \nu} e_R \tau^a \Phi_1 W^a_{\mu \nu}$ & $\mathcal{O}_{d W \Phi_1} = \overline{q}_L \sigma^{\mu \nu} d_R \tau^a \Phi_1 W^a_{\mu \nu}$ & $\mathcal{O}_{u W \Phi_2} = \overline{q}_L \sigma^{\mu \nu} u_R \tau^a \cdot \Phi_2^{\dagger} W^a_{\mu \nu}$ \\
       & & $\mathcal{O}_{d G \Phi_1} = \overline{q}_L \sigma^{\mu \nu} \lambda^a d_R \Phi_1 G^a_{\mu \nu}$ & $\mathcal{O}_{u G \Phi_2} = \overline{q}_L \sigma^{\mu \nu} \lambda^a u_R \cdot \Phi_2^{\dagger} G^a_{\mu \nu}$ \\ [1ex]
       \hline
       \end{tabular}
       \begin{tabular}{||c | c  c ||} 
        \hline
        $\psi^2 \phi^2 D$ & $\mathcal{O}_{\Phi u d}^{(21)} = (\Phi_2 i \cdot \overleftrightarrow{D}_{\mu} \Phi_1)(\overline{u}_R \gamma^{\mu} d_R)$ & \\
        \hline 
        $\psi^4$ & $\mathcal{O}_{ledq} = (\overline{l}_L e_R)(\overline{d}_R q_L)$ & $\mathcal{O}_{qqq} = \epsilon^{\alpha \beta \gamma} \epsilon_{jn} ((q_{L j}^{\alpha})^T \textbf{C} q_{L}^{\beta})\cdot ((q_{L}^{\gamma})^T \textbf{C} l_{L n})$ \\
        & & $\mathcal{O}_{duu} = \epsilon^{\alpha \beta \gamma} ((d_R^{\alpha})^T \textbf{C} u_R^{\beta})((u_R^{\gamma})^T \textbf{C} e_R)$ \\ [1ex]
       \hline
       \end{tabular}
    \caption{Type-II 2HDM-specific operators, each with a distinct Hermitian conjugate.}
    \label{tab:t2ops}
\end{table}

\begin{table}[H]
\centering
 \begin{tabular}{||c | c c c||} 
      \hline
       & & Type-X (26 + $h.c.$ = 52) & \\
      \hline
       & $\mathcal{O}_{l \Phi_1}^{(11)} = \overline{l}_L e_R \Phi_1 (\Phi_1^{\dagger} \Phi_1)$ & $\mathcal{O}_{d \Phi_1}^{(21)} = \overline{q}_L d_R \Phi_1 (\Phi_2^{\dagger} \Phi_1)$ & $\mathcal{O}_{u \Phi_1}^{(21)} = \overline{q}_L u_R \cdot \Phi_1^{\dagger} (\Phi_2^{\dagger} \Phi_1)$ \\
       $\psi^2 \phi^3$ & $\mathcal{O}_{l \Phi_1}^{(22)} = \overline{l}_L e_R \Phi_1 (\Phi_2^{\dagger} \Phi_2)$ & $\mathcal{O}_{d \Phi_1}^{(12)} = \overline{q}_L d_R \Phi_1 (\Phi_1^{\dagger} \Phi_2)$ & $\mathcal{O}_{u \Phi_1}^{(12)} = \overline{q}_L u_R \cdot \Phi_1^{\dagger} (\Phi_1^{\dagger} \Phi_2)$ \\
       & $\mathcal{O}_{l \Phi_2}^{(21)} = \overline{l}_L e_R \Phi_2 (\Phi_2^{\dagger} \Phi_1)$ & $\mathcal{O}_{d \Phi_2}^{(22)} = \overline{q}_L d_R \Phi_2 (\Phi_2^{\dagger} \Phi_2)$ & $\mathcal{O}_{u \Phi_2}^{(22)} = \overline{q}_L u_R \cdot \Phi_2^{\dagger} (\Phi_2^{\dagger} \Phi_2)$ \\
       & $\mathcal{O}_{l \Phi_2}^{(12)} = \overline{l}_L e_R \Phi_2 (\Phi_1^{\dagger} \Phi_2)$ & $\mathcal{O}_{d \Phi_2}^{(11)} = \overline{q}_L d_R \Phi_2 (\Phi_1^{\dagger} \Phi_1)$ & $\mathcal{O}_{u \Phi_2}^{(11)} = \overline{q}_L u_R \cdot \Phi_2^{\dagger} (\Phi_1^{\dagger} \Phi_1)$ \\ 
       \hline 
       & $\mathcal{O}_{l B \Phi_1} = \overline{l}_L \sigma^{\mu \nu} e_R \Phi_1 B_{\mu \nu}$ & $\mathcal{O}_{d B \Phi_2} = \overline{q}_L \sigma^{\mu \nu} d_R \Phi_2 B_{\mu \nu}$ & $\mathcal{O}_{u B \Phi_2} = \overline{q}_L \sigma^{\mu \nu} u_R \cdot \Phi_2^{\dagger} B_{\mu \nu}$ \\ 
       $\psi^2 X \phi$ & $\mathcal{O}_{l W \Phi_1} = \overline{l}_L \sigma^{\mu \nu} e_R \tau^a \Phi_1 W^a_{\mu \nu}$ & $\mathcal{O}_{d W \Phi_2} = \overline{q}_L \sigma^{\mu \nu} d_R \tau^a \Phi_2 W^a_{\mu \nu}$ & $\mathcal{O}_{u W \Phi_2} = \overline{q}_L \sigma^{\mu \nu} u_R \tau^a \cdot \Phi_2^{\dagger} W^a_{\mu \nu}$  \\
       & & $\mathcal{O}_{d G \Phi_2} = \overline{q}_L \sigma^{\mu \nu} \lambda^a d_R \Phi_2 G^a_{\mu \nu}$ & $\mathcal{O}_{u G \Phi_2} = \overline{q}_L \sigma^{\mu \nu} \lambda^a u_R \cdot \Phi_2^{\dagger} G^a_{\mu \nu}$ \\ [1ex]
       \hline
       \end{tabular}
       \begin{tabular}{||c | c c ||} 
        \hline
        $\psi^2 \phi^2 D$ & $\mathcal{O}_{\Phi u d}^{(11)} = (\Phi_1 \cdot i D_{\mu} \Phi_1)(\overline{u}_R \gamma^{\mu} d_R)$ & \ \ \ \ \ $\mathcal{O}_{\Phi u d}^{(22)} = (\Phi_2 \cdot i D_{\mu} \Phi_2)(\overline{u}_R \gamma^{\mu} d_R)$ \\
        \hline 
        $\psi^4$  & $\mathcal{O}_{quqd}^{(1)} =(\overline{q}_L u_R) \cdot (\overline{q}_L d_R)$ & \ \ \ \ \ $\mathcal{O}_{duq} = \epsilon^{\alpha \beta \gamma} \epsilon_{jk}((d_R^{\alpha})^T \textbf{C} u_R^{\beta})((q_{L j}^{\gamma})^T \textbf{C} l_{L k})$ \\
       & $\mathcal{O}_{quqd}^{(8)} = (\overline{q}_L \lambda^a u_R) \cdot (\overline{q}_L \lambda^a d_R)$ & \ \ \ \ \ $\mathcal{O}_{qqq} = \epsilon^{\alpha \beta \gamma} \epsilon_{jn} ((q_{L j}^{\alpha})^T \textbf{C} q_{L}^{\beta})\cdot ((q_{L}^{\gamma})^T \textbf{C} l_{L n})$ \\ [1ex]
       \hline
       \end{tabular}
    \caption{Type-X 2HDM-specific operators, each with a distinct Hermitian conjugate.}
    \label{tab:txops}
\end{table}

\begin{table}[H]
\centering
 \begin{tabular}{||c | c c c||} 
      \hline
       & & Type-Y (25 + $h.c.$ = 50) & \\
      \hline
       & $\mathcal{O}_{l \Phi_1}^{(21)} = \overline{l}_L e_R \Phi_1 (\Phi_2^{\dagger} \Phi_1)$ & $\mathcal{O}_{d \Phi_1}^{(11)} = \overline{q}_L d_R \Phi_1 (\Phi_1^{\dagger} \Phi_1)$ & $\mathcal{O}_{u \Phi_1}^{(21)} = \overline{q}_L u_R \cdot \Phi_1^{\dagger} (\Phi_2^{\dagger} \Phi_1)$ \\
       $\psi^2 \phi^3$ & $\mathcal{O}_{l \Phi_1}^{(12)} = \overline{l}_L e_R \Phi_1 (\Phi_1^{\dagger} \Phi_2)$ & $\mathcal{O}_{d \Phi_1}^{(22)} = \overline{q}_L d_R \Phi_1 (\Phi_2^{\dagger} \Phi_2)$ & $\mathcal{O}_{u \Phi_1}^{(12)} = \overline{q}_L u_R \cdot \Phi_1^{\dagger} (\Phi_1^{\dagger} \Phi_2)$ \\
       & $\mathcal{O}_{l \Phi_2}^{(22)} = \overline{l}_L e_R \Phi_2 (\Phi_2^{\dagger} \Phi_2)$ & $\mathcal{O}_{d \Phi_2}^{(21)} = \overline{q}_L d_R \Phi_2 (\Phi_2^{\dagger} \Phi_1)$ & $\mathcal{O}_{u \Phi_2}^{(22)} = \overline{q}_L u_R \cdot \Phi_2^{\dagger} (\Phi_2^{\dagger} \Phi_2)$ \\
       & $\mathcal{O}_{l \Phi_2}^{(11)} = \overline{l}_L e_R \Phi_2 (\Phi_1^{\dagger} \Phi_1)$ & $\mathcal{O}_{d \Phi_2}^{(12)} = \overline{q}_L d_R \Phi_2 (\Phi_1^{\dagger} \Phi_2)$ & $\mathcal{O}_{u \Phi_2}^{(11)} = \overline{q}_L u_R \cdot \Phi_2^{\dagger} (\Phi_1^{\dagger} \Phi_1)$ \\ 
       \hline 
       & $\mathcal{O}_{l B \Phi_2} = \overline{l}_L \sigma^{\mu \nu} e_R \Phi_2 B_{\mu \nu}$ & $\mathcal{O}_{d B \Phi_1} = \overline{q}_L \sigma^{\mu \nu} d_R \Phi_1 B_{\mu \nu}$ & $\mathcal{O}_{u B \Phi_2} = \overline{q}_L \sigma^{\mu \nu} u_R \cdot \Phi_2^{\dagger} B_{\mu \nu}$ \\ 
       $\psi^2 X \phi$ & $\mathcal{O}_{l W \Phi_2} = \overline{l}_L \sigma^{\mu \nu} e_R \tau^a \Phi_2 W^a_{\mu \nu}$ & $\mathcal{O}_{d W \Phi_1} = \overline{q}_L \sigma^{\mu \nu} d_R \tau^a \Phi_1 W^a_{\mu \nu}$ & $\mathcal{O}_{u W \Phi_2} = \overline{q}_L \sigma^{\mu \nu} u_R \tau^a \cdot \Phi_2^{\dagger} W^a_{\mu \nu}$  \\
       & & $\mathcal{O}_{d G \Phi_1} = \overline{q}_L \sigma^{\mu \nu} \lambda^a d_R \Phi_1 G^a_{\mu \nu}$ & $\mathcal{O}_{u G \Phi_2} = \overline{q}_L \sigma^{\mu \nu} \lambda^a u_R \cdot \Phi_2^{\dagger} G^a_{\mu \nu}$ \\ [1ex]
       \hline
       \end{tabular}
       \begin{tabular}{||c | c c ||} 
        \hline
        $\psi^2 \phi^2 D$ & $\mathcal{O}_{\Phi u d}^{(21)} = (\Phi_2 i \cdot \overleftrightarrow{D}_{\mu} \Phi_1)(\overline{u}_R \gamma^{\mu} d_R)$ & \\
        \hline 
        $\psi^4$ & $\mathcal{O}_{lequ}^{(1)} = (\overline{l}_L e_R) \cdot (\overline{q}_L u_R)$
        & \ \ \ \ \ $\mathcal{O}_{qqu} = \epsilon^{\alpha \beta \gamma} \epsilon_{jk}((q_{L j}^{\alpha})^T \textbf{C} q_{L k}^{\beta})((u_R^{\gamma})^T \textbf{C} e_R)$ \\
       & $\mathcal{O}_{lequ}^{(3)} = (\overline{l}_L \sigma^{\mu \nu} e_R) \cdot (\overline{q}_L \sigma_{\mu \nu} u_R)$ & \ \ \ \ \ $\mathcal{O}_{qqq} = \epsilon^{\alpha \beta \gamma} \epsilon_{jn} ((q_{L j}^{\alpha})^T \textbf{C} q_{L}^{\beta})\cdot ((q_{L}^{\gamma})^T \textbf{C} l_{L n})$ \\ [1ex]
       \hline
       \end{tabular}
    \caption{Type-Y 2HDM-specific operators, each with a distinct Hermitian conjugate.}
    \label{tab:tyops}
\end{table}

\subsection{Comments on $H_d$ and $H_u$ notation for the type-II 2HDM}

It is well known that the Higgs potential of the minimal supersymmetric SM (MSSM) is that of a type-II 2HDM (see \cite{Martin:1997ns} for definitions). In the MSSM, the doublets are defined in a more suggestive way by

\begin{equation}
    H_d = \begin{pmatrix}
        H_d^+ \\
        H_d^0
    \end{pmatrix} \equiv \Phi_1 = \begin{pmatrix}
        \Phi_1^+ \\
        \Phi_1^0
    \end{pmatrix}, \ \ \ \ \ H_u = \begin{pmatrix}
        H_u^0 \\
        H_u^-
    \end{pmatrix} \equiv i \sigma^2 \Phi_2^{\dagger} = \begin{pmatrix}
        \Phi_2^{0*} \\
        - \Phi_2^-
    \end{pmatrix},
    \label{eq:mssm_d}
\end{equation}
where the subscripts denote which doublet explicitly couples to the down $(H_d)$ and up $(H_u)$ sectors, and $i \sigma^2 = \epsilon$. The hypercharges are then $Y_{H_d} = + 1/2$ and $Y_{H_u} = -1/2$. For a type-II 2HDM, one can make the appropriate replacements in Table~\ref{tab:t2ops} to this notation. These notations and definitions in the context of a type-II 2HDM as a low energy effective field theory were previously used in \cite{Dermisek:2023tgq} and \cite{Dermisek:2023rvv}, the former in the context of generating electric and magnetic dipole moments of the muon after integrating out heavy vectorlike leptons and the latter for signals of a modified Yukawa coupling. The notation of the mass operators therein are slightly different than presented here, utilizing $H_{d,u}$ explicitly: $\mathcal{O}_{l \Phi_1}^{(11)} \rightarrow \mathcal{O}_{l H_d} = \overline{l}_L e_R H_d (H_d^{\dagger} H_d), \ \mathcal{O}_{l \Phi_1}^{(22)} \rightarrow \mathcal{O}_{l H_u}^{(1)} = \overline{l}_L e_R H_d (H_u^{\dagger} H_u), \ \mathcal{O}_{l \Phi_2}^{(21)} \rightarrow \mathcal{O}_{l H_u}^{(2)} = \overline{l}_L e_R \cdot H_u^{\dagger} (H_d \cdot H_u)$, and $\mathcal{O}_{l \Phi_2}^{(12)} \rightarrow \mathcal{O}_{l H_u}^{(3)} = \overline{l}_L e_R \cdot H_u^{\dagger} (H_d^{\dagger} \cdot H_u^{\dagger})$. By exploiting the $SU(2)$ algebra $\epsilon_{ij} \epsilon_{kl} = \delta_{ik} \delta_{jl} - \delta_{il} \delta_{jk}$ on the second and third operators, one can perform a basis change and define a new operator as a linear combination of the two:
\begin{equation}
    \mathcal{O}'_{l H_u} = \mathcal{O}_{l H_u}^{(1)} - \mathcal{O}_{l H_u}^{(2)} = \overline{l}_L e_R H_u (H_u^{\dagger} H_d),
\label{eq:other_op}
\end{equation}
where indices are contracted purely with $\delta_{ij}$. After EWSB, this operator does not contribute to the lepton's mass. Note that in the standard notation, this operator corresponds to
\begin{equation}
    \mathcal{O}_{l \Phi}' = \mathcal{O}_{l \Phi_1}^{(22)} - \mathcal{O}_{l \Phi_2}^{(21)} = \overline{l}_L e_R \cdot \Phi_2^{\dagger} \left(\Phi_1 \cdot \Phi_2 \right).
\end{equation}
One can also perform this basis change through the $SU(2)$ algebra on other operators in favor of new combinations. Considering different bases may be useful when considering specific UV completions, for example, different representations of vectorlike leptons restrict which kind of (mass) operator can be generated at tree level. In a model where \textit{only} the operator in Eq.~(\ref{eq:other_op}) is generated, one can immediately see that working in the basis that contains this operator shows that there is no contribution to the lepton's mass. Otherwise, in the standard basis, one would find a precise cancellation after writing all contributions. Note that in all kinds of models, it is expected that each mass operator can be generated via loop corrections \cite{Dermisek:2023rvv,Dermisek:2023tgq}. 

Furthermore, in the MSSM, the couplings to SM fermions may appear as a type-II; however, they are not enforced by a $Z_2$ symmetry but rather the condition that the superpotential is holomorphic with respect to doublets $H_d$ and $H_u$ in Eq.~(\ref{eq:mssm_d}). Holomorphicity in the operators will drastically reduce the general list of 2HDM EFT operators to 31 operators containing only $\Phi_1 \rightarrow H_d$ and $\Phi_2^{\dagger} \rightarrow H_u$.

\section{The 2HDM EFT in the Higgs Basis}
\label{sec:eft_hb}

From Eq.~(\ref{eq:components}), one can see that the SM degrees of freedom $h, G, G^{\pm}$ are mixed with new scalars $H, A, H^{\pm}$, which in general can have masses anywhere from the EW scale up to the scale of new physics $\Lambda$. It is convenient to work in the Higgs basis \cite{Lavoura:1994fv,Lavoura:1994yu,Botella:1994cs,Davidson:2005cw} where, in the alignment limit, the SM fields and additional scalars are separated into two doublets $H_1$ and $H_2$. We can rotate the doublets $(\Phi_1, \Phi_2)$ via angle $\beta$ by
\begin{equation}
    \begin{pmatrix}
        H_1 \\
        H_2 
    \end{pmatrix} = \begin{pmatrix}
        \cos \beta & \sin \beta \\
        - \sin \beta & \cos \beta
    \end{pmatrix} \begin{pmatrix}
        \Phi_1 \\ 
        \Phi_2
    \end{pmatrix},
\label{eq:hb_transform}
\end{equation}
 which leads to the following new doublets $H_1$ and $H_2$:
\begin{equation}
\begin{split}
     H_1 & = \begin{pmatrix}
         G^+ \cos (\beta - \hat{\beta}^{\pm}) + H^+ \sin (\beta - \hat{\beta}^{\pm}) \\
         v + \frac{1}{\sqrt{2}} (h \sin (\beta - \hat{\alpha}) + H \cos (\beta - \hat{\alpha}) + i G \cos (\beta - \hat{\beta}) + i A \sin (\beta - \hat{\beta}))
     \end{pmatrix} \\
     & \rightarrow \begin{pmatrix}
         G^+ \\
         v + \frac{1}{\sqrt{2}} (h + i G)
     \end{pmatrix} + \mathcal{O}\left(\frac{v^4}{\Lambda^2 M^2} \right), \\
    H_2 & = \begin{pmatrix}
         - G^+ \sin(\beta - \hat{\beta}^{\pm}) +  H^+ \cos(\beta - \hat{\beta}^{\pm}) \\
         \frac{1}{\sqrt{2}} (h \cos (\beta - \hat{\alpha}) - H \sin (\beta - \hat{\alpha})  - i G \sin (\beta - \hat{\beta}) +  i A \cos(\beta - \hat{\beta}))
     \end{pmatrix} \\
     & \rightarrow \begin{pmatrix}
         H^+ \\
         \frac{1}{\sqrt{2}} (- H + i A)
     \end{pmatrix} + \mathcal{O}\left(\frac{v^4}{\Lambda^2 M^2} \right),
\end{split}
\end{equation}
where the arrow denotes evaluating $\beta - \hat{\alpha} \rightarrow \pi/2$ in the alignment limit \cite{Gunion:2002zf}. The alignment limit also enforces that couplings to the light eigenstate $h$ are SM-like, and the masses of $H, A,$ and $H^{\pm}$ are comparable. In this new basis $\langle H_1^0 \rangle = v $ and $\langle H_2^0 \rangle = 0 $, where $H_1$ now only contains the SM degrees of freedom and $H_2$ the remaining Higgs fields, up to terms of order $v^4 / \Lambda^2 M^2$. Misalignment of the angles $\beta \neq \hat{\beta}^{(\pm)}$ is expected when additional higher-dimensional operators are present.
Contributions from the dimension-six operators are suppressed by $v^2 / \Lambda^2$ in addition to $v^2 / M^2$, where $M^2$ is $m_A^2$ or $m_{H^{\pm}}^2$ depending on which rotation angle is present in the approximation of $\cos(\beta - \hat{\beta}^{(\pm)}) \simeq 1 + \mathcal{O}((v^4 / \Lambda^2 M^2)^2)$ and $\sin(\beta - \hat{\beta}^{(\pm)}) \simeq \mathcal{O}(v^4 / \Lambda^2 M^2)$. Inverting the transformations in Eq.~(\ref{eq:hb_transform}), we have
\begin{equation}
\Phi_1 = \cos \beta H_1 - \sin \beta H_2,
\label{eq:phi1_trans}
\end{equation}
\begin{equation}
\Phi_2 = \sin \beta H_1 + \cos \beta H_2.
\label{eq:phi2_trans}
\end{equation}
The complete lists of operators in the Higgs basis are provided in Tables~\ref{tab:hb_mass_ops},\ref{tab:hb_dipole_ops},\ref{tab:hb_der_ops},\ref{tab:hb_der_b_ops},\ref{tab:hb_scalar_ops}, and \ref{tab:hb_mix_ops},
and their Wilson coefficients written in terms of the coefficients in the standard basis are provided in Appendixes~\ref{sec:hb_mass_ops_match},~\ref{sec:hb_dipole_ops_match},~\ref{sec:hb_der_ops_match},~\ref{sec:hb_derb_ops_match},~\ref{sec:hb_scalar_ops_match}, and~\ref{sec:hb_mixed_ops_match}, for each table, respectively. Since dimension-five and -six operators scale as $1 / \Lambda$ and $1/ \Lambda^2$, respectively, the effects of the mismatch between diagonalization angles are of higher order. There are exactly the same number of operators in the Higgs basis as the standard basis at dimension-six, 228 (including Hermitian conjugates). However, in a specified type of 2HDM, all of the operators except four-fermion operators are identical; the type of model only restricts which Wilson coefficients contribute to operators in the Higgs basis. For convenience, the notation is identical to the general 2HDM of the main text with the replacement of $\Phi \rightarrow H$ in the label.

\begin{table}[H]
\centering
 \begin{tabular}{||l | c | l||} 
    \hline
        $\mathcal{O}_{l H_1}^{(11)} = \overline{l}_L e_R H_1 (H_1^{\dagger} H_1)$ & $\mathcal{O}_{d H_1}^{(11)} = \overline{q}_L d_R H_1 (H_1^{\dagger} H_1)$ & $\mathcal{O}_{u H_1}^{(11)} = \overline{q}_L u_R \cdot H_1^{\dagger} (H_1^{\dagger} H_1)$ \\
        $\mathcal{O}_{l H_1}^{(22)} = \overline{l}_L e_R H_1 (H_2^{\dagger} H_2)$ & $\mathcal{O}_{d H_1}^{(22)} = \overline{q}_L d_R H_1 (H_2^{\dagger} H_2)$ & $\mathcal{O}_{u H_1}^{(22)} = \overline{q}_L u_R \cdot H_1^{\dagger} (H_2^{\dagger} H_2)$ \\
        $\mathcal{O}_{l H_1}^{(21)} = \overline{l}_L e_R H_1 (H_2^{\dagger} H_1)$ & $\mathcal{O}_{d H_1}^{(21)} = \overline{q}_L d_R H_1 (H_2^{\dagger} H_1)$ & $\mathcal{O}_{u H_1}^{(21)} = \overline{q}_L u_R \cdot H_1^{\dagger} (H_2^{\dagger} H_1)$ \\
        $\mathcal{O}_{l H_1}^{(12)} = \overline{l}_L e_R H_1 (H_1^{\dagger} H_2)$ & $\mathcal{O}_{d H_1}^{(12)} = \overline{q}_L d_R H_1 (H_1^{\dagger} H_2)$ & $\mathcal{O}_{u H_1}^{(12)} = \overline{q}_L u_R \cdot H_1^{\dagger} (H_1^{\dagger} H_2)$ \\
        $\mathcal{O}_{l H_2}^{(22)} = \overline{l}_L e_R H_2 (H_2^{\dagger} H_2)$ & $\mathcal{O}_{d H_2}^{(22)} = \overline{q}_L d_R H_2 (H_2^{\dagger} H_2)$ & $\mathcal{O}_{u H_2}^{(22)} = \overline{q}_L u_R \cdot H_2^{\dagger} (H_2^{\dagger} H_2)$ \\
        $\mathcal{O}_{l H_2}^{(11)} = \overline{l}_L e_R H_2 (H_1^{\dagger} H_1)$ & $\mathcal{O}_{d H_2}^{(11)} = \overline{q}_L d_R H_2 (H_1^{\dagger} H_1)$ & $\mathcal{O}_{u H_2}^{(11)} = \overline{q}_L u_R \cdot H_2^{\dagger} (H_1^{\dagger} H_1)$ \\
        $\mathcal{O}_{l H_2}^{(21)} = \overline{l}_L e_R H_2 (H_2^{\dagger} H_1)$ & $\mathcal{O}_{d H_2}^{(21)} = \overline{q}_L d_R H_2 (H_2^{\dagger} H_1)$ & $\mathcal{O}_{u H_2}^{(21)} = \overline{q}_L u_R \cdot H_2^{\dagger} (H_2^{\dagger} H_1)$ \\
        $\mathcal{O}_{l H_2}^{(12)} = \overline{l}_L e_R H_2 (H_1^{\dagger} H_2)$ & $\mathcal{O}_{d H_2}^{(12)} = \overline{q}_L d_R H_2 (H_1^{\dagger} H_2)$ & $\mathcal{O}_{u H_2}^{(12)} = \overline{q}_L u_R \cdot H_2^{\dagger} (H_1^{\dagger} H_2)$ \\ [1ex] 
 \hline    
\end{tabular}
    \caption{Mass operators in the Higgs basis, all of which have a distinct Hermitian conjugate.}
\label{tab:hb_mass_ops}
\end{table}

\begin{table}[H]
\centering
 \begin{tabular}{||l | l | l||} 
    \hline
    $\mathcal{O}_{l B H_1} = \overline{l}_L \sigma^{\mu \nu} e_R H_1 B_{\mu \nu}$ & $\mathcal{O}_{d B H_1} = \overline{q}_L \sigma^{\mu \nu} d_R H_1 B_{\mu \nu}$ & $\mathcal{O}_{u B H_1} = \overline{q}_L \sigma^{\mu \nu} u_R \cdot H_1^{\dagger} B_{\mu \nu}$ \\
    $\mathcal{O}_{l W H_1} = \overline{l}_L \sigma^{\mu \nu} e_R \tau^a H_1 W^a_{\mu \nu}$ & $\mathcal{O}_{d W H_1} = \overline{q}_L \sigma^{\mu \nu} d_R \tau^a H_1 W^a_{\mu \nu}$ & $\mathcal{O}_{u W H_1} = \overline{q}_L \sigma^{\mu \nu} u_R \tau^a \cdot H_1^{\dagger} W^a_{\mu \nu}$ \\ 
    $\mathcal{O}_{l B H_2} = \overline{l}_L \sigma^{\mu \nu} e_R H_2 B_{\mu \nu}$ & $\mathcal{O}_{d G H_1} = \overline{q}_L \sigma^{\mu \nu} \lambda^a d_R H_1 G^a_{\mu \nu}$ & $\mathcal{O}_{u G H_1} = \overline{q}_L \sigma^{\mu \nu} \lambda^a u_R \cdot H_1^{\dagger} G^a_{\mu \nu}$ \\
    $\mathcal{O}_{l W H_2} = \overline{l}_L \sigma^{\mu \nu} e_R \tau^a H_2 W^a_{\mu \nu}$ & $\mathcal{O}_{d B H_2} = \overline{q}_L \sigma^{\mu \nu} d_R H_2 B_{\mu \nu}$ & $\mathcal{O}_{u B H_2} = \overline{q}_L \sigma^{\mu \nu} u_R \cdot H_2^{\dagger} B_{\mu \nu}$ \\
    & $\mathcal{O}_{d W H_2} = \overline{q}_L \sigma^{\mu \nu} d_R \tau^a H_2 W^a_{\mu \nu}$ & $\mathcal{O}_{u W H_2} = \overline{q}_L \sigma^{\mu \nu} u_R \tau^a \cdot H_2^{\dagger} W^a_{\mu \nu}$ \\
    & $\mathcal{O}_{d G H_2} = \overline{q}_L \sigma^{\mu \nu} \lambda^a d_R H_2 G^a_{\mu \nu}$ & $\mathcal{O}_{u G H_2} = \overline{q}_L \sigma^{\mu \nu} \lambda^a u_R \cdot H_2^{\dagger} G^a_{\mu \nu}$ \\ [1ex] 
 \hline    
\end{tabular}
    \caption{Dipole operators in the Higgs basis, all of which have a distinct Hermitian conjugate.}
\label{tab:hb_dipole_ops}
\end{table}

\begin{table}[H]
\centering
 \begin{tabular}{||l | l||} 
    \hline
    $\mathcal{O}_{H e}^{(11)} = (H_1^{\dagger} i \overleftrightarrow{D}_{\mu} H_1)(\overline{e}_R \gamma^{\mu} e_R)$ & $\mathcal{O}_{H d}^{(11)} = (H_1^{\dagger} i \overleftrightarrow{D}_{\mu} H_1)(\overline{d}_R \gamma^{\mu} d_R)$ \\
    $\mathcal{O}_{H e}^{(22)} =(H_2^{\dagger} i \overleftrightarrow{D}_{\mu} H_2)(\overline{e}_R \gamma^{\mu} e_R)$ & $\mathcal{O}_{H d}^{(22)} =(H_2^{\dagger} i \overleftrightarrow{D}_{\mu} H_2)(\overline{d}_R \gamma^{\mu} d_R)$ \\
    $\mathcal{O}_{H e}^{(12)} =(H_1^{\dagger} i \overleftrightarrow{D}_{\mu} H_2)(\overline{e}_R \gamma^{\mu} e_R) + h.c.$ & $\mathcal{O}_{H d}^{(12)} =(H_1^{\dagger} i \overleftrightarrow{D}_{\mu} H_2)(\overline{d}_R \gamma^{\mu} d_R) + h.c.$ \\
    $\mathcal{O}_{H l}^{(11) [1]} =(H_1^{\dagger} i \overleftrightarrow{D}_{\mu} H_1)(\overline{l}_L \gamma^{\mu} l_L)$ & $\mathcal{O}_{H u}^{(11)} = (H_1^{\dagger} i \overleftrightarrow{D}_{\mu} H_1)(\overline{u}_R \gamma^{\mu} u_R)$ \\
    $\mathcal{O}_{H l}^{(22) [1]} =(H_2^{\dagger} i \overleftrightarrow{D}_{\mu} H_2)(\overline{l}_L \gamma^{\mu} l_L)$ & $\mathcal{O}_{H u}^{(22)} = (H_2^{\dagger} i \overleftrightarrow{D}_{\mu} H_2)(\overline{u}_R \gamma^{\mu} u_R)$ \\
    $\mathcal{O}_{H l}^{(12) [1]} =(H_1^{\dagger} i \overleftrightarrow{D}_{\mu} H_2)(\overline{l}_L \gamma^{\mu} l_L) + h.c.$ & $\mathcal{O}_{H u}^{(12)} = (H_1^{\dagger} i \overleftrightarrow{D}_{\mu} H_2)(\overline{u}_R \gamma^{\mu} u_R) + h.c.$ \\
    $\mathcal{O}_{H l}^{(11) [3]} =(H_1^{\dagger} i \overleftrightarrow{D}_{\mu}^a H_1)(\overline{l}_L \tau^a \gamma^{\mu} l_L)$ & $\mathcal{O}_{H q}^{(11) [1]} =(H_1^{\dagger} i \overleftrightarrow{D}_{\mu} H_1)(\overline{q}_L \gamma^{\mu} q_L)$ \\
    $\mathcal{O}_{H l}^{(22) [3]} =(H_2^{\dagger} i \overleftrightarrow{D}_{\mu}^a H_2)(\overline{l}_L \tau^a \gamma^{\mu} l_L)$ &  $\mathcal{O}_{H q}^{(22) [1]} =(H_2^{\dagger} i \overleftrightarrow{D}_{\mu} H_2)(\overline{q}_L \gamma^{\mu} q_L)$ \\
    $\mathcal{O}_{H l}^{(12) [3]} =(H_1^{\dagger} i \overleftrightarrow{D}_{\mu}^a H_2)(\overline{l}_L \tau^a \gamma^{\mu} l_L) + h.c.$ & $\mathcal{O}_{H q}^{(12) [1]} =(H_1^{\dagger} i \overleftrightarrow{D}_{\mu} H_2)(\overline{q}_L \gamma^{\mu} q_L) + h.c.$ \\
    & $\mathcal{O}_{H q}^{(11) [3]} =(H_1^{\dagger} i \overleftrightarrow{D}_{\mu}^a H_1)(\overline{q}_L \tau^a \gamma^{\mu} q_L)$ \\
    & $\mathcal{O}_{H q}^{(22) [3]} =(H_2^{\dagger} i \overleftrightarrow{D}_{\mu}^a H_2)(\overline{q}_L \tau^a \gamma^{\mu} q_L)$ \\
    & $\mathcal{O}_{H q}^{(12) [3]} = (H_1^{\dagger} i \overleftrightarrow{D}_{\mu}^a H_2)(\overline{q}_L \tau^a \gamma^{\mu} q_L) + h.c.$ \\
    & $\mathcal{O}_{H u d}^{(11)} = (H_1 \cdot i D_{\mu} H_1)(\overline{u}_R \gamma^{\mu} d_R) + h.c.$ \\
    & $\mathcal{O}_{H u d}^{(22)} = (H_2 \cdot i D_{\mu} H_2)(\overline{u}_R \gamma^{\mu} d_R) + h.c.$ \\
    & $\mathcal{O}_{H u d}^{(21)} = (H_2 i \cdot \overleftrightarrow{D}_{\mu} H_1)(\overline{u}_R \gamma^{\mu} d_R) + h.c.$ \\ [1ex] 
 \hline    
\end{tabular}
    \caption{Derivative operators with fermionic currents in the Higgs basis.}
\label{tab:hb_der_ops}
\end{table}

\begin{table}[H]
\centering
 \begin{tabular}{||l | l||} 
    \hline 
    $\mathcal{O}_{H \partial^2}^{(11)(11)} = \partial_{\mu} (H_1^{\dagger} H_1) \partial^{\mu} (H_1^{\dagger} H_1)$ & $\mathcal{O}_{H D}^{(11)(11)} = (H_1^{\dagger} \overleftrightarrow{D}_{\mu} H_1) (H_1^{\dagger} \overleftrightarrow{D}^{\mu} H_1)$ \\
    $\mathcal{O}_{ H \partial^2}^{(22)(22)} = \partial_{\mu}(H_2^{\dagger} H_2) \partial^{\mu} (H_2^{\dagger} H_2)$ & $\mathcal{O}_{H D}^{(22)(22)} = (H_2^{\dagger} \overleftrightarrow{D}_{\mu} H_2) (H_2^{\dagger} \overleftrightarrow{D}^{\mu} H_2)$ \\
    $\mathcal{O}_{H \partial^2}^{(11)(22)} = \partial_{\mu} (H_1^{\dagger} H_1) \partial^{\mu} (H_2^{\dagger} H_2)$ & $\mathcal{O}_{H D}^{(11)(22)} = (H_1^{\dagger} \overleftrightarrow{D}_{\mu} H_1) (H_2^{\dagger} \overleftrightarrow{D}^{\mu} H_2)$ \\
    $\mathcal{O}_{H \partial^2}^{(21)(21)} = \partial_{\mu} (H_2^{\dagger} H_1) \partial^{\mu} (H_2^{\dagger} H_1) + h.c.$ & $\mathcal{O}_{H D}^{(21)(21)} = (H_2^{\dagger} \overleftrightarrow{D}_{\mu} H_1) (H_2^{\dagger} \overleftrightarrow{D}^{\mu} H_1) + h.c.$ \\
    $\mathcal{O}_{H \partial^2}^{(21)(12)} = \partial_{\mu} (H_2^{\dagger} H_1) \partial^{\mu} (H_1^{\dagger} H_2) $ & $\mathcal{O}_{H D}^{(21)(12)} = (H_2^{\dagger} \overleftrightarrow{D}_{\mu} H_1) (H_1^{\dagger} \overleftrightarrow{D}^{\mu} H_2)$ \\
    $\mathcal{O}_{H \partial^2}^{(21)(11)} = \partial_{\mu} (H_2^{\dagger} H_1) \partial^{\mu} (H_1^{\dagger} H_1) + h.c.$ & $\mathcal{O}_{H D}^{(21)(11)} = (H_2^{\dagger} \overleftrightarrow{D}_{\mu} H_1) (H_1^{\dagger} \overleftrightarrow{D}^{\mu} H_1) + h.c.$ \\
    $\mathcal{O}_{H \partial^2}^{(21)(22)} = \partial_{\mu} (H_2^{\dagger} H_1) \partial^{\mu} (H_2^{\dagger} H_2) + h.c.$ & $\mathcal{O}_{H D}^{(21)(22)} = (H_2^{\dagger} \overleftrightarrow{D}_{\mu} H_1) (H_2^{\dagger} \overleftrightarrow{D}^{\mu} H_2) + h.c.$ \\ [1ex] 
 \hline    
\end{tabular}
    \caption{Derivative operators with scalar doublets in the Higgs basis.}
\label{tab:hb_der_b_ops}
\end{table}

\begin{table}[H]
\centering
 \begin{tabular}{||l | l||} 
    \hline
    $\mathcal{O}_{H}^{(11)(11)(11)} = (H_1^{\dagger} H_1)^3$ & $\mathcal{O}_{H}^{(11)(21)(21)} = (H_1^{\dagger} H_1) (H_2^{\dagger} H_1)^2 + h.c.$ \\
    $\mathcal{O}_{H}^{(11)(11)(22)} = (H_1^{\dagger} H_1)^2 (H_2^{\dagger} H_2)$ & $\mathcal{O}_{H}^{(11)(21)(12)} = (H_1^{\dagger} H_1) (H_2^{\dagger} H_1) (H_1^{\dagger} H_2)$ \\
    $\mathcal{O}_{H}^{(11)(22)(22)} = (H_1^{\dagger} H_1)(H_2^{\dagger} H_2)^2$ & $\mathcal{O}_{H}^{(22)(21)(21)} = (H_2^{\dagger} H_2) (H_2^{\dagger} H_1)^2 + h.c.$ \\
    $\mathcal{O}_{H}^{(11)(11)(21)} = (H_1^{\dagger} H_1)^2 (H_2^{\dagger} H_1) + h.c.$ & $\mathcal{O}_{H}^{(22)(21)(12)} = (H_2^{\dagger} H_2) (H_2^{\dagger} H_1) (H_1^{\dagger} H_2)$ \\
    $\mathcal{O}_{H}^{(22)(22)(21)} = (H_2^{\dagger} H_2)^2 (H_2^{\dagger} H_1) + h.c.$ & $\mathcal{O}_{H}^{(21)(21)(21)} = (H_2^{\dagger} H_1)^3 + h.c.$ \\
    $\mathcal{O}_{H}^{(22)(22)(22)} = (H_2^{\dagger} H_2)^3$ & $\mathcal{O}_{H}^{(21)(21)(12)} = (H_2^{\dagger} H_1)^2 (H_1^{\dagger} 
    H_2) + h.c.$ \\
    & $\mathcal{O}_{H}^{(11)(22)(21)} = (H_1^{\dagger} H_1) (H_2^{\dagger} H_2)(H_2^{\dagger} H_1) + h.c.$ \\ [1ex] 
 \hline    
\end{tabular}
    \caption{Scalar operators in the Higgs basis.}
\label{tab:hb_scalar_ops}
\end{table}

\begin{table}[H]
\centering
 \begin{tabular}{||l | l ||} 
    \hline
    $\mathcal{O}_{H G}^{(11)} = (H_1^{\dagger} H_1) G_{\mu \nu}^a G^{a \mu \nu}$ & $\mathcal{O}_{H W}^{(11)} = (H_1^{\dagger} H_1) W_{\mu \nu}^a W^{a \mu \nu}$ \\
    $\mathcal{O}_{H G}^{(22)} = (H_2^{\dagger} H_2) G_{\mu \nu}^a G^{a \mu \nu}$ & $\mathcal{O}_{H W}^{(22)} = (H_2^{\dagger} H_2) W_{\mu \nu}^a W^{a \mu \nu}$ \\
    $\mathcal{O}_{H G}^{(21)} = (H_2^{\dagger} H_1) G_{\mu \nu}^a G^{a \mu \nu} + h.c.$ & $\mathcal{O}_{H W}^{(21)} = (H_2^{\dagger} H_1) W_{\mu \nu}^a W^{a \mu \nu} + h.c.$ \\
    $\mathcal{O}_{H \widetilde{G}}^{(11)} = (H_1^{\dagger} H_1) \widetilde{G}_{\mu \nu}^a G^{a \mu \nu}$ & $\mathcal{O}_{H \widetilde{W}}^{(11)} = (H_1^{\dagger} H_1) \widetilde{W}_{\mu \nu}^a W^{a \mu \nu}$ \\
    $\mathcal{O}_{H \widetilde{G}}^{(22)} = (H_2^{\dagger} H_2) \widetilde{G}_{\mu \nu}^a G^{a \mu \nu}$ & $\mathcal{O}_{H \widetilde{W}}^{(22)} = (H_2^{\dagger} H_2) \widetilde{W}_{\mu \nu}^a W^{a \mu \nu}$ \\
    $\mathcal{O}_{H \widetilde{G}}^{(21)} = (H_2^{\dagger} H_1) \widetilde{G}_{\mu \nu}^a G^{a \mu \nu} + h.c.$ & $\mathcal{O}_{H \widetilde{W}}^{(21)} = (H_2^{\dagger} H_1) \widetilde{W}_{\mu \nu}^a W^{a \mu \nu} + h.c.$ \\
    \hline
    $\mathcal{O}_{H B}^{(11)} = (H_1^{\dagger} H_1) B_{\mu \nu} B^{\mu \nu}$ & $\mathcal{O}_{H W B}^{(11)} = (H_1^{\dagger} \tau^a H_1) W_{\mu \nu}^a B^{\mu \nu}$ \\
    $\mathcal{O}_{H B}^{(22)} = (H_2^{\dagger} H_2) B_{\mu \nu} B^{\mu \nu}$ & $\mathcal{O}_{H W B}^{(22)} = (H_2^{\dagger} \tau^a H_2) W_{\mu \nu}^a B^{\mu \nu}$ \\
    $\mathcal{O}_{H B}^{(21)} = (H_2^{\dagger} H_1) B_{\mu \nu} B^{\mu \nu} + h.c.$ & $\mathcal{O}_{H W B}^{(21)} = (H_2^{\dagger} \tau^a H_1) W_{\mu \nu}^a B^{\mu \nu} + h.c.$ \\
    $\mathcal{O}_{H \widetilde{B}}^{(11)} = (H_1^{\dagger} H_1) \widetilde{B}_{\mu \nu} B^{\mu \nu}$ & $\mathcal{O}_{H \widetilde{W} B}^{(11)} = (H_1^{\dagger} \tau^a H_1) \widetilde{W}_{\mu \nu}^a B^{\mu \nu}$ \\
    $\mathcal{O}_{H \widetilde{B}}^{(22)} = (H_2^{\dagger} H_2) \widetilde{B}_{\mu \nu} B^{\mu \nu}$ & $\mathcal{O}_{H \widetilde{W} B}^{(22)} = (H_2^{\dagger} \tau^a H_2) \widetilde{W}_{\mu \nu}^a B^{\mu \nu}$ \\
    $\mathcal{O}_{H \widetilde{B}}^{(21)} = (H_2^{\dagger} H_1) \widetilde{B}_{\mu \nu} B^{\mu \nu} + h.c.$ & $\mathcal{O}_{H \widetilde{W} B}^{(21)} = (H_2^{\dagger} \tau^a H_1) \widetilde{W}_{\mu \nu}^a B^{\mu \nu} + h.c.$ \\ [1ex] 
 \hline    
\end{tabular}
    \caption{Mixed bosonic operators in the Higgs basis.}
\label{tab:hb_mix_ops}
\end{table}

\subsection{Advantages of the Higgs basis}

The convenience of the Higgs basis originates from the separation of the SM fields (contained in $H_1$ with the characteristic scale $M_1 \simeq v$) from the  additional scalars, $H$, $A$, and $H^\pm$ contained in $H_2$ with masses at a different scale $M_2$. In addition to numerous calculations in the 2HDM, the Higgs basis has also been used to obtain SMEFT from the 2HDM in the alignment limit~\cite{Gorbahn:2015gxa,Belusca-Maito:2016dqe,DasBakshi:2024krs} by integrating out the $H_2$ doublet at the scale $M_2 \simeq \Lambda \gg v$. Here we will point out further advantages of the Higgs basis when working within the 2HDM EFT. These include the separation of operators that modify SM couplings and masses from operators that contribute to scattering processes only, transparent correlations between scattering processes resulting from the same operator, and derivation of  correlations between different operators in specific UV completions.  

\begin{figure}[t!]
\includegraphics[scale=0.8]{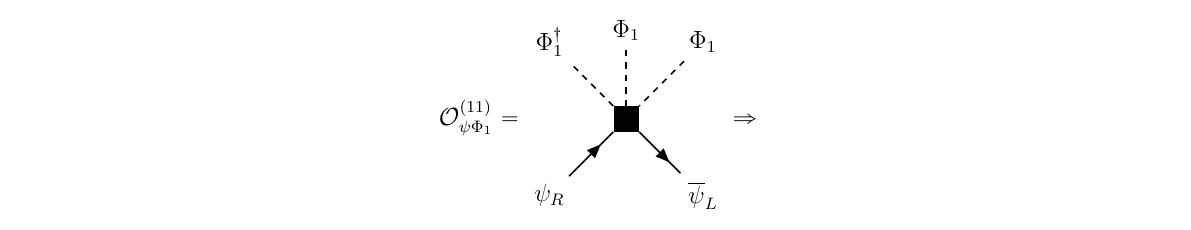}
\includegraphics[scale=0.8]{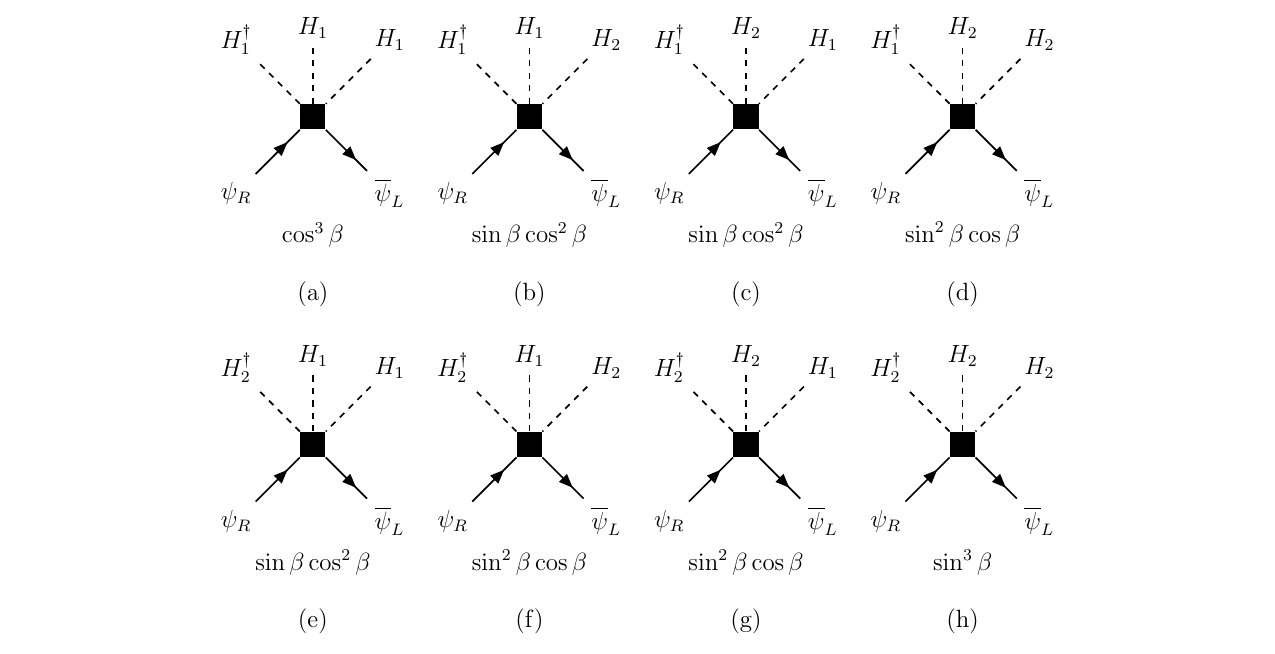}
\caption{Top: diagram of the $\mathcal{O}_{\psi \Phi_1}^{(11)}$ dimension-six mass operator. Bottom: the same operator decomposed in the Higgs basis. There are a total of eight diagrams (a)-(h), and their corresponding  Wilson coefficients are proportional to factors of $\cos \beta$ for each $H_1$ and $\sin \beta$ for every $H_2$ (indicated under the diagrams). }
\label{fig:hb_diagrams}
\end{figure}

An example of the separation of operators that modifies the SM couplings and masses is given in Fig.~\ref{fig:hb_diagrams}. In the standard basis, there are 8 mass operators, $\psi^2 \phi^3$, for each fermion type (see Table \ref{tab:gen2hdmops1}). Correspondingly, there are 8 mass operators in the Higgs basis. However, although all 8 operators in the standard basis modify the fermion mass and the Yukawa coupling, only one operator in the Higgs basis [Fig.~\ref{fig:hb_diagrams} (a)] does this. All other operators contribute only to scattering processes  involving at least one additional Higgs boson.

Furthermore, the correlations between different scattering processes resulting from the same operator are obvious in the Higgs basis. The coefficients under the diagrams in Fig.~\ref{fig:hb_diagrams} are proportionality factors assuming the operators result from the $\mathcal{O}_{\psi \Phi_1}^{(11)}$ operator in the standard basis. Proportionality factors for other cases can be easily read out from Appendix~\ref{sec:hb_mass_ops_match}. Thus, if the $\mathcal{O}_{\psi \Phi_1}^{(11)}$ is the dominant operator (as already mentioned, specific UV completions often generate only one mass operator at tree level~\cite{Dermisek:2023rvv,Dermisek:2023tgq}), we can immediately see a number of interesting correlations. For example, $\sigma(\psi \bar \psi \rightarrow HHH)/\sigma(\psi \bar \psi \rightarrow hhh) \propto \tan^6 \beta$, and similar $\tan^6 \beta$ enhancement is expected for other heavy tri-Higgs final states: $A A A$, $H A A$, $A H H$, $H H^+ H^-$, or $A H^+ H^-$. Correlations with many other processes featuring mixed $h$,  $H$, $A,$ and $H^\pm$  final states can be readily obtained. Enhancements of this type were recently studied in the connection with the sensitivity to the muon Yukawa coupling at a muon collider~\cite{Dermisek:2023rvv}, but are clearly not limited to muons.

\begin{figure}[t!]
\includegraphics[scale=0.8]{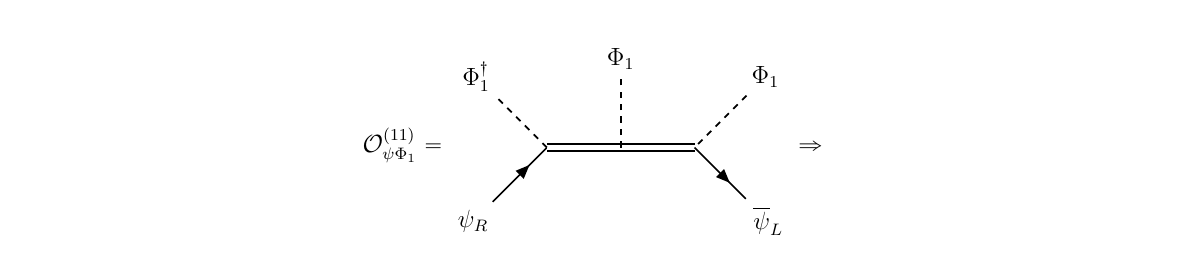}
\includegraphics[scale=0.8]{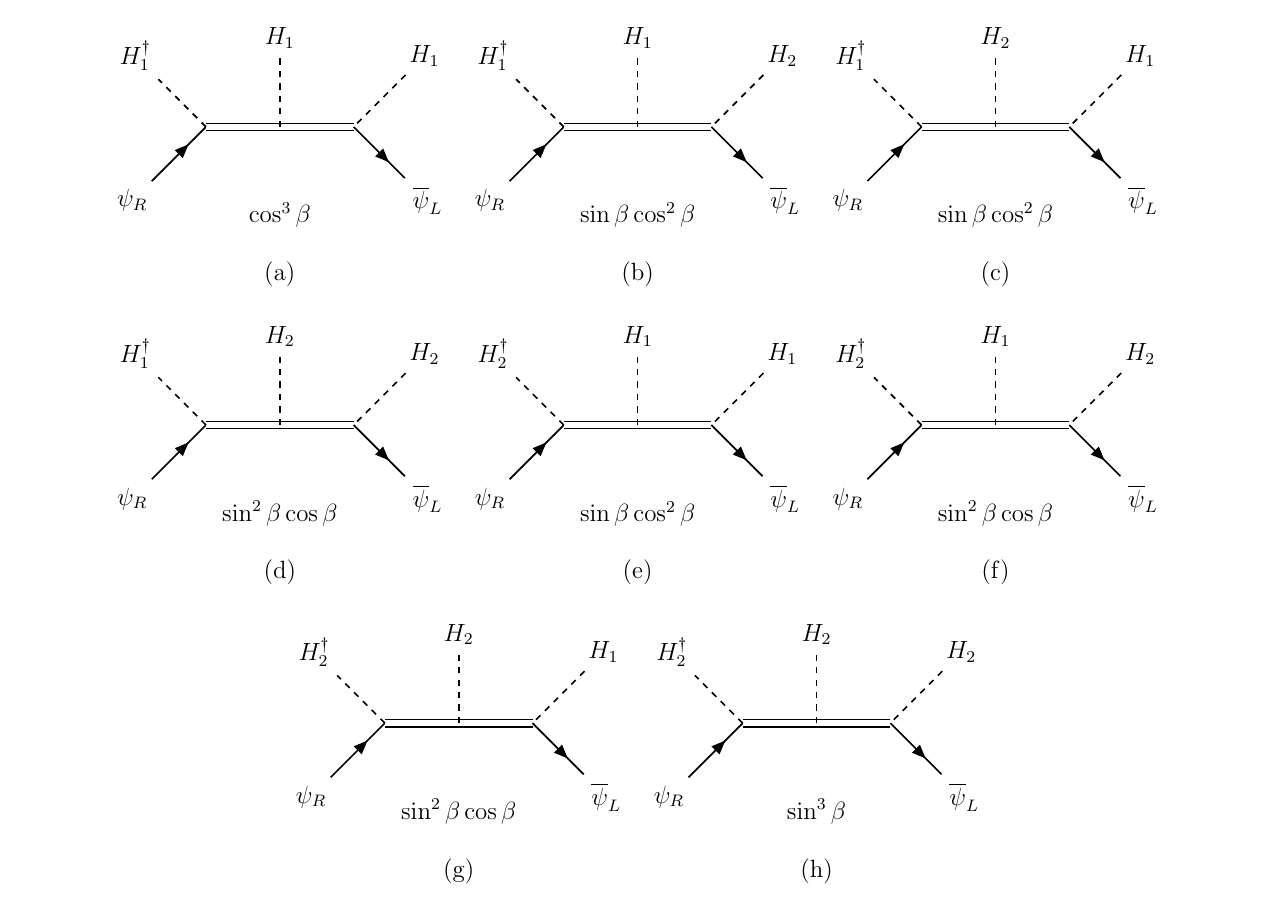}
\caption{Top: tree level diagram generating $\mathcal{O}_{\psi \Phi_1}^{(11)}$ mass operator in a UV completion with vectorlike fermions.  Bottom: The same operator decomposed in the Higgs basis. There are a total of eight diagrams (a-h), and their corresponding  Wilson coefficients are proportional to factors of $\cos \beta$ for each $H_1$ and $\sin \beta$ for every $H_2$ (indicated under the diagrams). }
\label{fig:hb_decomp}
\end{figure}

\begin{figure}[t!]
\includegraphics[scale=0.8]{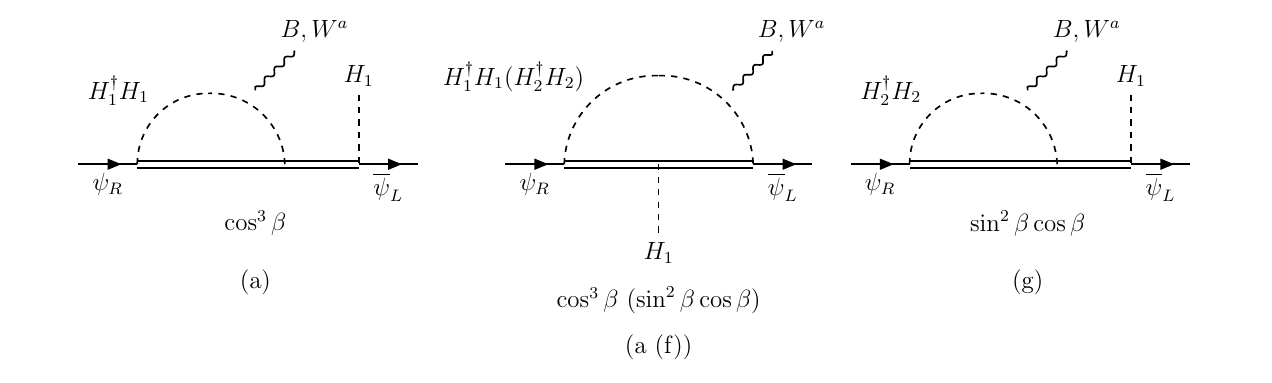}
\includegraphics[scale=0.8]{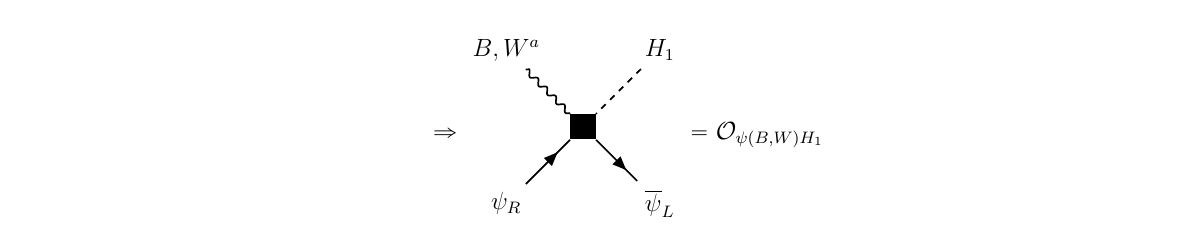}
\caption{A subset of diagrams in Fig.~\ref{fig:hb_decomp}, where $H_1^{\dagger} H_1$ or $H_2^{\dagger} H_2$ pairs are closed in $SU(2)$-invariant ways. The loops are dressed with a $B$ or $W^a$ fields, generating the dipole operators $\mathcal{O}_{\psi (B,W) H_1}$.  } 
\label{fig:match_dipole}
\end{figure}

Working in the Higgs basis is also crucial when deriving correlations between different operators in specific UV completions. To illustrate this, consider a specific UV completion that generates $\mathcal{O}_{\psi \Phi_1}^{(11)}$ via tree-level mixing of  SM fermions with new vectorlike fermion fields, depicted in Fig.~\ref{fig:hb_decomp}~\cite{Dermisek:2023nhe}. Such a UV completion will also generate dipole operators leading to dipole moments of the fermion that  mixes with new fermions. To calculate the dipole operators, consider a subset of the diagrams generating mass operators with $H_1$ and  close the loop on pairs of remaining $H_1^{\dagger} H_1$ and $H_2^{\dagger} H_2$ in an $SU(2)_L \times U(1)_Y$-invariant way  when possible. Then, dress the loop of each diagram with the $B$ and $W^a$ fields. The resulting diagrams are shown in Fig.~\ref{fig:match_dipole} where the labels indicate the diagrams from which they were obtained. These diagrams generate the dipole operators $\mathcal{O}_{\psi (B,W) H_1}$ in Table~\ref{tab:hb_dipole_ops}. Moreover, since dipole operators are obtained from the mass operators, their Wilson coefficients are correlated.\footnote{The loop functions can be evaluated thanks to the separation of SM fields from heavy Higgses. Note also, that it is straightforward to see that the loops involving $H_2$ are $\tan^2 \beta$ enhanced compared to loops involving $H_1$. The same results can be obtained by the complete calculation in the mass eigenstate basis in a given UV completion~\cite{Dermisek:2020cod,Dermisek:2021ajd,Dermisek:2023tgq} which is however much more complicated and does not provide a simple understanding of the results.}  This connection between mass and dipole operators, and the resulting correlations between the modification of the Yukawa coupling and electric and magnetic dipole moments was extensively studied for the muon in Refs.~\cite{Dermisek:2022aec,Dermisek:2023nhe,Dermisek:2023tgq}. Similarly, we may also consider the subset of mass diagrams containing $H_2$ and dress the closed loops of the remaining $H_1^{\dagger} H_1$ and $H_2^{\dagger} H_2$ pairs with $B$ and $W^a$ fields, generating the $\mathcal{O}_{\psi (B,W) H_2}$ operators in Table~\ref{tab:hb_dipole_ops}.
However, these operators will only affect scattering processes since $H_2$ does not acquire a VEV in this basis. This is yet another example of the separation of operators that illustrates the usefulness of the Higgs basis. \\ \\ \\ \\ \\ \\

\section{Conclusions}
\label{sec:conc}

We have constructed the general 2HDM EFT in a Warsaw-like basis and found 228 linearly independent operators of mass dimension-six. We have also constructed specific versions of 2HDM EFTs, corresponding to four types of 2HDMs: type-I, -II, -X, and -Y, distinguished by $Z_2$ symmetries that restrict the couplings of Higgs doublets to SM fermions.  We demonstrated that in all specific types of 2HDM EFTs, there are 76 common operators due to imposing a $Z_2$ symmetry on the Higgs doublets, $\Phi_1 \rightarrow - \Phi_1$ and  $\Phi_2 \rightarrow + \Phi_2$, while the $Z_2$ charges of fermions lead to a different number of model specific operators:  the type-I model predicts 62 operators,  type-II predicts 48 operators, type-X comes with 52 operators, and type-Y has 50 model specific operators. We have also discussed disagreements with previous works on 2HDM EFT~\cite{Crivellin:2016ihg, Karmakar:2017yek, Anisha:2019nzx}. These include most notably twice as many operators which modify  quark and lepton  masses, as well as operators involving covariant derivatives acting on either doublet contracted with right-handed quark currents in the case of a $CP$-conserving Higgs potential, compared to Ref.~\cite{Crivellin:2016ihg}, and disagreements with Ref.~\cite{Anisha:2019nzx} in constructing independent derivative operators involving  only the Higgs doublets.

Furthermore, we have transformed the general 2HDM EFT to the Higgs basis and provided matching of the Wilson coefficients between the two descriptions. In the Higgs basis, in the alignment limit favored in all types of 2HDMs~\cite{ATLAS:2024lyh}, the SM degrees of freedom are contained in one doublet, $H_1$,  and all additional Higgses are in another doublet, $H_2$. There is exactly the same number of independent dimension-six operators, 228, as in the general 2HDM in the standard basis. For specific types of 2HDM EFTs, all of the operators except four-fermion operators are now identical; the type of model only restricts which Wilson coefficients contribute to operators in the Higgs basis. On specific examples, we have illustrated the advantages of working with the 2HDM EFT in the Higgs basis that included the separation of operators that modify SM couplings and masses from operators that contribute to scattering processes only, transparent correlations between scattering processes resulting from the same operator, and derivation of correlations between different operators in specific UV completions. 

Generally, $\phi^6$ terms affect the scalar potential, and thus the vacuum and stability conditions are modified compared to those in the renormalizable 2HDM~\cite{Gunion:2002zf}.  For completeness, we have  derived general vacuum and stability conditions of the scalar potential in the presence of dimension-six terms. 

The SMEFT, parametrizing possible effects of new physics with characteristic scale far above the EW scale has led to a broad theoretical and experimental effort to constrain possible effects of new physics. New contact interactions resulting from nonrenormalizable operators~\cite{Dedes:2017zog}  could lead to numerous signals at the LHC and future colliders~\cite{Accettura:2023ked}. By power counting, processes mediated by these contact interactions grow with energy, and can supersede the SM background, until the theory breaks down by perturbativity and unitarity, constraining the involved Wilson coefficients~\cite{Maltoni:2001dc,Allwicher:2021jkr,DiLuzio:2016sur}. Traditional techniques of partial wave analysis and analyticity of the amplitude~\cite{Remmen:2019cyz,Remmen:2020uze,Remmen:2022orj} have been successful in limiting the parameter space of these energy-growing processes in the SMEFT  and could be tailored to 2HDM EFT that predicts a significanlty larger number of new contact interactions. Furthermore, studying the effects of renormalization group equations and one-loop mixing of the SMEFT operators~\cite{Jenkins:2013zja,Jenkins:2013wua,Alonso:2013hga,Dawson:2020oco,Aebischer:2020lsx,Baratella:2020lzz} could be also extended to the 2HDM EFT. 

\acknowledgments

We would like to thank Taegyu Lee, Navin McGinnis, and Sangsik Yoon for their collaboration on earlier projects that inspired this work.

\appendix

\section{Corrections to the Scalar Potential}
\label{sec:scalar_cor}
In this appendix, we provide details and conditions about how the presence of higher-dimensional terms affect the two vacua and stability of the scalar potential. The inclusion of higher-dimensional operators from the class $\phi^6$ in the theory will introduce corrections to the potential, which now reads as $V'(\Phi_1, \Phi_2) = V(\Phi_1, \Phi_2) - \sum_{i = \phi^6} C_{i} \mathcal{O}_i$. Extemizing the potential of Eq.~(\ref{eq:2hdm_pot}) while including all $\phi^6$ operators in Table~\ref{tab:gen2hdmops2} and defining $\sin \beta \equiv s_{\beta}, \cos \beta \equiv c_{\beta}$ and $\tan \beta \equiv t_{\beta}$, we find at the vacuum $v$ 
\begin{equation}
    \begin{split}
        \frac{1}{v_1} \frac{\partial V'}{\partial \Phi_1} \Bigg|_{v} & = m_1^2 + t_{\beta} (m_{12}^2)^* + v^2 \left(c^2_{\beta} \lambda_1 + s^2_{\beta} (\lambda_3 + \lambda_4 + \lambda_5^*) + s_{\beta} c_{\beta} (\lambda_6 + 2 \lambda_6^*) + s^2_{\beta} t_{\beta} \lambda_7^* \right)  \\
        & - v^4 \left(3 c^4_{\beta} C_{\Phi}^{(11)(11)(11)} + 3 s^3_{\beta} c_{\beta} C_{\Phi}^{(21)(21)(21)} + 2 s^3_{\beta} c_{\beta} C_{\Phi}^{(21)(21)(12)} + s^3_{\beta} c_{\beta} C_{\Phi}^{(21)(21)(12)*} \right. \\
        & \left. + 2 s^3_{\beta} c_{\beta} C_{\Phi}^{(11)(22)(21)}  + s^3_{\beta} c_{\beta} C_{\Phi}^{(11)(22)(21)*} + 2 s^2_{\beta} c^2_{\beta} C_{\Phi}^{(11)(11)(22)} + 3 s^2_{\beta} c^2_{\beta} C_{\Phi}^{(11)(21)(21)} \right. \\
        & \left. + s^2_{\beta} c^2_{\beta} C_{\Phi}^{(11)(21)(21)*} + 2 s^2_{\beta} c^2_{\beta} C_{\Phi}^{(11)(21)(12)} + s^4_{\beta} C_{\Phi}^{(11)(22)(22)} + 2 s^4_{\beta} C_{\Phi}^{(22)(21)(21)} \right. \\
        & \left. + s^4_{\beta} C_{\Phi}^{(22)(21)(12)} + 3 s_{\beta} c^3_{\beta} C_{\Phi}^{(11)(11)(21)} + 2 s_{\beta} c^3_{\beta}  C_{\Phi}^{(11)(11)(21)*} + s^4_{\beta} t_{\beta}  C_{\Phi}^{(22)(22)(21)} \right) = 0,
    \end{split}
\end{equation}
\begin{equation}
    \begin{split}
        \frac{1}{v_2} \frac{\partial V'}{\partial \Phi_2} \Bigg|_{v} & = m_2^2 + \frac{m_{12}^2}{t_{\beta}} + v^2 \left(s^2_{\beta} \lambda_2 + c^2_{\beta} (\lambda_3 + \lambda_4 + \lambda_5) + \frac{c^2_{\beta}}{t_{\beta}} \lambda_6 + s_{\beta} c_{\beta} (2 \lambda_7 + \lambda_7^*) \right) \\
        & - v^4 \left(3 s^4_{\beta} C_{\Phi}^{(22)(22)(22)} + 3 s_{\beta} c^3_{\beta} C_{\Phi}^{(21)(21)(21)*} + s_{\beta} c^3_{\beta} C_{\Phi}^{(21)(21)(12)} + 2 s_{\beta} c^3_{\beta} C_{\Phi}^{(21)(21)(12)*} \right. \\
        & \left. + s_{\beta} c^3_{\beta} C_{\Phi}^{(11)(22)(21)} + 2 s_{\beta} c^3_{\beta} C_{\Phi}^{(11)(22)(21)*} + c^4_{\beta} C_{\Phi}^{(11)(11)(22)} + 2 c^4_{\beta} C_{\Phi}^{(11)(21)(21)*} \right. \\
        & \left. + c^4_{\beta} C_{\Phi}^{(11)(21)(12)} + 2 s^2_{\beta} c^2_{\beta} C_{\Phi}^{(11)(22)(22)} + s^2_{\beta} c^2_{\beta} C_{\Phi}^{(22)(21)(21)} + 3 s^2_{\beta} c^2_{\beta} C_{\Phi}^{(22)(21)(21)*} \right. \\
        & \left. + 2 s^2_{\beta} c^2_{\beta} C_{\Phi}^{(22)(21)(12)} + \frac{c^4_{\beta}}{t_{\beta}} C_{\Phi}^{(11)(11)(21)*} + 2 s^3_{\beta} c_{\beta}  C_{\Phi}^{(22)(22)(21)} + 3 s^3_{\beta} c_{\beta}  C_{\Phi}^{(22)(22)(21)*} \right) = 0. 
    \end{split}
\end{equation}
Since the parameters $m_1^2, \ m_2^2,\ \lambda_1, \ \lambda_2, \ \lambda_3$, and $\lambda_4$ and Wilson coefficients $C_{\Phi}^{(11)(11)(11)}, \ C_{\Phi}^{(22)(22)(22)}, \ C_{\Phi}^{(11)(11)(22)}, \ C_{\Phi}^{(11)(21)(12)}, \ C_{\Phi}^{(11)(22)(22)},$ and $C_{\Phi}^{(22)(21)(12)}$ are Hermitian, the phases of the remaining contributions can be constrained from either extrema condition:
\begin{equation}
\begin{split}
    - \textrm{Im}[m_{12}^2] = v^2 & \left(s_{\beta} c_{\beta} \ \textrm{Im}[\lambda_5] + c^2_{\beta} \ \textrm{Im}[\lambda_6] + s^2_{\beta} \ \textrm{Im}[\lambda_7]  \right) \\
    + v^4 & \left(3 s^2_{\beta} c^2_{\beta} \ \textrm{Im}[C_{\Phi}^{(21)(21)(21)}] + s^2_{\beta} c^2_{\beta} \ \textrm{Im}[C_{\Phi}^{(21)(21)(12)}] + s^2_{\beta} c^2_{\beta} \ \textrm{Im}[C_{\Phi}^{(11)(22)(21)}] \right. \\
    & \left. + 2 s_{\beta} c^3_{\beta} \ \textrm{Im}[C_{\Phi}^{(11)(21)(21)}] + 2 s^3_{\beta} c_{\beta} \ \textrm{Im}[C_{\Phi}^{(22)(21)(21)}] + c^4_{\beta} \ \textrm{Im}[C_{\Phi}^{(11)(11)(21)}] \right. \\
    & \left. + s^4_{\beta} \ \textrm{Im}[C_{\Phi}^{(22)(22)(21)}] \right). 
\end{split}
\end{equation}
For a complex scalar function of two variables, the conditions for positive concavity needed to guarantee these extrema are truly minima are
\begin{equation}
\begin{split}
    & \begin{vmatrix}
        \frac{\partial^2 V'}{\partial \Phi_1^{\dagger} \partial \Phi_1} & \frac{\partial^2 V'}{\partial \Phi_1^{\dagger} \partial \Phi_2} \\
        \frac{\partial^2 V'}{\partial \Phi_2^{\dagger} \partial \Phi_1} & \frac{\partial^2 V'}{\partial \Phi_2^{\dagger} \partial \Phi_2}
    \end{vmatrix}\Bigg|_{v} = \left( \frac{\partial^2 V'}{\partial \Phi_1^{\dagger} \partial \Phi_1} \right) \left( \frac{\partial^2 V'}{\partial \Phi_2^{\dagger} \partial \Phi_2} \right) - \Bigg| \frac{\partial^2 V'}{\partial \Phi_1^{\dagger} \partial \Phi_2} \Bigg|^2 \Bigg|_{v} > 0, \\
    & \frac{\partial^2 V'}{\partial \Phi_1^{\dagger} \partial \Phi_2} \Bigg|_v = \left(\frac{\partial^2 V'}{\partial \Phi_2^{\dagger} \partial \Phi_1} \right)^* \Bigg|_v, \\
    & \frac{\partial^2 V'}{\partial \Phi_{1,2}^{\dagger} \partial \Phi_{1,2}} \Bigg|_v > 0.
\end{split}
\end{equation}
The partial second derivatives at the vacuum are
\begin{equation}
    \begin{split}
       \frac{\partial^2 V'}{\partial \Phi_1^{\dagger} \partial \Phi_1} \Bigg|_v & = m_1^2 + v^2 (2 c^2_{\beta} \lambda_1 + s^2_{\beta}(\lambda_3 + \lambda_4) + 4 s_{\beta} c_{\beta} \ \textrm{Re}[\lambda_6]) \\
       & - v^4 \left[ 9 c^4_{\beta} C_{\Phi}^{(11)(11)(11)} + 4 s^3_{\beta} c_{\beta} \ \textrm{Re}[C_{\Phi}^{(21)(21)(12)}] + 4 s^3_{\beta} c_{\beta} \ \textrm{Re}[C_{\Phi}^{(11)(22)(21)}] \right. \\
       & \left. + 4 s^2_{\beta} c^2_{\beta} C_{\Phi}^{(11)(11)(22)} + s^4_{\beta} C_{\Phi}^{(11)(22)(22)} + 6 s^2_{\beta} c^2_{\beta} \ \textrm{Re}[C_{\Phi}^{(11)(21)(21)}] \right. \\
       & \left. + 4 s^2_{\beta} c^2_{\beta} C_{\Phi}^{(11)(21)(12)} + s^4_{\beta} C_{\Phi}^{(22)(21)(12)} + 12 s_{\beta} c^3_{\beta} \ \textrm{Re}[C_{\Phi}^{(11)(11)(21)}] \right],
    \end{split}
\end{equation}
\begin{equation}
    \begin{split}
       \frac{\partial^2 V'}{\partial \Phi_2^{\dagger} \partial \Phi_2} \Bigg|_v & = m_2^2 + v^2 (2 s^2_{\beta} \lambda_2 + c^2_{\beta}(\lambda_3 + \lambda_4) + 4 s_{\beta} c_{\beta} \ \textrm{Re}[\lambda_7]) \\
       & - v^4 \left[ 9 s^4_{\beta} C_{\Phi}^{(22)(22)(22)} + 4 s_{\beta} c^3_{\beta} \ \textrm{Re}[C_{\Phi}^{(21)(21)(12)}] + 4 s_{\beta} c^3_{\beta} \ \textrm{Re}[C_{\Phi}^{(11)(22)(21)}] \right. \\
       & \left. + c^4_{\beta} C_{\Phi}^{(11)(11)(22)} + 4 s^2_{\beta} c^2_{\beta} C_{\Phi}^{(11)(22)(22)} + 6 s^2_{\beta} c^2_{\beta} \ \textrm{Re}[C_{\Phi}^{(22)(21)(21)}] \right. \\
       & \left. + c^4_{\beta} C_{\Phi}^{(11)(21)(12)} + 4 s^2_{\beta} c^2_{\beta} C_{\Phi}^{(22)(21)(12)} + 12 s^3_{\beta} c_{\beta} \ \textrm{Re}[C_{\Phi}^{(22)(22)(21)}] \right],
    \end{split}
\end{equation}
and 
\begin{equation}
    \begin{split}
       \frac{\partial^2 V'}{\partial \Phi_2^{\dagger} \partial \Phi_1} \Bigg|_v & = \left(\frac{\partial^2 V'}{\partial \Phi_1^{\dagger} \partial \Phi_2} \right)^* \Bigg|_v = (m_{12}^2)^* + v^2 ( s_{\beta} c_{\beta}(\lambda_3 + \lambda_4 + 2 \lambda_5^*) + 2 c^2_{\beta} \lambda_6^* + 2 s^2_{\beta} \lambda_7^*) \\
       & - v^4 \left[ 9 s^2_{\beta} c^2_{\beta} C_{\Phi}^{(21)(21)(21)} + 4 s^2_{\beta} c^2_{\beta} C_{\Phi}^{(21)(21)(12)} + s^2_{\beta} c^2_{\beta} C_{\Phi}^{(21)(21)(12)*} + 4 s^2_{\beta} c^2_{\beta} C_{\Phi}^{(11)(22)(21)} \right. \\
        & \left. + s^2_{\beta} c^2_{\beta} C_{\Phi}^{(11)(22)(21)*} + 2 s_{\beta} c^3_{\beta} C_{\Phi}^{(11)(11)(22)} + 2 s^3_{\beta} c_{\beta} C_{\Phi}^{(11)(22)(22)} + 6 s_{\beta} c^3_{\beta} C_{\Phi}^{(11)(21)(21)} \right. \\
        & \left. + 6 s^3_{\beta} c_{\beta} C_{\Phi}^{(22)(21)(21)} + 2 s_{\beta} c^3_{\beta} C_{\Phi}^{(11)(21)(12)} + 2 s^3_{\beta} c_{\beta} C_{\Phi}^{(22)(21)(12)} + 3 c^4_{\beta} C_{\Phi}^{(11)(11)(21)} \right. \\
        & \left. + 3 s^4_{\beta} c_{\beta} C_{\Phi}^{(22)(22)(21)} \right].
    \end{split}
\end{equation}
Finally, the presence of $\phi^6$ operators may asymptotically grow over the dimension-four terms in the potential, provided the Wilson coefficients of the operators are sufficiently large. Mandating the modified potential $V'$ is bounded below in all directions by large values of fields, we can marginalize over them in field space. Defining $A \equiv \Phi_1^{\dagger} \Phi_1, \ B \equiv \Phi_2^{\dagger} \Phi_2, \ C \equiv \textrm{Re}[\Phi_1^{\dagger} \Phi_2],$ and $D \equiv \textrm{Im}[\Phi_1^{\dagger} \Phi_2]$ whereby $C = \xi D$ and $AB \geq C^2 + D^2$ and keeping terms which grow asymptotically, we first set $C = D = 0$ and find 
\begin{equation}
    \begin{split}
      & \frac{1}{2}\lambda_1 A^2 + \frac{1}{2} \lambda_2 B^2 + \lambda_3 AB \\
      & - [A^3 C_{\Phi}^{(11)(11)(11)} + B^3 C_{\Phi}^{(22)(22)(22)} + A^2 B C_{\Phi}^{(11)(11)(22)} + A B^2 C_{\Phi}^{(11)(22)(22)}] > 0.  
    \end{split}
\end{equation}
If we define $A' \equiv \lambda_1^{1/2} A \equiv \sqrt{A'^2 + B'^2} s_{\zeta}$, $B' \equiv \lambda_2^{1/2} B \equiv \sqrt{A'^2 + B'^2} c_{\zeta}$ by parametrizing on a right triangle, this condition can be written as
\begin{equation}
\begin{split}
    (\lambda_1 \lambda_2)^{3/2} + 2 \lambda_1 \lambda_2 \lambda_3 s_{\zeta} c_{\zeta} - 2 \sqrt{A'^2 + B'^2} & \left[ s^3_{\zeta} \lambda_2^{3/2} C_{\Phi}^{(11)(11)(11)} + c^3_{\zeta} \lambda_1^{3/2} C_{\Phi}^{(22)(22)(22)} \right. \\
    & \left. + s^2_{\zeta} c_{\zeta}\lambda_1^{1/2} \lambda_2 C_{\Phi}^{(11)(11)(22)} + s_{\zeta} c^2_{\zeta} \lambda_1 \lambda_2^{1/2}  C_{\Phi}^{(11)(22)(22)} \right] > 0.
\end{split}
\end{equation}
By setting $\zeta = \pi/2$ and $0$, respectively, we find 
\begin{equation}
    \lambda_1 - 2 A C_{\Phi}^{(11)(11)(11)} > 0, \ \ \ \ \ \lambda_2 - 2 B C_{\Phi}^{(22)(22)(22)} > 0, 
\end{equation}
and by setting $\zeta = \pi/4$, we have $ A' = B'$ and the above condition can be written in terms of either $A' = \lambda_1^{1/2} A$ or $B' = \lambda_2^{1/2} B$. For all values of $A$ and $B$, we find
\begin{equation}
\begin{split}
        (\lambda_1 \lambda_2)^{3/2} + \lambda_1 \lambda_2 \lambda_3 - A & \left[C_{\Phi}^{(11)(11)(11)} \lambda_1^{1/2} \lambda_2^{3/2} + C_{\Phi}^{(22)(22)(22)} \lambda_1^2 \right. \\
        & \left. + C_{\Phi}^{(11)(11)(22)} \lambda_1 \lambda_2 + C_{\Phi}^{(11)(22)(22)} \lambda_1^{3/2} \lambda_2^{1/2}  \right] > 0, \\
        (\lambda_1 \lambda_2)^{3/2} + \lambda_1 \lambda_2 \lambda_3 - B & \left[C_{\Phi}^{(11)(11)(11)} \lambda_2^2 + C_{\Phi}^{(22)(22)(22)} \lambda_1^{3/2} \lambda_2^{1/2} \right. \\
        & \left. + C_{\Phi}^{(11)(11)(22)} \lambda_1^{1/2} \lambda_2^{3/2} + C_{\Phi}^{(11)(22)(22)} \lambda_1 \lambda_2 \right] > 0.
\end{split}
\end{equation}
Marginalizing over other values of field space when $AB = C^2 + D^2$ for $\lambda_1^{1/2} A = \lambda_2^{1/2} B$ leads to a complicated polynomial in $\xi$ that cannot be solved analytically for all values of $\xi$. Notice that in the limit all Wilson coefficients are zero, we recover the same conditions on the dimension-four scalar potential \cite{Gunion:2002zf}. 

\section{Matching to the Higgs Basis}
\label{sec:higgs_basis}

In this appendix, we list the translation of all terms from the standard basis to the Higgs basis using Eqs.~(\ref{eq:phi1_trans}) and (\ref{eq:phi2_trans}). For dimension-six terms, once the type of 2HDM is specified in Table~\ref{tab:2hdmQ_numbers} and Tables~\ref{tab:t1ops},~\ref{tab:t2ops},~\ref{tab:txops}, or~\ref{tab:tyops}, one can set the irrelevant Wilson coefficients to zero in this basis. For compactness, we suppress flavor indices for higher-dimensional operators, where we define $\overline{\psi} C_i \psi \equiv \overline{\psi}_a (C_i)_{ab} \psi_b$ and $\overline{\psi} C_i^* \psi \equiv \overline{\psi}_a(C_i^*)_{ba} \psi_b$. Note that for flavor-diagonal terms below, the combination of Wilson coefficients reduces to $C_i + C_i^* = 2 \ \textrm{Re}[C_i]$. For convenience, the notation is identical to the general 2HDM of the main text with the replacement of $\Phi \rightarrow H$ in the label.

\subsection{Dimension-four terms}
The scalar potential [Eq.~(\ref{eq:2hdm_pot})] translated to the Higgs basis is
\begin{equation}
\begin{split}
    V(H_1, H_2) & = M_1^2 \left( H_1^{\dagger} H_1 \right) + M_2^2 \left( H_2^{\dagger} H_2 \right) + \left(  M_{12}^2 H_1^{\dagger} H_2 + h.c. \right) \\
    & + \frac{1}{2} \Lambda_1 \left(H_1^{\dagger} H_1 \right)^2 + \frac{1}{2} \Lambda_2 \left(H_2^{\dagger} H_2 \right)^2 + \Lambda_3 \left(H_1^{\dagger} H_1 \right) \left(H_2^{\dagger} H_2 \right) + \Lambda_4 \left(H_1^{\dagger} H_2 \right)  \left(H_2^{\dagger} H_1 \right) \\
    & + \left( \frac{1}{2} \Lambda_5 \left(H_1^{\dagger} H_2 \right)^2 + \Lambda_6 (H_1^{\dagger} H_1) H_1^{\dagger} H_2 + \Lambda_7 (H_2^{\dagger} H_2) H_1^{\dagger} H_2 + h.c.\right),
\label{eq:hb_pot}
\end{split}
\end{equation}
where the new quartic couplings are, if we use the the shorthand notation $\lambda_{345} \equiv \lambda_3 + \lambda_4 + \textrm{Re} [\lambda_5]$,
\begin{equation}
    \begin{split}
        & \Lambda_1 = \lambda_1 c^4_{\beta} + \lambda_2 s^4_{\beta} + 2 \lambda_{345} s^2_{\beta} c^2_{\beta} + 4 s_{\beta} c_{\beta} \left( c^2_{\beta} \ \textrm{Re}[\lambda_6] + s^2_{\beta} \ \textrm{Re}[\lambda_7] \right), \\
        & \Lambda_2 = \lambda_1 s^4_{\beta} + \lambda_2 c^4_{\beta} + 2 \lambda_{345} s^2_{\beta} c^2_{\beta} - 4 s_{\beta} c_{\beta} \left( s^2_{\beta} \ \textrm{Re}[\lambda_6] + c^2_{\beta} \ \textrm{Re}[\lambda_7] \right), \\
        & \Lambda_3 = (\lambda_1 + \lambda_2 - 2 \lambda_{345}) s^2_{\beta} c^2_{\beta} + \lambda_3 - 2 s_{\beta} c_{\beta} (c^2_{\beta} - s^2_{\beta}) \ \textrm{Re}[\lambda_6 - \lambda_7], \\
        & \Lambda_4 = (\lambda_1 + \lambda_2 - 2 \lambda_{345}) s^2_{\beta} c^2_{\beta} + \lambda_4 - 2 s_{\beta} c_{\beta} (c^2_{\beta} - s^2_{\beta}) \ \textrm{Re}[\lambda_6 - \lambda_7], \\
        & \Lambda_5 = (\lambda_1 + \lambda_2 - 2 \lambda_{345}) s^2_{\beta} c^2_{\beta} + c^2_{\beta} \lambda_5 + s^2_{\beta} \lambda_5^* - 2 s_{\beta} c_{\beta} (c^2_{\beta} (\lambda_6 - \lambda_7) - s^2_{\beta} (\lambda_6^* - \lambda_7^*)) , \\
        & \Lambda_6 = - (\lambda_1 - \lambda_3 - \lambda_4 - \lambda_5) s_{\beta} c^3_{\beta} + (\lambda_2 - \lambda_3 - \lambda_4 - \lambda_5^*) s^3_{\beta} c_{\beta} \\
        & + c^2_{\beta} ((c^2_{\beta} - s^2_{\beta}) \lambda_6 - 2 s^2_{\beta} \lambda_6^*) + s^2_{\beta} ((c^2_{\beta} - s^2_{\beta}) \lambda_7^* + 2 c^2_{\beta} \lambda_7), \\ 
        & \Lambda_7 = - (\lambda_1 - \lambda_3 - \lambda_4 - \lambda_5^*) s^3_{\beta} c_{\beta} + (\lambda_2 - \lambda_3 - \lambda_4 - \lambda_5) s_{\beta} c^3_{\beta} \\
        & + s^2_{\beta} ((c^2_{\beta} - s^2_{\beta}) \lambda_6^* + 2 c^2_{\beta} \lambda_6) + c^2_{\beta} ((c^2_{\beta} - s^2_{\beta}) \lambda_7 - 2 s^2_{\beta} \lambda_7^*).
    \end{split}
\label{eq:hbasis_couplings}
\end{equation}
Note that in general, each doublet $\Phi_{1,2}$ can contain possible phases and by performing a $U(1)_Y$ rotation on $\Phi_2$, the phase $\xi$ can be placed on the VEV of $\Phi_2: v_2 e^{i \xi}$. Additional symmetries of the scalar potential in the Higgs basis include $U(1)$ transformations of Higgs flavors $H_1 \rightarrow e^{-i\chi}$ and $H_2 \rightarrow e^{i \chi} H_2$, introducing another arbitrary phase $\chi$. For our discussion, it suffices to neglect all possible phase factors. Otherwise, these would introduce additional phase factors on $m_{12}^2, \ \lambda_5, \ \lambda_6, \ \lambda_7$, and $M_{12}^2, \ \Lambda_5, \ \Lambda_6,$ and $\Lambda_7$ (see \cite{Davidson:2005cw} for a detailed discussion). The dimensionful parameters listed below can be simplified as
\begin{equation}
    \begin{split}
        & M_1^2 = m_1^2 c^2_{\beta} + m_2^2 s^2_{\beta} + 2 s_{\beta} c_{\beta} \ \textrm{Re}[m_{12}^2] = - v^2 \Lambda_1, \\
        & M_2^2 = m_1^2 s^2_{\beta} + m_2^2 c^2_{\beta} - 2 s_{\beta} c_{\beta} \ \textrm{Re}[m_{12}^2] \\
        & M_{12}^2 = - (m_1^2 - m_2^2) s_{\beta} c_{\beta} + m_{12}^2 c^2_{\beta} - (m_{12}^2)^* s^2_{\beta} = - v^2 \Lambda_6.
    \end{split}
\label{eq:hbasis_mass}
\end{equation}
Kinetic terms involving the doublets $H_1$ and $H_2$ are
\begin{equation}
    \begin{split}
      \mathcal{L} \supset & (1 + 2 s_{\beta} c_{\beta} \ \textrm{Re}[\eta]) \left(D_{\mu} H_1 \right)^{\dagger} D^{\mu} H_1 + (1 - 2 s_{\beta} c_{\beta} \ \textrm{Re}[\eta]) \left(D_{\mu} H_2 \right)^{\dagger} D^{\mu} H_2 \\
      & + (\eta c^2_{\beta} - \eta^* s^2_{\beta}) \left(D_{\mu} H_1 \right)^{\dagger} D^{\mu} H_2 + h.c..
    \end{split}
\end{equation}
If the $Z_2$ symmetry is enforced in the standard basis, $\eta \rightarrow 0$. The Yukawa interactions in this basis are
\begin{equation}
\begin{split}
    \mathcal{L} \supset & - \left(y_e^{(1)} c_{\beta} + y_e^{(2)} s_{\beta} \right) \overline{l}_L e_R H_1 - \left(- y_e^{(1)} s_{\beta} + y_e^{(2)} c_{\beta} \right)  \overline{l}_L e_R H_2 \\
    & - \left(y_d^{(1)} c_{\beta} + y_d^{(2)} s_{\beta} \right) \overline{q}_L d_R H_1 - \left(- y_d^{(1)} s_{\beta} + y_d^{(2)} c_{\beta} \right)  \overline{q}_L d_R H_2 \\
    & - \left(y_u^{(1)} c_{\beta} + y_u^{(2)} s_{\beta} \right) \overline{q}_L u_R \cdot H_1^{\dagger} - \left(- y_u^{(1)} s_{\beta} + y_u^{(2)} c_{\beta} \right)  \overline{q}_L u_R \cdot H_2^{\dagger} + h.c..
\end{split}
\end{equation}

\subsection{Dimension-five terms}
The Weinberg-like operators in the Higgs basis become
\begin{equation}
    \mathcal{O}_{\nu \nu H}^{(11)} = (H_1 \cdot l_L)^T \textbf{C} (H_1 \cdot l_L),
\end{equation}
\begin{equation}
    \mathcal{O}_{\nu \nu H}^{(22)} = (H_2 \cdot l_L)^T \textbf{C} (H_2 \cdot l_L),
\end{equation}
\begin{equation}
    \mathcal{O}_{\nu \nu H}^{(12)} = (H_1 \cdot l_L)^T \textbf{C} (H_2 \cdot l_L),
\end{equation}
whose Wilson coefficients are
\begin{equation}
    C_{\nu \nu H}^{(11)} = c^2_{\beta} C_{\nu \nu \Phi}^{(11)} + s^2_{\beta} C_{\nu \nu \Phi}^{(22)} + s_{\beta} c_{\beta} C_{\nu \nu \Phi}^{(12)},
\end{equation}
\begin{equation}
    C_{\nu \nu H}^{(22)} = s^2_{\beta} C_{\nu \nu \Phi}^{(11)} + c^2_{\beta} C_{\nu \nu \Phi}^{(22)} - s_{\beta} c_{\beta} C_{\nu \nu \Phi}^{(12)}, 
\end{equation}
\begin{equation}
    C_{\nu \nu H}^{(12)} = - s_{\beta} c_{\beta} C_{\nu \nu \Phi}^{(11)} + s_{\beta} c_{\beta} C_{\nu \nu \Phi}^{(22)} + (c^2_{\beta} - s^2_{\beta}) C_{\nu \nu \Phi}^{(12)}.
\end{equation}

\subsection{Dimension-six terms - class $\psi^2 \phi^3$ operators}
\label{sec:hb_mass_ops_match}
Operators in the class of $\psi^2 \phi^3$ translated to the Higgs basis are given in Table~\ref{tab:hb_mass_ops}, whose Wilson coefficients are
\begin{equation}
\begin{split}
        C_{(l,d,u) H_1}^{(11)} & = c^3_{\beta} C_{(l,d,u) \Phi_1}^{(11)} + s^2_{\beta} c_{\beta} C_{(l,d,u) \Phi_1}^{(22)} + s_{\beta} c^2_{\beta} C_{(l,d,u) \Phi_1}^{(21)} + s_{\beta} c^2_{\beta} C_{(l,d,u) \Phi_1}^{(12)} \\
        & + s^3_{\beta} C_{(l,d,u) \Phi_2}^{(22)} + s_{\beta} c^2_{\beta} C_{(l,d,u) \Phi_2}^{(11)} + s^2_{\beta} c_{\beta} C_{(l,d,u) \Phi_2}^{(21)} + s^2_{\beta} c_{\beta} C_{(l,d,u) \Phi_2}^{(12)},
\end{split} 
\label{eq:hb_massopH1}
\end{equation}
\begin{equation}
\begin{split}
    C_{(l,d,u) H_1}^{(22)} & = s^2_{\beta} c_{\beta} C_{(l,d,u) \Phi_1}^{(11)} + c^3_{\beta} C_{(l,d,u) \Phi_1}^{(22)} - s_{\beta} c^2_{\beta} C_{(l,d,u) \Phi_1}^{(21)} - s_{\beta} c^2_{\beta} C_{(l,d,u) \Phi_1}^{(12)} \\
    & + s_{\beta} c^2_{\beta} C_{(l,d,u) \Phi_2}^{(22)} + s^3_{\beta} C_{(l,d,u) \Phi_2}^{(11)} - s^2_{\beta} c_{\beta} C_{(l,d,u) \Phi_2}^{(21)} - s^2_{\beta} c_{\beta} C_{(l,d,u) \Phi_2}^{(12)},
\end{split}
\end{equation}
\begin{equation}
\begin{split}
    C_{(l,d,u) H_1}^{(21)} & = - s_{\beta} c^2_{\beta} C_{(l,d,u) \Phi_1}^{(11)} + s_{\beta} c^2_{\beta} C_{(l,d,u) \Phi_1}^{(22)} + c^3_{\beta} C_{(l,d,u) \Phi_1}^{(21)} - s^2_{\beta} c_{\beta} C_{(l,d,u) \Phi_1}^{(12)} \\
    & + s^2_{\beta} c_{\beta} C_{(l,d,u) \Phi_2}^{(22)} - s^2_{\beta} c_{\beta} C_{(l,d,u) \Phi_2}^{(11)} + s_{\beta} c^2_{\beta} C_{(l,d,u) \Phi_2}^{(21)} -s^3_{\beta} C_{(l,d,u) \Phi_2}^{(12)},
\end{split}
\end{equation}
\begin{equation}
\begin{split}
    C_{(l,d,u) H_1}^{(12)} & = - s_{\beta} c^2_{\beta} C_{(l,d,u) \Phi_1}^{(11)} + s_{\beta} c^2_{\beta} C_{(l,d,u) \Phi_1}^{(22)} - s^2_{\beta} c_{\beta} C_{(l,d,u) \Phi_1}^{(21)} + c^3_{\beta} C_{(l,d,u) \Phi_1}^{(12)} \\
    & + s^2_{\beta} c_{\beta} C_{(l,d,u) \Phi_2}^{(22)} - s^2_{\beta} c_{\beta} C_{(l,d,u) \Phi_2}^{(11)} - s^3_{\beta} C_{(l,d,u) \Phi_2}^{(21)} + s_{\beta} c^2_{\beta} C_{(l,d,u) \Phi_2}^{(12)},
\end{split}
\end{equation}
\begin{equation}
\begin{split}
    C_{(l,d,u) H_2}^{(22)} & = - s^3_{\beta} C_{(l,d,u) \Phi_1}^{(11)} - s_{\beta} c^2_{\beta} C_{(l,d,u) \Phi_1}^{(22)} + s^2_{\beta} c_{\beta} C_{(l,d,u) \Phi_1}^{(21)} + s^2_{\beta} c_{\beta} C_{(l,d,u) \Phi_1}^{(12)} \\
    & + c^3_{\beta} C_{(l,d,u) \Phi_2}^{(22)} + s^2_{\beta} c_{\beta} C_{(l,d,u) \Phi_2}^{(11)} - s_{\beta} c^2_{\beta} C_{(l,d,u) \Phi_2}^{(21)} - s_{\beta} c^2_{\beta} C_{(l,d,u) \Phi_2}^{(12)},
\label{eq:hb_massopH2}
\end{split} 
\end{equation}
\begin{equation}
\begin{split}
    C_{(l,d,u) H_2}^{(11)} & = - s_{\beta} c^2_{\beta} C_{(l,d,u) \Phi_1}^{(11)} - s^3_{\beta} C_{(l,d,u) \Phi_1}^{(22)} - s^2_{\beta} c_{\beta} C_{(l,d,u) \Phi_1}^{(21)} - s^2_{\beta} c_{\beta} C_{(l,d,u) \Phi_1}^{(12)} \\
    & + s^2_{\beta} c_{\beta} C_{(l,d,u) \Phi_2}^{(22)} + c^3_{\beta} C_{(l,d,u) \Phi_2}^{(11)} + s_{\beta} c^2_{\beta} C_{(l,d,u) \Phi_2}^{(21)} + s_{\beta} c^2_{\beta} C_{(l,d,u) \Phi_2}^{(12)},
\end{split}
\end{equation}
\begin{equation}
\begin{split}
    C_{(l,d,u) H_2}^{(21)} & = s^2_{\beta} c_{\beta} C_{(l,d,u) \Phi_1}^{(11)} - s^2_{\beta} c_{\beta} C_{(l,d,u) \Phi_1}^{(22)} - s_{\beta} c^2_{\beta} C_{(l,d,u) \Phi_1}^{(21)} + s^3_{\beta} C_{(l,d,u) \Phi_1}^{(12)} \\
    & + s_{\beta} c^2_{\beta} C_{(l,d,u) \Phi_2}^{(22)} - s_{\beta} c^2_{\beta} C_{(l,d,u) \Phi_2}^{(11)} + c^3_{\beta} C_{(l,d,u) \Phi_2}^{(21)} - s^2_{\beta} c_{\beta} C_{(l,d,u) \Phi_2}^{(12)}, 
\end{split}
\end{equation}
and 
\begin{equation}
\begin{split}
        C_{(l,d,u) H_2}^{(12)} & = s^2_{\beta} c_{\beta} C_{(l,d,u) \Phi_1}^{(11)} - s^2_{\beta} c_{\beta} C_{(l,d,u) \Phi_1}^{(22)} + s^3_{\beta} C_{(l,d,u) \Phi_1}^{(21)} - s_{\beta} c^2_{\beta} C_{(l,d,u) \Phi_1}^{(12)} \\
    & + s_{\beta} c^2_{\beta} C_{(l,d,u) \Phi_2}^{(22)} - s_{\beta} c^2_{\beta} C_{(l,d,u) \Phi_2}^{(11)} - s^2_{\beta} c_{\beta} C_{(l,d,u) \Phi_2}^{(21)} + c^3_{\beta} C_{(l,d,u) \Phi_2}^{(12)}. 
\end{split}
\end{equation}

\subsection{Class $\psi^2 X \phi$ operators}
\label{sec:hb_dipole_ops_match}
Operators in the $\psi^2 X \phi$ class translated to the Higgs basis are given in Table~\ref{tab:hb_dipole_ops}, whose Wilson coefficients are
\begin{equation}
    C_{(l,d,u) B H_1} = c_{\beta} C_{(l,d,u) B \Phi_1} + s_{\beta} C_{(l,d,u) B \Phi_2},
\end{equation}
\begin{equation}
    C_{(l,d,u) B H_2} = - s_{\beta} C_{(l,d,u) B \Phi_1} + c_{\beta} C_{(l,d,u) B \Phi_2},
\end{equation}
\begin{equation}
    C_{(l,d,u) W H_1} = c_{\beta} C_{(l,d,u) W \Phi_1} + s_{\beta} C_{(l,d,u) W \Phi_2},
\end{equation}
\begin{equation}
    C_{(l,d,u) W H_2} = - s_{\beta} C_{(l,d,u) W \Phi_1} + c_{\beta} C_{(l,d,u) W \Phi_2},
\end{equation}
\begin{equation}
    C_{(d,u) G H_1} = c_{\beta} C_{(d,u) G \Phi_1} + s_{\beta} C_{(d,u) G \Phi_2}, \\
\end{equation}
and
\begin{equation}
    C_{(d,u) G H_2} = - s_{\beta} C_{(d,u) G \Phi_1} + c_{\beta} C_{(d,u) G \Phi_2}.
\end{equation}

\subsection{Class $\psi^2 \phi^2 D $ operators}
\label{sec:hb_der_ops_match}
Operators in the $\psi^2 \phi^2 D$ class defined in the Higgs basis are provided in Table~\ref{tab:hb_der_ops}, with the following Wilson coefficients
\begin{equation}
    C_{H (e,d,u)}^{(11)} = c^2_{\beta} C_{\Phi (e,d,u)}^{(11)} + s^2_{\beta} C_{\Phi (e,d,u)}^{(22)} + s_{\beta} c_{\beta} (C_{\Phi (e,d,u)}^{(12)} + C_{\Phi (e,d,u)}^{(12)*}),    
\end{equation}
\begin{equation}
    C_{H (e,d,u)}^{(22)} = s^2_{\beta} C_{\Phi (e,d,u)}^{(11)} + c^2_{\beta} C_{\Phi (e,d,u)}^{(22)} - s_{\beta} c_{\beta} (C_{\Phi (e,d,u)}^{(12)} + C_{\Phi (e,d,u)}^{(12)*}),    
\end{equation}
\begin{equation}
    C_{H (e,d,u)}^{(12)} = - s_{\beta} c_{\beta} C_{\Phi (e,d,u)}^{(11)} + s_{\beta} c_{\beta} C_{\Phi (e,d,u)}^{(22)} + c^2_{\beta} C_{\Phi (e,d,u)}^{(12)} - s^2_{\beta} C_{\Phi (e,d,u)}^{(12)*},    
\end{equation}
\begin{equation}
    C_{H (l,q)}^{(11)[1,3]} = c^2_{\beta} C_{\Phi (l,q)}^{(11)[1,3]} + s^2_{\beta} C_{\Phi (l,q)}^{(22)[1,3]} + s_{\beta} c_{\beta} (C_{\Phi (l,q)}^{(12)[1,3]} + C_{\Phi (l,q)}^{(12)[1,3]*}),
\end{equation}
\begin{equation}
    C_{H (l,q)}^{(22)[1,3]} = s^2_{\beta} C_{\Phi (l,q)}^{(11)[1,3]} + c^2_{\beta} C_{\Phi (l,q)}^{(22)[1,3]} - s_{\beta} c_{\beta} (C_{\Phi (l,q)}^{(12)[1,3]} + C_{\Phi (l,q)}^{(12)[1,3]*}), 
\end{equation}
\begin{equation}
    C_{H (l,q)}^{(12)[1,3]} = - s_{\beta} c_{\beta} C_{\Phi (l,q)}^{(11)[1,3]} + s_{\beta} c_{\beta} C_{\Phi (l,q)}^{(22)[1,3]} + c^2_{\beta} C_{\Phi (l,q)}^{(12)[1,3]} - s^2_{\beta} C_{\Phi (l,q)}^{(12)[1,3]*},
\end{equation}
\begin{equation}
    C_{H ud}^{(11)} = c^2_{\beta} C_{\Phi u d}^{(11)} + s^2_{\beta} C_{\Phi u d}^{(22)} + 2 s_{\beta} c_{\beta} C_{\Phi u d}^{(21)},
\end{equation}
\begin{equation}
    C_{H ud}^{(22)} = s^2_{\beta} C_{\Phi u d}^{(11)} + c^2_{\beta} C_{\Phi u d}^{(22)} - 2 s_{\beta} c_{\beta} C_{\Phi u d}^{(21)},
\end{equation}
and
\begin{equation}
    C_{H ud}^{(21)} = - s_{\beta} c_{\beta} C_{\Phi u d}^{(11)} + s_{\beta} c_{\beta} C_{\Phi u d}^{(22)} + (c^2_{\beta} - s^2_{\beta}) C_{\Phi u d}^{(21)}.
\end{equation}

\subsection{Class $\phi^4 D^2$ operators}
\label{sec:hb_derb_ops_match}
The $\phi^4 D^2$ class of operators translated to the Higgs basis are listed in Table~\ref{tab:hb_der_b_ops}, with Wilson coefficients
\begin{equation}
\begin{split}
    C_{H (\partial^2, D)}^{(11)(11)} & = c^4_{\beta} C_{\Phi (\partial^2, D)}^{(11)(11)} + s^4_{\beta} C_{\Phi (\partial^2, D)}^{(22)(22)} + 2 s^2_{\beta} c^2_{\beta} C_{\Phi (\partial^2, D)}^{(11)(22)} \\
    & + s^2_{\beta} c^2_{\beta} (C_{\Phi (\partial^2, D)}^{(21)(21)} + C_{\Phi (\partial^2, D)}^{(21)(21)*}) + s^2_{\beta} c^2_{\beta} (C_{\Phi (\partial^2, D)}^{(21)(12)} + C_{\Phi (\partial^2, D)}^{(21)(12)*}) \\
    & + 2 s_{\beta} c^3_{\beta} (C_{\Phi (\partial^2, D)}^{(21)(11)} + C_{\Phi (\partial^2, D)}^{(21)(11)*}) + 2 s^3_{\beta} c_{\beta} (C_{\Phi (\partial^2, D)}^{(21)(22)} + C_{\Phi (\partial^2, D)}^{(21)(22)*}),   
\end{split}
\end{equation}
\begin{equation}
\begin{split}
    C_{H (\partial^2, D)}^{(22)(22)} & = s^4_{\beta} C_{\Phi (\partial^2, D)}^{(11)(11)} + c^4_{\beta} C_{\Phi (\partial^2, D)}^{(22)(22)} + 2 s^2_{\beta} c^2_{\beta} C_{\Phi (\partial^2, D)}^{(11)(22)} \\
    & + s^2_{\beta} c^2_{\beta} (C_{\Phi (\partial^2, D)}^{(21)(21)} + C_{\Phi (\partial^2, D)}^{(21)(21)*}) + s^2_{\beta} c^2_{\beta} (C_{\Phi (\partial^2, D)}^{(21)(12)} + C_{\Phi (\partial^2, D)}^{(21)(12)*}) \\
    & - 2 s^3_{\beta} c_{\beta} (C_{\Phi (\partial^2, D)}^{(21)(11)} + C_{\Phi (\partial^2, D)}^{(21)(11)*}) - 2 s_{\beta} c^3_{\beta} (C_{\Phi (\partial^2, D)}^{(21)(22)} + C_{\Phi (\partial^2, D)}^{(21)(22)*}),  
\end{split}
\end{equation}
\begin{equation}
\begin{split}
    C_{H (\partial^2, D)}^{(11)(22)} & = s^2_{\beta} c^2_{\beta} C_{\Phi (\partial^2, D)}^{(11)(11)} + s^2_{\beta} c^2_{\beta} C_{\Phi (\partial^2, D)}^{(22)(22)} + (c^4_{\beta} + s^4_{\beta}) C_{\Phi (\partial^2, D)}^{(11)(22)} \\
    & - s^2_{\beta} c^2_{\beta} (C_{\Phi (\partial^2, D)}^{(21)(21)} + C_{\Phi (\partial^2, D)}^{(21)(21)*}) - 2 s^2_{\beta} c^2_{\beta} C_{\Phi (\partial^2, D)}^{(21)(12)}\\
    & + (s^3_{\beta} c_{\beta} - s_{\beta} c^3_{\beta}) (C_{\Phi (\partial^2, D)}^{(21)(11)} + C_{\Phi (\partial^2, D)}^{(21)(11)*}) + (s_{\beta} c^3_{\beta} - s^3_{\beta} c_{\beta}) (C_{\Phi (\partial^2, D)}^{(21)(22)} + C_{\Phi (\partial^2, D)}^{(21)(22)*}),
\end{split}
\end{equation}
\begin{equation}
\begin{split}
    C_{H (\partial^2, D)}^{(21)(21)} & = s^2_{\beta} c^2_{\beta} C_{\Phi (\partial^2, D)}^{(11)(11)} + s^2_{\beta} c^2_{\beta} C_{\Phi (\partial^2, D)}^{(22)(22)} - 2 s^2_{\beta} c^2_{\beta} C_{\Phi (\partial^2, D)}^{(11)(22)} \\
    & + c^4_{\beta} C_{\Phi (\partial^2, D)}^{(21)(21)} + s^4_{\beta} C_{\Phi (\partial^2, D)}^{(21)(21)*} - 2 s^2_{\beta} c^2_{\beta} C_{\Phi (\partial^2, D)}^{(21)(12)} \\
    & - 2 s_{\beta} c^3_{\beta} C_{\Phi (\partial^2, D)}^{(21)(11)} + 2 s^3_{\beta} c_{\beta} C_{\Phi (\partial^2, D)}^{(21)(11)*} + 2 s_{\beta} c^3_{\beta} C_{\Phi (\partial^2, D)}^{(21)(22)} - 2 s^3_{\beta} c_{\beta} C_{\Phi (\partial^2, D)}^{(21)(22)*},
\end{split}
\end{equation}
\begin{equation}
\begin{split}
    C_{H (\partial^2, D)}^{(21)(12)} & = s^2_{\beta} c^2_{\beta} C_{\Phi (\partial^2, D)}^{(11)(11)} + s^2_{\beta} c^2_{\beta} C_{\Phi (\partial^2, D)}^{(22)(22)} - 2 s^2_{\beta} c^2_{\beta} C_{\Phi (\partial^2, D)}^{(11)(22)} \\
    & - s^2_{\beta} c^2_{\beta} (C_{\Phi (\partial^2, D)}^{(21)(21)} + C_{\Phi (\partial^2, D)}^{(21)(21)*}) + (c^4_{\beta} + s^4_{\beta}) C_{\Phi (\partial^2, D)}^{(21)(12)} \\
    & + (s^3_{\beta} c_{\beta} - s_{\beta} c^3_{\beta}) (C_{\Phi (\partial^2, D)}^{(21)(11)} + C_{\Phi (\partial^2, D)}^{(21)(11)*})  \\
    & + (s_{\beta} c^3_{\beta} - s^3_{\beta} c_{\beta}) (C_{\Phi (\partial^2, D)}^{(21)(22)} + C_{\Phi (\partial^2, D)}^{(21)(22)*}),
\end{split}
\end{equation}
\begin{equation}
\begin{split}
    C_{H(\partial^2, D)}^{(21)(11)} & = - s_{\beta} c^3_{\beta} C_{\Phi (\partial^2, D)}^{(11)(11)} + s^3_{\beta} c_{\beta} C_{\Phi (\partial^2, D)}^{(22)(22)} + (s_{\beta} c^3_{\beta} - s^3_{\beta} c_{\beta}) C_{\Phi (\partial^2, D)}^{(11)(22)} \\
    & + s_{\beta} c^3_{\beta} C_{\Phi (\partial^2, D)}^{(21)(21)} - s^3_{\beta} c_{\beta} C_{\Phi (\partial^2, D)}^{(21)(21)*} + (s_{\beta} c^3_{\beta} - s^3_{\beta} c_{\beta}) C_{\Phi (\partial^2, D)}^{(21)(12)} \\
    & + (c^4_{\beta} - s^2_{\beta} c^2_{\beta}) C_{\Phi (\partial^2, D)}^{(21)(11)} - 2 s^2_{\beta} c^2_{\beta} C_{\Phi (\partial^2, D)}^{(21)(11)*} + 2 s^2_{\beta} c^2_{\beta} C_{\Phi (\partial^2, D)}^{(21)(22)} + (s^2_{\beta} c^2_{\beta} - s^4_{\beta}) C_{\Phi (\partial^2, D)}^{(21)(22)*},
\end{split}
\end{equation}
\begin{equation}
\begin{split}
    C_{H(\partial^2, D)}^{(21)(22)} & = - s^3_{\beta} c_{\beta} C_{\Phi (\partial^2, D)}^{(11)(11)} + s_{\beta} c^3_{\beta} C_{\Phi (\partial^2, D)}^{(22)(22)} + (s^3_{\beta} c_{\beta} - s_{\beta} c^3_{\beta}) C_{\Phi (\partial^2, D)}^{(11)(22)} \\
    & - s_{\beta} c^3_{\beta} C_{\Phi (\partial^2, D)}^{(21)(21)} + s^3_{\beta} c_{\beta} C_{\Phi (\partial^2, D)}^{(21)(21)*} + (s^3_{\beta} c_{\beta} - s_{\beta} c^3_{\beta}) C_{\Phi (\partial^2, D)}^{(21)(12)} \\
    & + 2 s^2_{\beta} c^2_{\beta} C_{\Phi (\partial^2, D)}^{(21)(11)} + (s^2_{\beta} c^2_{\beta} - s^4_{\beta}) C_{\Phi (\partial^2, D)}^{(21)(11)*} + (c^4_{\beta} - s^2_{\beta} c^2_{\beta}) C_{\Phi (\partial^2, D)}^{(21)(22)} - 2 s^2_{\beta} c^2_{\beta} C_{\Phi (\partial^2, D)}^{(21)(22)*}.
\end{split}
\end{equation}

\subsection{Class $\phi^6$ operators}
\label{sec:hb_scalar_ops_match}
The $\phi^6$ class of operators in the Higgs basis are listed in Table~\ref{tab:hb_scalar_ops}, with Wilson coefficients
\begin{equation}
\begin{split}
    C_{H}^{(11)(11)(11)} & = c^6_{\beta} C_{\Phi}^{(11)(11)(11)} + s^2_{\beta} c^4_{\beta} C_{\Phi}^{(11)(11)(22)} + s^4_{\beta} c^2_{\beta} C_{\Phi}^{(11)(22)(22)} \\
    & + s_{\beta} c^5_{\beta} (C_{\Phi}^{(11)(11)(21)} + C_{\Phi}^{(11)(11)(21)*}) + s^5_{\beta} c_{\beta} (C_{\Phi}^{(22)(22)(21)} + C_{\Phi}^{(22)(22)(21)*})  + s^6_{\beta} C_{\Phi}^{(22)(22)(22)} \\
    & + s^2_{\beta} c^4_{\beta} (C_{\Phi}^{(11)(21)(21)} + C_{\Phi}^{(11)(21)(21)*})  + s^2_{\beta} c^4_{\beta} C_{\Phi}^{(11)(21)(12)} \\
    & + s^4_{\beta} c^2_{\beta} (C_{\Phi}^{(22)(21)(21)} + C_{\Phi}^{(22)(21)(21)*}) + s^4_{\beta} c^2_{\beta} C_{\Phi}^{(22)(21)(12)} \\
    & + s^3_{\beta} c^3_{\beta} (C_{\Phi}^{(21)(21)(21)} + C_{\Phi}^{(21)(21)(21)*}) \\
    & + s^3_{\beta} c^3_{\beta} (C_{\Phi}^{(21)(21)(12)} + C_{\Phi}^{(21)(21)(12)*}) + s^3_{\beta} c^3_{\beta} (C_{\Phi}^{(11)(22)(21)} + C_{\Phi}^{(11)(22)(21)*}),   
\end{split}
\end{equation}
\begin{equation}
\begin{split}
    C_{H}^{(11)(11)(22)} & = 3 s^2_{\beta} c^4_{\beta} C_{\Phi}^{(11)(11)(11)} + (c^6_{\beta} + 2 s^4_{\beta} c^2_{\beta}) C_{\Phi}^{(11)(11)(22)} + (s^6_{\beta} + 2 s^2_{\beta} c^4_{\beta}) C_{\Phi}^{(11)(22)(22)} \\
    & + (-s_{\beta} c^5_{\beta} + 2 s^3_{\beta} c^3_{\beta}) (C_{\Phi}^{(11)(11)(21)} + C_{\Phi}^{(11)(11)(21)*}) \\
    & + (-s^5_{\beta} c_{\beta} + 2 s^3_{\beta} c^3_{\beta}) (C_{\Phi}^{(22)(22)(21)} + C_{\Phi}^{(22)(22)(21)*}) + 3 s^4_{\beta} c^2_{\beta} C_{\Phi}^{(22)(22)(22)} \\
    & + ( s^4_{\beta} c^2_{\beta} - 2 s^2_{\beta} c^4_{\beta}) (C_{\Phi}^{(11)(21)(21)} + C_{\Phi}^{(11)(21)(21)*}) + (s^4_{\beta} c^2_{\beta} - 2 s^2_{\beta} c^4_{\beta})C_{\Phi}^{(11)(21)(12)} \\
    & + (s^2_{\beta} c^4_{\beta} - 2 s^4_{\beta} c^2_{\beta}) (C_{\Phi}^{(22)(21)(21)} + C_{\Phi}^{(22)(21)(21)*}) + (s^2_{\beta} c^4_{\beta} - 2 s^4_{\beta} c^2_{\beta}) C_{\Phi}^{(22)(21)(12)} \\
    & - 3 s^3_{\beta} c^3_{\beta} (C_{\Phi}^{(21)(21)(21)} + C_{\Phi}^{(21)(21)(21)*}) - 3 s^3_{\beta} c^3_{\beta} (C_{\Phi}^{(21)(21)(12)} + C_{\Phi}^{(21)(21)(12)*}) \\
    & + (-s^3_{\beta} c^3_{\beta} + s^5_{\beta} c_{\beta} + s_{\beta} c^5_{\beta}) (C_{\Phi}^{(11)(22)(21)} + C_{\Phi}^{(11)(22)(21)*}),
\end{split}
\end{equation}
\begin{equation}
\begin{split}
    C_{H}^{(11)(22)(22)} & = 3 s^4_{\beta} c^2_{\beta} C_{\Phi}^{(11)(11)(11)} + (s^6_{\beta} + 2 s^2_{\beta} c^4_{\beta}) C_{\Phi}^{(11)(11)(22)} + (c^6_{\beta} + 2 s^4_{\beta} c^2_{\beta}) C_{\Phi}^{(11)(22)(22)} \\
    & + (s^5_{\beta} c_{\beta} - 2 s^3_{\beta} c^3_{\beta}) (C_{\Phi}^{(11)(11)(21)} + C_{\Phi}^{(11)(11)(21)*}) \\
    & + (s_{\beta} c^5_{\beta} - 2 s^3_{\beta} c^3_{\beta}) (C_{\Phi}^{(22)(22)(21)} + C_{\Phi}^{(22)(22)(21)*}) + 3 s^2_{\beta} c^4_{\beta} C_{\Phi}^{(22)(22)(22)} \\
    & + ( s^2_{\beta} c^4_{\beta} - 2 s^4_{\beta} c^2_{\beta}) (C_{\Phi}^{(11)(21)(21)} + C_{\Phi}^{(11)(21)(21)*}) + (s^2_{\beta} c^4_{\beta} - 2 s^4_{\beta} c^2_{\beta}) C_{\Phi}^{(11)(21)(12)} \\
    & + (s^4_{\beta} c^2_{\beta} - 2 s^2_{\beta} c^4_{\beta}) (C_{\Phi}^{(22)(21)(21)} + C_{\Phi}^{(22)(21)(21)*}) + (s^4_{\beta} c^2_{\beta} - 2 s^2_{\beta} c^4_{\beta}) C_{\Phi}^{(22)(21)(12)} \\
    & + 3 s^3_{\beta} c^3_{\beta} (C_{\Phi}^{(21)(21)(21)} + C_{\Phi}^{(21)(21)(21)*}) + 3 s^3_{\beta} c^3_{\beta} (C_{\Phi}^{(21)(21)(12)}  + C_{\Phi}^{(21)(21)(12)*}) \\
    & + (s^3_{\beta} c^3_{\beta} - s^5_{\beta} c_{\beta} - s_{\beta} c^5_{\beta}) (C_{\Phi}^{(11)(22)(21)} + C_{\Phi}^{(11)(22)(21)*}),    
\end{split}
\end{equation}
\begin{equation}
\begin{split}
    C_{H}^{(11)(11)(21)} & = -3 s_{\beta} c^5_{\beta} C_{\Phi}^{(11)(11)(11)} + (s_{\beta} c^5_{\beta} - 2 s^3_{\beta} c^3_{\beta}) C_{\Phi}^{(11)(11)(22)} \\
    & + (-s^5_{\beta} c_{\beta} + 2 s^3_{\beta} c^3_{\beta}) C_{\Phi}^{(11)(22)(22)} + (c^6_{\beta} - 2 s^2_{\beta} c^4_{\beta}) C_{\Phi}^{(11)(11)(21)} - 3 s^2_{\beta} c^4_{\beta} C_{\Phi}^{(11)(11)(21)*} \\
    & + 3 s^4_{\beta} c^2_{\beta} C_{\Phi}^{(22)(22)(21)} + (-s^6_{\beta} + 2 s^4_{\beta} c^2_{\beta}) C_{\Phi}^{(22)(22)(21)*} + 3 s^5_{\beta} c_{\beta} C_{\Phi}^{(22)(22)(22)} \\
    & + (-s^3_{\beta} c^3_{\beta} + 2 s_{\beta} c^5_{\beta}) C_{\Phi}^{(11)(21)(21)} - 3 s^3_{\beta} c^3_{\beta} C_{\Phi}^{(11)(21)(21)*} \\
    & + (-2 s^3_{\beta} c^3_{\beta} + s_{\beta} c^5_{\beta}) C_{\Phi}^{(11)(21)(12)} \\
    & + 3 s^3_{\beta} c^3_{\beta} C_{\Phi}^{(22)(21)(21)} + (s^3_{\beta} c^3_{\beta} - 2 s^5_{\beta} c_{\beta}) C_{\Phi}^{(22)(21)(21)*} \\
    & + (2 s^3_{\beta} c^3_{\beta} - s^5_{\beta} c_{\beta}) C_{\Phi}^{(22)(21)(12)} \\
    & + 3 s^2_{\beta} c^4_{\beta} C_{\Phi}^{(21)(21)(21)} - 3 s^4_{\beta} c^2_{\beta} C_{\Phi}^{(21)(21)(21)*} \\
    & + (- s^4_{\beta} c^2_{\beta} + 2 s^2_{\beta} c^4_{\beta}) C_{\Phi}^{(21)(21)(12)} + (s^2_{\beta} c^4_{\beta} - 2 s^4_{\beta} c^2_{\beta}) C_{\Phi}^{(21)(21)(12)*} \\
    & + (2 s^2_{\beta} c^4_{\beta} - s^4_{\beta} c^2_{\beta}) C_{\Phi}^{(11)(22)(21)} + (-2 s^4_{\beta} c^2_{\beta} + s^2_{\beta} c^4_{\beta}) C_{\Phi}^{(11)(22)(21)*},    
\end{split}
\end{equation}
\begin{equation}
\begin{split}
    C_{H}^{(22)(22)(21)} & = -3 s^5_{\beta} c_{\beta} C_{\Phi}^{(11)(11)(11)} + (s^5_{\beta} c_{\beta} - 2 s^3_{\beta} c^3_{\beta}) C_{\Phi}^{(11)(11)(22)} \\
    & + (-s_{\beta} c^5_{\beta} + 2 s^3_{\beta} c^3_{\beta}) C_{\Phi}^{(11)(22)(22)} + 3 s^4_{\beta} c^2_{\beta} C_{\Phi}^{(11)(11)(21)} + (-s^6_{\beta} + 2 s^4_{\beta} c^2_{\beta}) C_{\Phi}^{(11)(11)(21)*} \\
    & + (c^6_{\beta} - 2 s^2_{\beta} c^4_{\beta}) C_{\Phi}^{(22)(22)(21)} - 3 s^2_{\beta} c^4_{\beta} C_{\Phi}^{(22)(22)(21)*} + 3 s_{\beta} c^5_{\beta} C_{\Phi}^{(22)(22)(22)} \\
    & - 3 s^3_{\beta} c^3_{\beta} C_{\Phi}^{(11)(21)(21)} +( - s^3_{\beta} c^3_{\beta}  + 2 s^5_{\beta} c_{\beta}) C_{\Phi}^{(11)(21)(21)*} \\
    & + (-2 s^3_{\beta} c^3_{\beta} + s^5_{\beta} c_{\beta}) C_{\Phi}^{(11)(21)(12)} \\
    & + (s^3_{\beta} c^3_{\beta} - 2 s_{\beta} c^5_{\beta}) C_{\Phi}^{(22)(21)(21)} + 3 s^3_{\beta} c^3_{\beta} C_{\Phi}^{(22)(21)(21)*} \\
    & + (2s^3_{\beta} c^3_{\beta} - s_{\beta} c^5_{\beta}) C_{\Phi}^{(22)(21)(12)} \\
    & + 3 s^2_{\beta} c^4_{\beta} C_{\Phi}^{(21)(21)(21)} - 3 s^4_{\beta} c^2_{\beta} C_{\Phi}^{(21)(21)(21)*} \\
    & + (- s^4_{\beta} c^2_{\beta} + 2 s^2_{\beta} c^4_{\beta}) C_{\Phi}^{(21)(21)(12)} + (s^2_{\beta} c^4_{\beta} - 2 s^4_{\beta} c^2_{\beta}) C_{\Phi}^{(21)(21)(12)*} \\
    & + (2 s^2_{\beta} c^4_{\beta} - s^4_{\beta} c^2_{\beta}) C_{\Phi}^{(11)(22)(21)} + (-2 s^4_{\beta} c^2_{\beta} + s^2_{\beta} c^4_{\beta}) C_{\Phi}^{(11)(22)(21)*},   
\end{split}
\end{equation}
\begin{equation}
\begin{split}
    C_{H}^{(22)(22)(22)} & = s^6_{\beta} C_{\Phi}^{(11)(11)(11)} + s^4_{\beta} c^2_{\beta} C_{\Phi}^{(11)(11)(22)} + s^2_{\beta} c^4_{\beta} C_{\Phi}^{(11)(22)(22)} \\
    & - s^5_{\beta} c_{\beta} (C_{\Phi}^{(11)(11)(21)} + C_{\Phi}^{(11)(11)(21)*}) - s_{\beta} c^5_{\beta} (C_{\Phi}^{(22)(22)(21)} + C_{\Phi}^{(11)(11)(21)*}) + c^6_{\beta} C_{\Phi}^{(22)(22)(22)} \\
    & + s^4_{\beta} c^2_{\beta} (C_{\Phi}^{(11)(21)(21)} + C_{\Phi}^{(11)(21)(21)*}) + s^4_{\beta} c^2_{\beta} C_{\Phi}^{(11)(21)(12)} \\
    & + s^2_{\beta} c^4_{\beta} (C_{\Phi}^{(22)(21)(21)} + C_{\Phi}^{(22)(21)(21)*}) + s^2_{\beta} c^4_{\beta} C_{\Phi}^{(22)(21)(12)} \\
    & - s^3_{\beta} c^3_{\beta} (C_{\Phi}^{(21)(21)(21)} + C_{\Phi}^{(21)(21)(21)*}) \\
    & - s^3_{\beta} c^3_{\beta} (C_{\Phi}^{(21)(21)(12)} + C_{\Phi}^{(21)(21)(12)*}) - s^3_{\beta} c^3_{\beta} (C_{\Phi}^{(11)(22)(21)} + C_{\Phi}^{(11)(22)(21)*}),  
\end{split}
\end{equation}
\begin{equation}
\begin{split}
    C_{H}^{(11)(21)(21)} & = 3 s^2_{\beta} c^4_{\beta} C_{\Phi}^{(11)(11)(11)} + (- 2 s^2_{\beta} c^4_{\beta} + s^4_{\beta} c^2_{\beta}) C_{\Phi}^{(11)(11)(22)} + (-2 s^4_{\beta} c^2_{\beta} + s^2_{\beta} c^4_{\beta}) C_{\Phi}^{(11)(22)(22)} \ \ \ \ \ \ \ \ \ \ \\
    & + (s^3_{\beta} c^3_{\beta} - 2s_{\beta} c^5_{\beta}) C_{\Phi}^{(11)(11)(21)} + 3 s^3_{\beta} c^3_{\beta} C_{\Phi}^{(11)(11)(21)*} \\
    & + 3 s^3_{\beta} c^3_{\beta} C_{\Phi}^{(22)(22)(21)} + ( s^3_{\beta} c^3_{\beta}  - 2 s^5_{\beta} c_{\beta}) C_{\Phi}^{(22)(22)(21)*} \\
    & + 3 s^4_{\beta} c^2_{\beta} C_{\Phi}^{(22)(22)(22)} \\
    & + (c^6_{\beta} - 2 s^2_{\beta} c^4_{\beta}) C_{\Phi}^{(11)(21)(21)} + 3 s^4_{\beta} c^2_{\beta} C_{\Phi}^{(11)(21)(21)*} \\
    & + (-2 s^2_{\beta} c^4_{\beta} + s^4_{\beta} c^2_{\beta}) C_{\Phi}^{(11)(21)(12)} \\
    & + 3 s^2_{\beta} c^4_{\beta} C_{\Phi}^{(22)(21)(21)} + (s^6_{\beta} - 2 s^4_{\beta} c^2_{\beta}) C_{\Phi}^{(22)(21)(21)*} \\
    & + ( s^2_{\beta} c^4_{\beta} - 2 s^4_{\beta} c^2_{\beta}) C_{\Phi}^{(22)(21)(12)} \\
    & + 3 s_{\beta} c^5_{\beta} C_{\Phi}^{(21)(21)(21)} + 3 s^5_{\beta} c_{\beta} C_{\Phi}^{(21)(21)(21)*} \\
    & + (s_{\beta} c^5_{\beta} - 2 s^3_{\beta} c^3_{\beta}) C_{\Phi}^{(21)(21)(12)} + (s^5_{\beta} c_{\beta} - 2 s^3_{\beta} c^3_{\beta}) C_{\Phi}^{(21)(21)(12)*} \\
    & + (-2 s^3_{\beta} c^3_{\beta} + s_{\beta} c^5_{\beta}) C_{\Phi}^{(11)(22)(21)} + (-2 s^3_{\beta} c^3_{\beta} + s^5_{\beta} c_{\beta}) C_{\Phi}^{(11)(22)(21)*},    
\end{split}
\end{equation}
\begin{equation}
\begin{split}
    C_{H}^{(11)(21)(12)} & = 6 s^2_{\beta} c^4_{\beta} C_{\Phi}^{(11)(11)(11)} + (-4 s^2_{\beta} c^4_{\beta} + 2 s^4_{\beta} c^2_{\beta}) C_{\Phi}^{(11)(11)(22)} + (-4 s^4_{\beta} c^2_{\beta} + 2 s^2_{\beta} c^4_{\beta}) C_{\Phi}^{(11)(22)(22)} \\
    & + (4 s^3_{\beta} c^3_{\beta} - 2 s_{\beta} c^5_{\beta}) (C_{\Phi}^{(11)(11)(21)} + C_{\Phi}^{(11)(11)(21)*}) \\
    & + (4 s^3_{\beta} c^3_{\beta} - 2 s^5_{\beta} c_{\beta}) (C_{\Phi}^{(22)(22)(21)} + C_{\Phi}^{(22)(22)(21)*}) + 6 s^4_{\beta} c^2_{\beta} C_{\Phi}^{(22)(22)(22)} \\
    & + (2 s^4_{\beta} c^2_{\beta} - 4 s^2_{\beta} c^4_{\beta}) (C_{\Phi}^{(11)(21)(21)} + C_{\Phi}^{(11)(21)(21)*}) + (c^6_{\beta} + 3 s^4_{\beta} c^2_{\beta} - 2 s^2_{\beta} c^4_{\beta}) C_{\Phi}^{(11)(21)(12)} \\
    & + (-4 s^4_{\beta} c^2_{\beta} + 2 s^2_{\beta} c^4_{\beta}) (C_{\Phi}^{(22)(21)(21)} + C_{\Phi}^{(22)(21)(21)*}) + (s^6_{\beta} + 3 s^2_{\beta} c^4_{\beta} - 2 s^4_{\beta} c^2_{\beta}) C_{\Phi}^{(22)(21)(12)} \\
    & - 6 s^3_{\beta} c^3_{\beta} (C_{\Phi}^{(21)(21)(21)} + C_{\Phi}^{(21)(21)(21)*}) \\
    & + (2 s^5_{\beta} c_{\beta} + 2 s_{\beta} c^5_{\beta} - 2 s^3_{\beta} c^3_{\beta}) (C_{\Phi}^{(21)(21)(12)} + C_{\Phi}^{(21)(21)(12)*}) \\
    & + (-4 s^3_{\beta} c^3_{\beta} + s_{\beta} c^5_{\beta} + s^5_{\beta} c_{\beta}) (C_{\Phi}^{(11)(22)(21)} + C_{\Phi}^{(11)(22)(21)*}),    
\end{split}
\end{equation}
\begin{equation}
\begin{split}
    C_{H}^{(22)(21)(21)} & = 3 s^4_{\beta} c^2_{\beta} C_{\Phi}^{(11)(11)(11)} + (- 2 s^4_{\beta} c^2_{\beta} + s^2_{\beta} c^4_{\beta}) C_{\Phi}^{(11)(11)(22)} + (-2 s^2_{\beta} c^4_{\beta} + s^4_{\beta} c^2_{\beta}) C_{\Phi}^{(11)(22)(22)} \\
    & - 3 s^3_{\beta} c^3_{\beta} C_{\Phi}^{(11)(11)(21)} + (- s^3_{\beta} c^3_{\beta} + 2 s^5_{\beta} c_{\beta}) C_{\Phi}^{(11)(11)(21)*} \\
    & + (- s^3_{\beta} c^3_{\beta} + 2 s_{\beta} c^5_{\beta}) C_{\Phi}^{(22)(22)(21)} - 3 s^3_{\beta} c^3_{\beta} C_{\Phi}^{(22)(22)(21)*} \\
    & + 3 s^2_{\beta} c^4_{\beta} C_{\Phi}^{(22)(22)(22)} \\
    & + 3 s^2_{\beta} c^4_{\beta} C_{\Phi}^{(11)(21)(21)} + (s^6_{\beta} - 2 s^4_{\beta} c^2_{\beta}) C_{\Phi}^{(11)(21)(21)*} \\
    & + (-2 s^4_{\beta} c^2_{\beta} + s^2_{\beta} c^4_{\beta}) C_{\Phi}^{(11)(21)(12)} \\
    & + (c^6_{\beta} - 2 s^2_{\beta} c^4_{\beta}) C_{\Phi}^{(22)(21)(21)} +  3 s^4_{\beta} c^2_{\beta} C_{\Phi}^{(22)(21)(21)*} \\
    & + ( - 2 s^2_{\beta} c^4_{\beta} + s^4_{\beta} c^2_{\beta}) C_{\Phi}^{(22)(21)(12)} \\
    & - 3 s_{\beta} c^5_{\beta} C_{\Phi}^{(21)(21)(21)} - 3 s^5_{\beta} c_{\beta} C_{\Phi}^{(21)(21)(21)*} \\
    & + (-s_{\beta} c^5_{\beta} + 2 s^3_{\beta} c^3_{\beta}) C_{\Phi}^{(21)(21)(12)} + (-s^5_{\beta} c_{\beta} + 2 s^3_{\beta} c^3_{\beta}) C_{\Phi}^{(21)(21)(12)*} \\
    & + (2 s^3_{\beta} c^3_{\beta} - s_{\beta} c^5_{\beta}) C_{\Phi}^{(11)(22)(21)} + (2 s^3_{\beta} c^3_{\beta} - s^5_{\beta} c_{\beta}) C_{\Phi}^{(11)(22)(21)*},    
\end{split}
\end{equation}
\begin{equation}
\begin{split}
    C_{H}^{(22)(21)(12)} & = 6 s^4_{\beta} c^2_{\beta} C_{\Phi}^{(11)(11)(11)} + (-4 s^4_{\beta} c^2_{\beta} + 2 s^2_{\beta} c^4_{\beta}) C_{\Phi}^{(11)(11)(22)} + (-4 s^2_{\beta} c^4_{\beta} + 2 s^4_{\beta} c^2_{\beta}) C_{\Phi}^{(11)(22)(22)} \\
    & + (- 4 s^3_{\beta} c^3_{\beta} + 2 s^5_{\beta} c_{\beta}) (C_{\Phi}^{(11)(11)(21)} + C_{\Phi}^{(11)(11)(21)*}) \\
    & + (- 4 s^3_{\beta} c^3_{\beta} + 2 s_{\beta} c^5_{\beta}) (C_{\Phi}^{(22)(22)(21)} + C_{\Phi}^{(22)(22)(21)*}) + 6 s^2_{\beta} c^4_{\beta} C_{\Phi}^{(22)(22)(22)} \\
    & + (2 s^2_{\beta} c^4_{\beta} - 4 s^4_{\beta} c^2_{\beta}) (C_{\Phi}^{(11)(21)(21)} + C_{\Phi}^{(11)(21)(21)*}) + (s^6_{\beta} + 3 s^2_{\beta} c^4_{\beta} - 2 s^4_{\beta} c^2_{\beta}) C_{\Phi}^{(11)(21)(12)} \\
    & + (-4 s^2_{\beta} c^4_{\beta} + 2 s^4_{\beta} c^2_{\beta}) (C_{\Phi}^{(22)(21)(21)} + C_{\Phi}^{(22)(21)(21)*}) + (c^6_{\beta} + 3 s^4_{\beta} c^2_{\beta} - 2 s^2_{\beta} c^4_{\beta}) C_{\Phi}^{(22)(21)(12)} \\
    & + 6 s^3_{\beta} c^3_{\beta} (C_{\Phi}^{(21)(21)(21)} + C_{\Phi}^{(21)(21)(21)*}) \\
    & + (- 2 s^5_{\beta} c_{\beta} - 2 s_{\beta} c^5_{\beta} + 2 s^3_{\beta} c^3_{\beta}) (C_{\Phi}^{(21)(21)(12)} + C_{\Phi}^{(21)(21)(12)*}) \\
    & + (4 s^3_{\beta} c^3_{\beta} - s_{\beta} c^5_{\beta} - s^5_{\beta} c_{\beta}) (C_{\Phi}^{(11)(22)(21)} + C_{\Phi}^{(11)(22)(21)*}) ,
\end{split}
\end{equation}
\begin{equation}
\begin{split}
    C_{H}^{(21)(21)(21)} & = - s^3_{\beta} c^3_{\beta} C_{\Phi}^{(11)(11)(11)} + s^3_{\beta} c^3_{\beta} C_{\Phi}^{(11)(11)(22)} - s^3_{\beta} c^3_{\beta} C_{\Phi}^{(11)(22)(22)} \\
    & + s^2_{\beta} c^4_{\beta} C_{\Phi}^{(11)(11)(21)} - s^4_{\beta} c^2_{\beta} C_{\Phi}^{(11)(11)(21)*} \\
    & + s^2_{\beta} c^4_{\beta} C_{\Phi}^{(22)(22)(21)} - s^4_{\beta} c^2_{\beta} C_{\Phi}^{(22)(22)(21)*} + s^3_{\beta} c^3_{\beta} C_{\Phi}^{(22)(22)(22)} \\
    & - s_{\beta} c^5_{\beta} C_{\Phi}^{(11)(21)(21)} - s^5_{\beta} c_{\beta} C_{\Phi}^{(11)(21)(21)*} + s^3_{\beta} c^3_{\beta} C_{\Phi}^{(11)(21)(12)} \\
    & + s_{\beta} c^5_{\beta} C_{\Phi}^{(22)(21)(21)} + s^5_{\beta} c_{\beta} C_{\Phi}^{(22)(21)(21)*} - s^3_{\beta} c^3_{\beta} C_{\Phi}^{(22)(21)(12)} \\
    & + c^6_{\beta} C_{\Phi}^{(21)(21)(21)} - s^6_{\beta} C_{\Phi}^{(21)(21)(21)*} - s^2_{\beta} c^4_{\beta} C_{\Phi}^{(21)(21)(12)} + s^4_{\beta} c^2_{\beta} C_{\Phi}^{(21)(21)(12)*} \\
    & - s^2_{\beta} c^4_{\beta} C_{\Phi}^{(11)(22)(21)} + s^4_{\beta} c^2_{\beta} C_{\Phi}^{(11)(22)(21)*},    
\end{split}
\end{equation}
\begin{equation}
\begin{split}
    C_{H}^{(21)(21)(12)} & = - s^3_{\beta} c^3_{\beta} C_{\Phi}^{(11)(11)(11)} + 3 s^3_{\beta} c^3_{\beta} C_{\Phi}^{(11)(11)(22)} - 3 s^3_{\beta} c^3_{\beta} C_{\Phi}^{(11)(22)(22)} \\
    & + (-s^4_{\beta} c^2_{\beta} + 2 s^2_{\beta} c^4_{\beta}) C_{\Phi}^{(11)(11)(21)} + (s^2_{\beta} c^4_{\beta} - 2 s^4_{\beta} c^2_{\beta}) C_{\Phi}^{(11)(11)(21)*} \\
    & + (- s^4_{\beta} c^2_{\beta} + 2 s^2_{\beta} c^4_{\beta}) C_{\Phi}^{(22)(22)(21)} + (s^2_{\beta} c^4_{\beta} - 2 s^4_{\beta} c^2_{\beta}) C_{\Phi}^{(22)(22)(21)*} \\
    & + 3 s^3_{\beta} c^3_{\beta} C_{\Phi}^{(22)(22)(22)} \\
    & + (- s_{\beta} c^5_{\beta} + 2 s^3_{\beta} c^3_{\beta}) C_{\Phi}^{(11)(21)(21)} + (- s^5_{\beta} c_{\beta} + 2 s^3_{\beta} c^3_{\beta}) C_{\Phi}^{(11)(21)(21)*} \\
    & + (s^3_{\beta} c^3_{\beta} - s^5_{\beta} c_{\beta} - s_{\beta} c^5_{\beta}) C_{\Phi}^{(11)(21)(12)} \\
    & + (s_{\beta} c^5_{\beta} - 2 s^3_{\beta} c^3_{\beta}) C_{\Phi}^{(22)(21)(21)} + (s^5_{\beta} c_{\beta} - 2 s^3_{\beta} c^3_{\beta}) C_{\Phi}^{(11)(22)(22)*} \\
    & + (- s^3_{\beta} c^3_{\beta} + s^5_{\beta} c_{\beta} + s_{\beta} c^5_{\beta}) C_{\Phi}^{(22)(21)(12)} \\
    & - 3 s^2_{\beta} c^4_{\beta} C_{\Phi}^{(21)(21)(21)} + 3 s^4_{\beta} c^2_{\beta} C_{\Phi}^{(21)(21)(21)*} \\ 
    & + (c^6_{\beta} + 2 s^4_{\beta} c^2_{\beta}) C_{\Phi}^{(21)(21)(12)} + (- s^6_{\beta} - 2 s^2_{\beta} c^4_{\beta}) C_{\Phi}^{(21)(21)(12)*} \\
    & + (s^4_{\beta} c^2_{\beta} - 2 s^2_{\beta} c^4_{\beta}) C_{\Phi}^{(11)(22)(21)} + (- s^2_{\beta} c^4_{\beta} + 2 s^4_{\beta} c^2_{\beta}) C_{\Phi}^{(11)(22)(21)*},
\end{split}
\end{equation}
and 
\begin{equation}
\begin{split}
    C_{H}^{(11)(22)(21)} & = - 6 s^3_{\beta} c^3_{\beta} C_{\Phi}^{(11)(11)(11)} + (- 2 s_{\beta} c^5_{\beta} - 2 s^5_{\beta} c_{\beta} + 2 s^3_{\beta} c^3_{\beta}) C_{\Phi}^{(11)(11)(22)} \\
    & + (2 s^5_{\beta} c_{\beta} + s_{\beta} c^5_{\beta} - 2 s^3_{\beta} c^3_{\beta}) C_{\Phi}^{(11)(22)(22)} \\
    & + (4 s^2_{\beta} c^4_{\beta} - 2 s^4_{\beta} c^2_{\beta}) C_{\Phi}^{(11)(11)(21)} + (2 s^2_{\beta} c^4_{\beta} - 4 s^4_{\beta} c^2_{\beta}) C_{\Phi}^{(11)(11)(21)*} \\
    & + (4 s^2_{\beta} c^4_{\beta} - 2 s^4_{\beta} c^2_{\beta}) C_{\Phi}^{(22)(22)(21)} + (-4 s^4_{\beta} c^2_{\beta} + 2 s^2_{\beta} c^4_{\beta}) C_{\Phi}^{(22)(22)(21)*} \\
    & + 6 s^3_{\beta} c^3_{\beta} C_{\Phi}^{(22)(22)(22)} \\
    & + (4 s^3_{\beta} c^3_{\beta} - 2 s_{\beta} c^5_{\beta}) C_{\Phi}^{(11)(21)(21)} + (4 s^3_{\beta} c^3_{\beta} - 2 s^5_{\beta} c_{\beta}) C_{\Phi}^{(11)(21)(21)*} \\
    & + (- s^5_{\beta} c_{\beta} + 4 s^3_{\beta} c^3_{\beta} - s_{\beta} c^5_{\beta}) C_{\Phi}^{(11)(21)(12)} \\
    & + (- 4 s^3_{\beta} c^3_{\beta} + 2 s_{\beta} c^5_{\beta}) C_{\Phi}^{(22)(21)(21)} + (- 4 s^3_{\beta} c^3_{\beta} + 2 s^5_{\beta} c_{\beta}) C_{\Phi}^{(22)(21)(21)*} \\
    & + (s_{\beta} c^5_{\beta} - 4 s^3_{\beta} c^3_{\beta} + s^5_{\beta} c_{\beta}) C_{\Phi}^{(22)(21)(12)} \\
    & - 6 s^2_{\beta} c^4_{\beta} C_{\Phi}^{(21)(21)(21)} + 6 s^4_{\beta} c^2_{\beta} C_{\Phi}^{(21)(21)(21)*} \\
    & + (2 s^4_{\beta} c^2_{\beta} - 4 s^2_{\beta} c^4_{\beta}) C_{\Phi}^{(21)(21)(12)} + (-2 s^2_{\beta} c^4_{\beta} + 4 s^4_{\beta} c^2_{\beta}) C_{\Phi}^{(21)(21)(12)*} \\
    & + (c^6_{\beta} + 3 s^4_{\beta} c^2_{\beta} - 2 s^2_{\beta} c^4_{\beta}) C_{\Phi}^{(11)(22)(21)} + (- s^6_{\beta} - 3 s^2_{\beta} c^4_{\beta} + 2 s^4_{\beta} c^2_{\beta}) C_{\Phi}^{(11)(22)(21)*}.
\end{split}
\end{equation}

\subsection{Class $X^2 \phi^2$ operators}
\label{sec:hb_mixed_ops_match}
Finally, for $X^2 \phi^2$ operators listed in Table~\ref{tab:hb_mix_ops}, the Wilson coefficients are
\begin{equation}
    C_{H (G,W,B)}^{(11)} = c^2_{\beta} C_{\Phi (G,W,B)}^{(11)} + s^2_{\beta} C_{\Phi (G,W,B)}^{(22)} + s_{\beta} c_{\beta} (C_{\Phi (G,W,B)}^{(21)} + C_{\Phi (G,W,B)}^{(21)*}), 
\end{equation}
\begin{equation}
    C_{H (G,W,B)}^{(22)} = s^2_{\beta} C_{\Phi (G,W,B)}^{(11)} + c^2_{\beta} C_{\Phi (G,W,B)}^{(22)} - s_{\beta} c_{\beta} (C_{\Phi (G,W,B)}^{(21)} + C_{\Phi (G,W,B)}^{(21)*}),
\end{equation}
\begin{equation}
    C_{H (G,W,B)}^{(21)} = - s_{\beta} c_{\beta} C_{\Phi (G,W,B)}^{(11)} + s_{\beta} c_{\beta} C_{\Phi (G,W,B)}^{(22)} + c^2_{\beta} C_{\Phi (G,W,B)}^{(21)} - s^2_{\beta} C_{\Phi (G,W,B)}^{(21)*},
\end{equation}
\begin{equation}
    C_{H (\widetilde{G},\widetilde{W},\widetilde{B} )}^{(11)} = c^2_{\beta} C_{\Phi (\widetilde{G},\widetilde{W},\widetilde{B} )}^{(11)} + s^2_{\beta} C_{\Phi (\widetilde{G},\widetilde{W},\widetilde{B} )}^{(22)} + s_{\beta} c_{\beta} (C_{\Phi (\widetilde{G},\widetilde{W},\widetilde{B} )}^{(21)} + C_{\Phi (\widetilde{G},\widetilde{W},\widetilde{B} )}^{(21)*}),
\end{equation}
\begin{equation}
    C_{H (\widetilde{G},\widetilde{W},\widetilde{B} )}^{(22)} = s^2_{\beta} C_{\Phi (\widetilde{G},\widetilde{W},\widetilde{B} )}^{(11)} + c^2_{\beta} C_{\Phi (\widetilde{G},\widetilde{W},\widetilde{B} )}^{(22)} - s_{\beta} c_{\beta} (C_{\Phi (\widetilde{G},\widetilde{W},\widetilde{B} )}^{(21)} + C_{\Phi (\widetilde{G},\widetilde{W},\widetilde{B} )}^{(21)*}),
\end{equation}
\begin{equation}
    C_{H (\widetilde{G},\widetilde{W},\widetilde{B} )}^{(21)} = - s_{\beta} c_{\beta} C_{\Phi (\widetilde{G},\widetilde{W},\widetilde{B} )}^{(11)} + s_{\beta} c_{\beta} C_{\Phi (\widetilde{G},\widetilde{W},\widetilde{B} )}^{(22)} + c^2_{\beta} C_{\Phi (\widetilde{G},\widetilde{W},\widetilde{B} )}^{(21)} - s^2_{\beta} C_{\Phi (\widetilde{G},\widetilde{W},\widetilde{B} )}^{(21)*},
\end{equation}
\begin{equation}
    C_{H (WB, \widetilde{W}B)}^{(11)} = c^2_{\beta} C_{\Phi  (WB, \widetilde{W}B)}^{(11)} + s^2_{\beta} C_{\Phi  (WB, \widetilde{W}B)}^{(22)} + s_{\beta} c_{\beta} (C_{\Phi  (WB, \widetilde{W}B)}^{(21)} + C_{\Phi  (WB, \widetilde{W}B)}^{(21)*}),
\end{equation}
\begin{equation}
    C_{H (WB, \widetilde{W}B)}^{(22)} = s^2_{\beta} C_{\Phi  (WB, \widetilde{W}B)}^{(11)} + c^2_{\beta} C_{\Phi  (WB, \widetilde{W}B)}^{(22)} - s_{\beta} c_{\beta} (C_{\Phi  (WB, \widetilde{W}B)}^{(21)} + C_{\Phi (WB, \widetilde{W}B)}^{(21)*}),
\end{equation}
and
\begin{equation}
    C_{H  (WB, \widetilde{W}B)}^{(21)} = - s_{\beta} c_{\beta} C_{\Phi  (WB, \widetilde{W}B)}^{(11)} + s_{\beta} c_{\beta} C_{\Phi  (WB, \widetilde{W}B)}^{(22)} + c^2_{\beta} C_{\Phi  (WB, \widetilde{W}B)}^{(21)} - s^2_{\beta} C_{\Phi  (WB, \widetilde{W}B)}^{(21)*}.
\end{equation}


\begin{thebibliography}{99}
\bibitem{Appelquist:1974tg}
T.~Appelquist and J.~Carazzone,
Phys. Rev. D \textbf{11}, 2856 (1975)
doi:10.1103/PhysRevD.11.2856

\bibitem{Grzadkowski:2010es}
B.~Grzadkowski, M.~Iskrzynski, M.~Misiak and J.~Rosiek,
JHEP \textbf{10}, 085 (2010)
doi:10.1007/JHEP10(2010)085
[arXiv:1008.4884 [hep-ph]].

\bibitem{Murphy:2020rsh}
C.~W.~Murphy,
JHEP \textbf{10}, 174 (2020)
doi:10.1007/JHEP10(2020)174
[arXiv:2005.00059 [hep-ph]].

\bibitem{Gildener:1976ih}
E.~Gildener and S.~Weinberg,
Phys. Rev. D \textbf{13}, 3333 (1976)
doi:10.1103/PhysRevD.13.3333

\bibitem{Lavoura:1994fv}
L.~Lavoura and J.~P.~Silva,
Phys. Rev. D \textbf{50}, 4619-4624 (1994)
doi:10.1103/PhysRevD.50.4619
[arXiv:hep-ph/9404276 [hep-ph]].

\bibitem{Lavoura:1994yu}
L.~Lavoura,
Phys. Rev. D \textbf{50}, 7089-7092 (1994)
doi:10.1103/PhysRevD.50.7089
[arXiv:hep-ph/9405307 [hep-ph]].

\bibitem{Botella:1994cs}
F.~J.~Botella and J.~P.~Silva,
Phys. Rev. D \textbf{51}, 3870-3875 (1995)
doi:10.1103/PhysRevD.51.3870
[arXiv:hep-ph/9411288 [hep-ph]].

\bibitem{Davidson:2005cw}
S.~Davidson and H.~E.~Haber,
Phys. Rev. D \textbf{72}, 035004 (2005)
[erratum: Phys. Rev. D \textbf{72}, 099902 (2005)]
doi:10.1103/PhysRevD.72.099902
[arXiv:hep-ph/0504050 [hep-ph]].

\bibitem{Crivellin:2016ihg}
A.~Crivellin, M.~Ghezzi and M.~Procura,
JHEP \textbf{09}, 160 (2016)
doi:10.1007/JHEP09(2016)160
[arXiv:1608.00975 [hep-ph]].

\bibitem{Karmakar:2017yek}
S.~Karmakar and S.~Rakshit,
JHEP \textbf{10}, 048 (2017)
doi:10.1007/JHEP10(2017)048
[arXiv:1707.00716 [hep-ph]].

\bibitem{Anisha:2019nzx}
Anisha, S.~Das Bakshi, J.~Chakrabortty and S.~Prakash,
JHEP \textbf{09}, 035 (2019)
doi:10.1007/JHEP09(2019)035
[arXiv:1905.11047 [hep-ph]].

\bibitem{Feruglio:1992wf}
F.~Feruglio,
Int. J. Mod. Phys. A \textbf{8}, 4937-4972 (1993)
doi:10.1142/S0217751X93001946
[arXiv:hep-ph/9301281 [hep-ph]].

\bibitem{Contino:2010mh}
R.~Contino, C.~Grojean, M.~Moretti, F.~Piccinini and R.~Rattazzi,
JHEP \textbf{05}, 089 (2010)
doi:10.1007/JHEP05(2010)089
[arXiv:1002.1011 [hep-ph]].

\bibitem{Brivio:2017vri}
I.~Brivio and M.~Trott,
Phys. Rept. \textbf{793}, 1-98 (2019)
doi:10.1016/j.physrep.2018.11.002
[arXiv:1706.08945 [hep-ph]].

\bibitem{Buchalla:2023hqk}
G.~Buchalla, F.~K\"onig, C.~M\"uller-Salditt and F.~Pandler,
[arXiv:2312.13885 [hep-ph]].

\bibitem{Gorbahn:2015gxa}
M.~Gorbahn, J.~M.~No and V.~Sanz,
JHEP \textbf{10}, 036 (2015)
doi:10.1007/JHEP10(2015)036
[arXiv:1502.07352 [hep-ph]].

\bibitem{Belusca-Maito:2016dqe}
H.~B\'elusca-Ma\"\i{}to, A.~Falkowski, D.~Fontes, J.~C.~Rom\~ao and J.~P.~Silva,
Eur. Phys. J. C \textbf{77}, no.3, 176 (2017)
doi:10.1140/epjc/s10052-017-4745-5
[arXiv:1611.01112 [hep-ph]].

\bibitem{DasBakshi:2024krs}
S.~Das Bakshi, S.~Dawson, D.~Fontes and S.~Homiller,
Phys. Rev. D \textbf{109}, no.7, 075022 (2024)
doi:10.1103/PhysRevD.109.075022
[arXiv:2401.12279 [hep-ph]].

\bibitem{Branco:2011iw}
G.~C.~Branco, P.~M.~Ferreira, L.~Lavoura, M.~N.~Rebelo, M.~Sher and J.~P.~Silva,
Phys. Rept. \textbf{516}, 1-102 (2012)
doi:10.1016/j.physrep.2012.02.002
[arXiv:1106.0034 [hep-ph]].

\bibitem{Dermisek:2023tgq}
R.~Dermisek, K.~Hermanek, N.~McGinnis and S.~Yoon,
Phys. Rev. D \textbf{108}, no.5, 055019 (2023)
doi:10.1103/PhysRevD.108.055019
[arXiv:2306.13212 [hep-ph]].

\bibitem{Gunion:2002zf}
J.~F.~Gunion and H.~E.~Haber,
Phys. Rev. D \textbf{67}, 075019 (2003)
doi:10.1103/PhysRevD.67.075019
[arXiv:hep-ph/0207010 [hep-ph]].

\bibitem{Krause:2016oke}
M.~Krause, R.~Lorenz, M.~Muhlleitner, R.~Santos and H.~Ziesche,
JHEP \textbf{09}, 143 (2016)
doi:10.1007/JHEP09(2016)143
[arXiv:1605.04853 [hep-ph]].

\bibitem{Bijnens:2018rqw}
J.~Bijnens, J.~Oredsson and J.~Rathsman,
Phys. Lett. B \textbf{792}, 238-243 (2019)
doi:10.1016/j.physletb.2019.03.051
[arXiv:1810.04483 [hep-ph]].

\bibitem{Weinberg:1979sa}
S.~Weinberg,
Phys. Rev. Lett. \textbf{43}, 1566-1570 (1979)
doi:10.1103/PhysRevLett.43.1566

\bibitem{Liao:2010ny}
Y.~Liao,
Phys. Lett. B \textbf{698}, 288-292 (2011)
doi:10.1016/j.physletb.2011.03.025
[arXiv:1010.5326 [hep-ph]].

\bibitem{Dedes:2017zog}
A.~Dedes, W.~Materkowska, M.~Paraskevas, J.~Rosiek and K.~Suxho,
JHEP \textbf{06}, 143 (2017)
doi:10.1007/JHEP06(2017)143
[arXiv:1704.03888 [hep-ph]].

\bibitem{Haber:1978jt}
H.~E.~Haber, G.~L.~Kane and T.~Sterling,
Nucl. Phys. B \textbf{161}, 493-532 (1979)
doi:10.1016/0550-3213(79)90225-6

\bibitem{Gunion:1989we}
J.~F.~Gunion, H.~E.~Haber, G.~L.~Kane and S.~Dawson,
Front. Phys. \textbf{80}, 1-404 (2000)
SCIPP-89/13.

\bibitem{Fayet:1974pd}
P.~Fayet,
Nucl. Phys. B \textbf{90}, 104-124 (1975)
doi:10.1016/0550-3213(75)90636-7

\bibitem{Aoki:2008av}
M.~Aoki, S.~Kanemura and O.~Seto,
Phys. Rev. Lett. \textbf{102}, 051805 (2009)
doi:10.1103/PhysRevLett.102.051805
[arXiv:0807.0361 [hep-ph]].

\bibitem{Goh:2009wg}
H.~S.~Goh, L.~J.~Hall and P.~Kumar,
JHEP \textbf{05}, 097 (2009)
doi:10.1088/1126-6708/2009/05/097
[arXiv:0902.0814 [hep-ph]].

\bibitem{Broggio:2014mna}
A.~Broggio, E.~J.~Chun, M.~Passera, K.~M.~Patel and S.~K.~Vempati,
JHEP \textbf{11}, 058 (2014)
doi:10.1007/JHEP11(2014)058
[arXiv:1409.3199 [hep-ph]].

\bibitem{Abe:2015oca}
T.~Abe, R.~Sato and K.~Yagyu,
JHEP \textbf{07}, 064 (2015)
doi:10.1007/JHEP07(2015)064
[arXiv:1504.07059 [hep-ph]].

\bibitem{Dermisek:2020cod}
R.~Dermisek, K.~Hermanek, N.~McGinnis and N.~McGinnis,
Phys. Rev. Lett. \textbf{126}, no.19, 191801 (2021)
doi:10.1103/PhysRevLett.126.191801
[arXiv:2011.11812 [hep-ph]].

\bibitem{Dermisek:2021ajd}
R.~Dermisek, K.~Hermanek and N.~McGinnis,
Phys. Rev. D \textbf{104}, no.5, 055033 (2021)
doi:10.1103/PhysRevD.104.055033
[arXiv:2103.05645 [hep-ph]].

\bibitem{Barnett:1983mm}
R.~M.~Barnett, G.~Senjanovic, L.~Wolfenstein and D.~Wyler,
Phys. Lett. B \textbf{136}, 191-195 (1984)
doi:10.1016/0370-2693(84)91179-1

\bibitem{Barnett:1984zy}
R.~M.~Barnett, G.~Senjanovic and D.~Wyler,
Phys. Rev. D \textbf{30}, 1529 (1984)
doi:10.1103/PhysRevD.30.1529

\bibitem{Martin:1997ns}
S.~P.~Martin,
Adv. Ser. Direct. High Energy Phys. \textbf{18}, 1-98 (1998)
doi:10.1142/9789812839657\_0001
[arXiv:hep-ph/9709356 [hep-ph]].

\bibitem{Dermisek:2023rvv}
R.~Dermisek, K.~Hermanek, T.~Lee, N.~McGinnis and S.~Yoon,
Phys. Rev. D \textbf{109}, no.9, 095003 (2024)
doi:10.1103/PhysRevD.109.095003
[arXiv:2311.05078 [hep-ph]].

\bibitem{Dermisek:2023nhe}
R.~Dermisek, K.~Hermanek, N.~McGinnis and S.~Yoon,
Phys. Rev. D \textbf{107}, no.9, 095043 (2023)
doi:10.1103/PhysRevD.107.095043
[arXiv:2302.14144 [hep-ph]].

\bibitem{Dermisek:2022aec}
R.~Dermisek, K.~Hermanek, N.~McGinnis and S.~Yoon,
Phys. Rev. Lett. \textbf{129}, no.22, 221801 (2022)
doi:10.1103/PhysRevLett.129.221801
[arXiv:2205.14243 [hep-ph]].

\bibitem{ATLAS:2024lyh}
G.~Aad \textit{et al.} [ATLAS],
[arXiv:2402.05742 [hep-ex]].

\bibitem{Accettura:2023ked}
C.~Accettura, D.~Adams, R.~Agarwal, C.~Ahdida, C.~Aim\`e, N.~Amapane, D.~Amorim, P.~Andreetto, F.~Anulli and R.~Appleby, \textit{et al.}
Eur. Phys. J. C \textbf{83}, no.9, 864 (2023)
[erratum: Eur. Phys. J. C \textbf{84}, no.1, 36 (2024)]
doi:10.1140/epjc/s10052-023-11889-x
[arXiv:2303.08533 [physics.acc-ph]].

\bibitem{Maltoni:2001dc}
F.~Maltoni, J.~M.~Niczyporuk and S.~Willenbrock,
Phys. Rev. D \textbf{65}, 033004 (2002)
doi:10.1103/PhysRevD.65.033004
[arXiv:hep-ph/0106281 [hep-ph]].

\bibitem{Allwicher:2021jkr}
L.~Allwicher, L.~Di Luzio, M.~Fedele, F.~Mescia and M.~Nardecchia,
Phys. Rev. D \textbf{104}, no.5, 055035 (2021)
doi:10.1103/PhysRevD.104.055035
[arXiv:2105.13981 [hep-ph]].

\bibitem{DiLuzio:2016sur}
L.~Di Luzio, J.~F.~Kamenik and M.~Nardecchia,
Eur. Phys. J. C \textbf{77}, no.1, 30 (2017)
doi:10.1140/epjc/s10052-017-4594-2
[arXiv:1604.05746 [hep-ph]].

\bibitem{Remmen:2019cyz}
G.~N.~Remmen and N.~L.~Rodd,
JHEP \textbf{12}, 032 (2019)
doi:10.1007/JHEP12(2019)032
[arXiv:1908.09845 [hep-ph]].

\bibitem{Remmen:2020uze}
G.~N.~Remmen and N.~L.~Rodd,
Phys. Rev. D \textbf{105}, no.3, 036006 (2022)
doi:10.1103/PhysRevD.105.036006
[arXiv:2010.04723 [hep-ph]].

\bibitem{Remmen:2022orj}
G.~N.~Remmen and N.~L.~Rodd,
JHEP \textbf{09}, 030 (2022)
doi:10.1007/JHEP09(2022)030
[arXiv:2206.13524 [hep-ph]].

\bibitem{Jenkins:2013zja}
E.~E.~Jenkins, A.~V.~Manohar and M.~Trott,
JHEP \textbf{10}, 087 (2013)
doi:10.1007/JHEP10(2013)087
[arXiv:1308.2627 [hep-ph]].

\bibitem{Jenkins:2013wua}
E.~E.~Jenkins, A.~V.~Manohar and M.~Trott,
JHEP \textbf{01}, 035 (2014)
doi:10.1007/JHEP01(2014)035
[arXiv:1310.4838 [hep-ph]].

\bibitem{Alonso:2013hga}
R.~Alonso, E.~E.~Jenkins, A.~V.~Manohar and M.~Trott,
JHEP \textbf{04}, 159 (2014)
doi:10.1007/JHEP04(2014)159
[arXiv:1312.2014 [hep-ph]].

\bibitem{Dawson:2020oco}
S.~Dawson, S.~Homiller and S.~D.~Lane,
Phys. Rev. D \textbf{102}, no.5, 055012 (2020)
doi:10.1103/PhysRevD.102.055012
[arXiv:2007.01296 [hep-ph]].

\bibitem{Aebischer:2020lsx}
J.~Aebischer and J.~Kumar,
JHEP \textbf{09}, 187 (2020)
doi:10.1007/JHEP09(2020)187
[arXiv:2005.12283 [hep-ph]].

\bibitem{Baratella:2020lzz}
P.~Baratella, C.~Fernandez and A.~Pomarol,
Nucl. Phys. B \textbf{959}, 115155 (2020)
doi:10.1016/j.nuclphysb.2020.115155
[arXiv:2005.07129 [hep-ph]].

\end{thebibliography}
\end{document}